\documentclass[12pt]{article}

\usepackage{amssymb,amsmath,amsfonts,eurosym,geometry,ulem,graphicx,caption,color,setspace,sectsty,comment,caption,pdflscape,subfigure,array,float}
\usepackage[round]{natbib}

\usepackage[english]{babel}	 
\usepackage{tabularx} 	     
\usepackage{longtable} 
\usepackage{multicol}	
\usepackage[bottom]{footmisc}
\usepackage[hidelinks]{hyperref}
\definecolor{navyblue}{rgb}{0.0,0.3576  ,  0.628}\definecolor{CustomColor}{rgb}{0.5,0.04,0.04}
\hypersetup{linkcolor=navyblue, citecolor=navyblue, urlcolor=navyblue, colorlinks=true}

\usepackage{multirow}
\usepackage{booktabs}
\usepackage{tablefootnote}	
 
\usepackage[utf8]{inputenc}

\usepackage[T1]{fontenc} 
\usepackage{amsthm}
\usepackage{comment}
\makeatletter
\renewcommand{\paragraph}{%
  \@startsection{paragraph}{4}%
  {\z@}{2ex \@plus 1ex \@minus .2ex}{-1em}%
  {\normalfont\normalsize\bfseries}%
}
\makeatother
\usepackage{soul}
\usepackage{titlesec}
\titlespacing*{\section}
{0pt}{2.5ex plus 1ex minus .2ex}{1ex plus .2ex}
\titlespacing*{\subsection}
{0pt}{2.5ex plus 1ex minus .2ex}{1ex plus .2ex}

\newtheorem{theorem}{Theorem}
\newtheorem{corollary}[theorem]{Corollary}
\newtheorem{proposition}{Proposition}
\newtheorem{lemma}[theorem]{Lemma}

\newcolumntype{L}[1]{>{\raggedright\let\newline\\arraybackslash\hspace{0pt}}m{#1}}
\newcolumntype{C}[1]{>{\centering\let\newline\\arraybackslash\hspace{0pt}}m{#1}}
\newcolumntype{R}[1]{>{\raggedleft\let\newline\\arraybackslash\hspace{0pt}}m{#1}}

\allowdisplaybreaks

\begin{document}

\title{Inflation - who cares? \\ Monetary Policy in Times of Low Attention}
\author{Oliver Pfäuti\thanks{Department of Economics, University of Bern, Schanzeneckstrasse 1, 3012 Bern, Germany, \href{mailto:pfaeuti.oliver@gmail.com}{\textcolor{navyblue}{pfaeuti.oliver@gmail.com}}.\\
I especially thank Klaus Adam for his invaluable support and guidance while working on this project. Further, I thank George-Marios Angeletos, Luca Benati, Chris Boehm, Antoine Camous, Antonio Ciccone, Oli Coibion, Laura G\'{a}ti, Cristina Griffa, Andreas Gulyas, Zhen Huo, Matthias Meier, Mykola Ryzhenkov (discussant), Fabian Seyrich, Efthymis Smyrniotis (discussant) and Andrei Zaloilo (discussant) as well as numerous seminar and conference participants. 
The author gratefully acknowledges financial support by the Deutsche Forschungsgemeinschaft (DFG, German Research Foundation) through CRC TR 224 (Project C02) and Stiftung Geld \& Währung. First version: April 2021. Declarations of interest: none.}}
\date{October 23, 2023\\\href{https://opfaeuti.github.io/website/Attention_OP_1.pdf}{\textcolor{navyblue}{Link to most recent version \vspace{-0.2cm}}}}
\maketitle
\begin{large}
\begin{centering}

\end{centering}
\end{large}

\begin{abstract}
I propose an approach to quantify \textit{attention to inflation} in the data and show that the decrease in the volatility and persistence of U.S.\ inflation after the Great Inflation period was accompanied by a decline in the public's attention to inflation. This decline in attention has important implications (positive and normative) for monetary policy as it renders managing inflation expectations more difficult and 
can lead to \textit{inflation-attention traps}: prolonged periods of a binding lower bound and low inflation due to slowly-adjusting inflation expectations. As attention declines the optimal policy response is to increase the inflation target. 
Accounting for the lower bound fundamentally changes the normative implications of declining attention. While lower attention raises welfare absent the lower-bound constraint, it decreases welfare when accounting for the lower bound. 
%
\\\vspace*{0.2cm}\\
\footnotesize{\textbf{JEL Codes:} E31, E52, E58, E71}

\end{abstract}

\clearpage
\newpage

\section{Introduction}
Managing inflation expectations is an important instrument for monetary policy. It offers a powerful substitute to conventional tools if the nominal interest rate is constrained by the lower bound. By making promises about the future conduct of its policy, the monetary authority can shape inflation expectations, and thus, steer the real interest rate even when the nominal rate is constrained. At least this is how it works in theory.\footnote{For optimal monetary policy under full-information rational expectations, see e.g., \citet{ClaridaGaliGertler1999}, \citet{woodford2003interest} or \citet{gali2015monetary}. See \citet{EggertssonWoodford2003}, \citet{adam2006optimal} or \citet{coibion2012optimal} for early analyses including the lower bound on nominal interest rates.} But while traditional analyses assume that agents are perfectly informed and have rational expectations, recent empirical evidence suggests that the general public is usually poorly informed about and inattentive to monetary policy and inflation.\footnote{\citet{candia2021inflation} and \citet{coibion2020average} show that U.S. firms as well as households are usually poorly informed about and quite inattentive to monetary policy. \citet{coibion2012can,coibion2015information}, for example, show that models of limited attention more closely align with empirical patterns of inflation expectations, compared to models with full-information rational expectations. In line with limited attention, \citet{d2020managing} show that forward guidance is quite ineffective in stimulating inflation expectations.} What do these low levels of attention imply for the conduct of monetary policy? And how has the public's attention to inflation changed over the last fifty years?

To tackle these questions, I propose an approach to quantify \textit{attention to inflation} in the data. This approach is based on a model of optimal attention choice subject to information acquisition costs. The result is a law of motion for inflation expectations in which attention governs how strongly agents update their expectations following an inflation surprise. The optimal degree of attention depends positively on how volatile and persistent inflation is. 

Using micro survey data of professional forecasters and consumers in the U.S., I then use this approach to estimate attention to inflation in the data and show that attention was very low in the years just before the Covid-19 crisis. In the 1970s and 1980s, on the other hand, attention to inflation was substantially higher. Consistent with the underlying model, times of higher inflation volatility and persistence are characterized by higher attention to inflation. 

How does this decline in attention matter for the conduct of optimal monetary policy? To answer this question, I solve for the Ramsey optimal monetary policy in a standard New Keynesian model augmented with an effective lower bound (ELB) on the nominal interest rate and with inflation expectations that are characterized by limited attention. Both of these ingredients matter greatly for optimal monetary policy and the normative implications of declining attention. Lower attention has a stabilizing effect on inflation expectations and actual inflation, resembling more anchored (short-run) expectations. These stabilization benefits imply that lower attention is welfare improving---if we do not account for the ELB (similar to the findings in \citet{paciello2014exogenous}). Accounting for the occasionally-binding ELB, however, completely overturns the normative implications of lower attention. Lower levels of attention lead to a decline in welfare when we account for the lower-bound constraint. The reason is that lower attention renders managing expectations more difficult, which is particularly relevant if the nominal interest rate is constrained by the lower bound.\footnote{\citet{coibion2019monetary} and \citet{coibion2020average} show that managing expectations by the central bank is indeed a difficult task and the effects of monetary policy are much smaller than in most theoretical models. \citet{d2020managing} find small effects of forward guidance on inflation expectations and durable consumption.}


To better understand these results, I first show that under sub-optimal policy, i.e., if monetary policy follows an \textit{ad-hoc} Taylor rule, limited attention can lead to substantially longer periods at the ELB. Even though lower levels of attention attenuate the initial response of inflation expectations to a given shock, the decline in expectations becomes more persistent which hinders actual inflation from recovering quickly. Due to the persistently-low inflation, the monetary authority keeps the interest rate at the ELB for longer. I refer to these periods of long spells at the ELB and persistent declines in inflation and inflation expectations as \textit{inflation-attention traps}. The response of the output gap, on the other hand, is very similar to the one under rational expectations. Thus, low attention offers a potential explanation for why inflation was relatively stable during the Great Recession but was persistently low during the subsequent recovery, seemingly disconnected from output (as documented in \citet{del2020s}).

When replacing the \textit{ad hoc} Taylor rule with the Ramsey optimal policy, I find that it is optimal to induce a higher average inflation rate to deal with these attention traps and the overall inability to manage inflation expectations at low levels of attention. With a higher average inflation rate, the nominal interest is also higher on average. This provides additional space when cutting the interest rate following adverse shocks, thus, mitigates the drawbacks of lower attention, and therefore helps to prevent long spells at the ELB.\footnote{Thus, the reason for the higher inflation rate is different from earlier papers that also consider an occasionally-binding ELB (e.g., \citet{adam2006optimal}, \citet{adam2022subjective}). In these papers, the higher average inflation rate arises due to promises the policymaker makes \textit{at the lower bound}. In the present paper, on the other hand, the higher inflation rate arises due to considerations \textit{before the lower bound} binds, \textit{foreseeing what will happen} at the lower bound.} Such an increase in the average inflation rate is not optimal if we abstract from the ELB, as then the nominal rate can go negative. Given the estimates of attention just before the Covid-19 crisis, the average inflation rate under Ramsey optimal policy is about 2-3 percentage points higher than under rational expectations. 
This increase in the level of inflation, however, is costly from a welfare perspective and it turns out that this level effect dominates the stabilization benefits. Thus, lower attention is welfare deteriorating when accounting for the ELB.


Another instrument to mitigate the drawbacks of limited attention are negative interest rate policies. Allowing for negative interest rates up to $-0.5\%$ (annualized) lowers the necessary increase in the optimal inflation target. As attention declines, however, the effectiveness of negative interest rate policies decreases and the optimal increase in the inflation target is close to the one without negative rates.

\paragraph{Related literature.}
\cite{bracha2023inflation} show that attention to inflation increases, when inflation increases. Using Google search data, \cite{korenok2022inflation} find that people's attention to inflation increases with inflation only after inflation exceeds a threshold of around 2-4\%. \cite{cavallo2017inflation} and \cite{weber2023tell} use randomized information treatments and show that attention of households and firms is higher in times of high inflation. \cite{link2023attention} also find that attention to inflation is higher in times of high and volatile inflation. My key contribution relative to these papers is that I provide estimates of attention in a way that directly maps into otherwise standard macroeconomic models. 

The empirical part of this paper is related to recent findings in \citet{jorgensen2019anchored} who show that inflation expectations have become more anchored over the last decades.\footnote{Similarly, \citet{gati2020monetary} documents that anchoring of long-run inflation expectations is time varying and has substantially increased recently.}  My measure of attention is inversely related to their definition of anchoring, but attention is concerned with short-run expectations whereas anchoring usually refers to the stabilization of long-run expectations. I complement their empirical analysis along several dimensions which I detail more closely in the empirical Section \ref{sec:data}. Additionally, I offer new insights in how stabilized expectations matter when nominal interest rates are constrained by a lower-bound constraint and what this implies for optimal monetary policy. 

My limited-attention model of inflation expectations is closely related to the general information choice problem in \citet{mackowiak2020rational}. In contrast to their model, and the rational inattention literature more generally, agents in my model have a perceived law of motion of inflation that can potentially differ from the actual law of motion.\footnote{\citet{mackowiak2020rational} provide a recent overview of this literature, which was inspired by the seminal paper \citet{sims2003implications}. For further developments in this literature, see, among others, \citet{mackowiak2009optimal}, \citet{paciello2014exogenous}, \citet{mackowiak2018dynamic}, \citet{afrouzi2021dynamic}, and see \citet{gabaix2019behavioral} for an overview of \textit{behavioral} inattention.} The reduced-form of the model that I bring to the data is close to the one in \citet{vellekoop2019inflation}. In contrast to their paper, I focus on how attention changed over the last fifty years and show that attention tends to be higher in times of volatile and persistent inflation.

\citet{ball2005monetary}, \citet{adam2007optimal}, \citet{paciello2014exogenous} and \citet{gati2020monetary} characterize optimal monetary policy in models with different forms of limited attention. I contribute to this literature by allowing for an occasionally-binding lower-bound constraint. I show that accounting for the lower bound can lead to \textit{qualitatively} different welfare implications due to changes in the optimal \textit{level} of inflation.
\citet{wiederholt2015empirical} and \citet{gabaix2020behavioral} examine how information rigidities and inattention matter at the zero lower bound. \citet{angeletos2018forward} study the implications of relaxing the common knowledge assumption for forward guidance and show that the effects of forward guidance are attenuated in such a setting. My paper complements these three papers by studying the Ramsey optimal policy in a fully stochastic setup and focuses on the implications for the optimal inflation target. To the best of my knowledge, the present paper is the first to study the trade off of lower attention in a fully stochastic model with an occasionally binding lower-bound constraint and to characterize the Ramsey optimal monetary policy in such a setting. 


\paragraph{Outline.}
The rest of the paper is structured as follows. The empirical strategy to quantify attention, the description of the data and the empirical results are presented in Section \ref{sec:data}. In Section \ref{sec:nkm}, I show how limited attention can lead to inflation-attention traps, before I then study optimal policy in Section \ref{sec:omp}. Section \ref{sec:conclusion} concludes. An online appendix provides all the derivations, several extensions and robustness checks.

\section{Quantifying Attention}\label{sec:data}
In this section, I derive an expectations-formation process under limited attention that provides a straightforward approach to measure attention to inflation empirically. The model is an application of \citet{mackowiak2020rational}, who study a general problem of optimal information acquisition. I relegate all the details and derivations to appendix A.


The main difference to \citet{mackowiak2020rational} is that agents in my model do not exactly know the underlying process of inflation but have a simplified view of how inflation evolves.\footnote{In the monetary model, later on, I will disentangle the effects from this potentially misperceived law of motion from the assumption of costly attention and show that most of the results are driven by limited attention and not the misperceived law of motion (see section \ref{sec:full_attention}).}
In particular, the agent believes that (demeaned) inflation tomorrow, $\pi'$, depends on (demeaned) inflation today, $\pi$, as follows
\begin{equation}
\pi' =   \rho_{\pi} \pi + \nu, \notag
\end{equation}
where $\rho_{\pi}\in[0,1]$ denotes the perceived persistence of inflation and $\nu \sim i.i.N.(0,\sigma^2_{\nu})$. This assumption is supported by empirical evidence (see e.g., \citet{faust2013forecasting} or \citet{canova2007g}).\footnote{\citet{fulton2021forecasting} show that simple models such as AR(1) models are hard to beat when forecasting inflation in real time.} Note, that the perceived volatility and persistence do not need to be the same as their actual counterparts, consistent with the empirical evidence on inflation expectations (see Table 7 in the Appendix). The agent wants to minimize her expected forecast error but inflation in the current period is unobservable and acquiring and processing information is costly. The agent thus faces a trade off how attentive she wants to be. 
I follow the literature on rational inattention and assume that the loss arising from making mistakes in her forecasts is quadratic with a scaling factor $r$, and the cost of information acquisition and processing is linear in mutual information with a scaling parameter $\lambda$. 

In this setup and with a normal prior, the optimal signal takes the form
\begin{equation}
s = \pi+\varepsilon, \notag
\end{equation}
with $\varepsilon\sim i.i.N.(0,\sigma^2_{\varepsilon})$ (see \citet{matvejka2015rational}).

The optimal forecast is given by $\pi^e = \rho_{\pi} E\left[\pi|s\right]$, and Bayesian updating implies
\begin{equation}
\pi^e = \rho_{\pi}\left(1-\gamma\right)\hat{\pi}+\rho_{\pi}\gamma s,\label{updating_theo1}
\end{equation}
where $\gamma = 1-\frac{\sigma^2_{\pi|s}}{\sigma^2_{\pi}}\in[0,1]$ measures how much attention the agent pays to inflation, and $\hat{\pi}$ denotes the prior mean of $\pi$.

Solving for the optimal $\gamma$ and writing the cost of information relative to the stakes, $\tilde{\lambda}\equiv\frac{\lambda}{r}$, yields the \textit{optimal} level of attention, summarized in the following Lemma.

\begin{lemma}\label{lemma_optgain1}
The optimal level of attention is given by
\begin{equation}
\gamma = max\left(0,1-\frac{\tilde{\lambda}}{2\rho_{\pi}^2\sigma^2_{\pi}}\right), \label{opt_gamma1}
\end{equation}
which shows that the optimal level of attention $\gamma$ is
\begin{itemize}
\item[(i)] decreasing in the relative cost of information acquisition, $\tilde{\lambda} \equiv \frac{\lambda}{r}$,
\item[(ii)] increasing in inflation volatility, $\sigma_{\pi}$, and
\item[(iii)] increasing in inflation persistence, $\rho_{\pi}$.
\end{itemize}
\end{lemma}

From Lemma \ref{lemma_optgain1}, we see that aside from the relative information cost, $\tilde{\lambda}$, the persistence, $\rho_{\pi}$, and the volatility of inflation, $\sigma_{\pi}$, are crucial drivers of attention. The model predicts a positive relationship between attention and $\sigma_{\pi}$, as well as between attention and $\rho_{\pi}$. In the following, I will first estimate attention $\gamma$, asses how it has changed over time and then test whether there is indeed evidence for these positive relations.

\subsection{Bringing the Model to the Data}
To estimate attention in the data, I extend the law of motion of inflation expectations, equation \eqref{updating_theo1}, to a dynamic setup.
The agent believes that inflation $\pi$ follows 
\begin{equation}
\pi_{t} = (1-\rho_{\pi})\bar{\pi}+\rho_{\pi}\pi_{t-1}+\nu_t, \notag
\end{equation}
where $\bar{\pi}$ is the agent's long-run belief about inflation and $\rho_{\pi}$ is the perceived persistence of inflation. I assume that the error term $\nu_t$ is normally distributed with mean zero and variance $\sigma^2_{\nu}$.\footnote{In Appendix B.1, I discuss the case in which the agent believes that inflation follows an AR(2) process.}

The agent receives a signal about inflation of the form
\begin{equation}
s_{it} = \pi_t + \varepsilon_{it}, \notag
\end{equation}
where the noise $\varepsilon_{it}$ is assumed to be normally distributed with variance $\sigma^2_{\varepsilon}$.

Given these assumptions, it follows from the (steady state) Kalman filter that optimal updating is given by
\begin{equation}
\pi^e_{t+1|t,i} = (1-\rho_{\pi})\bar{\pi}+\rho_{\pi}\pi^e_{t|t-1,i} + \rho_{\pi}\gamma\left(\pi_t - \pi^e_{t|t-1,i}\right)+u_{i,t}, \label{updating}
\end{equation}
where the updating gain $\gamma$ captures the agent's level of attention, and $u_{i,t} = \rho_{\pi}\gamma \varepsilon_{i,t}$ is a scaled version of the i.i.d.\ noise term $\varepsilon_{i,t}$.
From equation \eqref{updating}, we observe that lower attention, i.e., a lower $\gamma$, implies that the agent updates her expectations to a given forecast error, $(\pi_t - \pi^e_{t|t-1,i})$, less strongly. Lower attention is reflected in more noisy signals, and more noise means the agent trusts her received signals less and thus, puts less weight on these signals. Hence, her expectations remain more strongly anchored at her prior beliefs.

In the estimation of equation \eqref{updating}, I allow for individual-specific intercepts. This can either reflect a mean bias in the perceived inflation rate, $\bar{\pi}_i \neq \bar{\pi}$, or that the agent believes her signals are biased on average, as in \citet{vellekoop2019inflation}.

\subsection{Data}
I focus on the Survey of Professional Forecasters (SPF) from the Federal Reserve Bank of Philadelphia, as well as the Survey of Consumers from the University of Michigan (SoC). In the Appendix, I show that the findings extend to other data sets as well. For the SPF, I consider individual and aggregate forecasts. The main focus is on expectations about the quarter-on-quarter percentage change in the GDP deflator, which is available since 1969. I drop forecasters for which I have less than eight observations. As a robustness check, I will show that the results are robust to using expectations about the consumer price index, CPI. This data series, however, is only available since 1979.

While the SPF provides data on expectations about the next quarter, the SoC only provides one-year-ahead expectations. Therefore, I will compare them to the actual year-on-year changes in the CPI.\footnote{The question in the SoC is not explicitly about the CPI but about ``prices''. Using GDP deflator inflation instead of CPI inflation barely affects the results.}
As the SoC does not have a panel dimension, I consider average (and median) expectations. Additionally, I estimate attention using the Survey of Consumer Expectations from the Federal Reserve Bank of New York (SCE). The SCE, launched in 2013, has a panel structure and thus allows me to estimate attention using individual-consumer data, at least for the period after 2013. Data on actual inflation comes from the FRED database from the Federal Reserve Bank of St. Louis. Throughout, I focus on the pre-Covid period and end the sample in 2019Q4.\footnote{In a follow-up paper, I study the implications of changes in people's attention to inflation in driving high and persistent inflation and show that this offers a potential explanation for the inflation surge that occurred after the Covid-19 pandemic \citep{pfauti2023inflation}.} Appendix B provides summary statistics and plots the discussed time series.

\citet{jorgensen2019anchored} estimate how strongly anchored inflation expectations in the U.S. are and how this changed over the last fifty years. I extend their empirical strategy in several dimensions.  First, I allow the persistence of perceived inflation to change over time and do not restrict it to follow a random walk. Second, I show that not only aggregate professional forecasters' expectations have become more anchored, but also consider individual-specific expectations, as well as consumers' inflation expectations. Third, I do not impose the structure of the New Keynesian Phillips Curve on the data but directly estimate attention simply based on the proposed law of motion for inflation expectations. That said, we will see that my results are consistent with the results in \citet{jorgensen2019anchored}.

\subsection{Estimation Results}

Before estimating attention,
I rewrite the updating equation \eqref{updating} as
\begin{equation}
\pi^e_{t+1|t,i} = \beta_i + \beta_1 \pi^e_{t|t-1,i} +\beta_2 \left(\pi_t - \pi^e_{t|t-1,i}\right)+u_{i,t}, \label{reg1}
\end{equation}
where $\beta_i = (1-\rho_{\pi})\bar{\pi}_i$, $\beta_1 = \rho_{\pi}$ and $\tfrac{\beta_2}{\beta_1} = \gamma$. For the SPF, where I can use individual forecasts, I estimate \eqref{reg1} using a forecaster-fixed-effects regression. Since the dependent variable shows up with a lag on the right-hand side, however, a standard fixed-effects regression introduces a bias \citep{nickell1981biases}. Therefore, I apply the estimator proposed by \citet{blundell1998initial} (BB for short) and I use all available lags of the dependent variable as instruments.\footnote{Appendix B.1 shows that the results are robust to using fewer lags.} All reported standard errors are robust with respect to heteroskedasticity and serial correlation. As an alternative to the BB estimator, I also estimate \eqref{reg1} using pooled OLS. For the Survey of Consumers I cannot include individual fixed effects as I use average and median expectations and thus, I apply the Newey-West estimator using four lags (\citet{newey1987simple}).

To examine how attention changed over time, I run regression \eqref{reg1} for the period before and after 1990, separately. The results are robust to different split points (see Appendix B.1). Later on, I will estimate \eqref{reg1} using rolling-windows of ten years each and show that the general patterns I document are robust.

\begin{table}[ht]

\caption{Regression Results of Equation \eqref{reg1}}
\centering 
\vspace{-0.2cm}

\begin{tabular}{lcccc}
\hline\hline\vspace{-0.45cm}\\
 &  \multicolumn{2}{c}{Professional Forecasters} &  \multicolumn{2}{c}{Consumers}  \\ \cline{2-3} \cline{4-5}\\\vspace*{-0.9cm}\\
 & \multicolumn{1}{c}{Blundell Bond} & \multicolumn{1}{c}{Pooled OLS} & \multicolumn{1}{c}{Averages} & \multicolumn{1}{c}{Median} \\
 \cline{2-2} \cline{3-3} \cline{4-4} \cline{5-5}\\\vspace*{-0.9cm}\\
$\widehat{\gamma}_{pre}$ &0.70   & 0.44  & 0.75  & 0.43   \\
s.e. & (0.1005)  & (0.0397)  & (0.1574)  & (0.0970)  \\
$\widehat{\gamma}_{post}$ &0.41  & 0.22  & 0.31  & 0.24   \\
s.e. & (0.0522)  & (0.0290)  & (0.0881)  & (0.0601)  \\
\hline\\\vspace*{-0.9cm}\\
$N$ &3566 & 3566 & 120 & 120
\\\hline\hline
\end{tabular}%
\label{tab:reg1}
\vspace{0.2cm} \\\begin{minipage}{1\textwidth}

 \footnotesize{
Note: This table shows the results from regression \eqref{reg1} for professional forecasters (SPF) as well as for consumers. For the SPF, I use the \citet{blundell1998initial} (BB) estimator (first two columns), as well as pooled OLS (columns 3-4). For the Survey of Consumer, I consider average expectations (columns 5-6) and median expectations (columns 7-8). $\widehat{\gamma}_{pre}$ and $\widehat{\gamma}_{post}$ denote the estimated attention parameters for the period pre 1990 and post 1990, respectively. The standard errors, reported in parentheses, are robust with respect to heteroskedasticity and serial correlation.
 }%
 \end{minipage}
\end{table}
%
%
%
Table \ref{tab:reg1} shows the results. Here, $\widehat{\gamma}_{pre}$ and $\widehat{\gamma}_{post}$ denote the estimated attention parameters for the period pre 1990 and post 1990, respectively. We see that attention is substantially lower after 1990 compared to the period before 1990.\footnote{I test the validity of the instruments in the Blundell-Bond estimation by testing for autocorrelation of order one and two in the first-differenced error terms. The respective $p$-values are 0.000 (order 1) and 0.973 (order 2) for the period before 1990 and 0.000 (order 1) and 0.737 (order 2) for the period after 1990. This indicates that the instruments used in the estimation are valid.} This is true for professional forecasters and for consumers, and as I show in  Appendix B.1, also for other datasets. The point estimates after 1990 are basically half of what they were before the 1990s and these differences are statistically significant at the 1\% level. 

The decline in attention is even more pronounced when focusing on the most recent decade. To show this, I run regression \eqref{reg1} for the period between 2010 and 2020. Additionally, I also use data from the New York Fed Survey of Consumer Expectations, starting in 2013. The advantage of this survey compared to the Michigan Survey is that it surveys the same consumers up to twelve times in a row, providing a much larger sample size and a panel dimension. Table \ref{tab:reg12010} shows the results. We see that overall, attention declined substantially compared to earlier periods and is between 0.04 and 0.17 during this period of low and stable inflation. Furthermore, the results from the SCE lie in the same ballpark as the ones from the Michigan Survey, which indicates that using average (or median) consumer expectations does not fundamentally affect the results. In fact, the estimated attention parameter for the average expectations from the Michigan Survey is 0.04 when restricting the sample to 2013-2020, which is exactly the same as the estimate obtained from the New York Fed Survey.

\begin{table}[ht]

\caption{Attention since 2010}
\centering 
\vspace{-0.2cm}

\begin{tabular}{lccccc}
\hline\hline\vspace{-0.45cm}\\
 &  \multicolumn{2}{c}{Professional Forecasters} &  \multicolumn{2}{c}{Consumers} &   \multicolumn{1}{c}{NY Fed Survey}  \\ \cline{2-3} \cline{4-5} \cline{6-6}\\\vspace*{-0.9cm}\\
 & Blundell Bond & Pooled OLS & Averages & Median &  Pooled OLS \\
\hline\\\vspace*{-0.9cm}\\
$\widehat{\gamma}$ & 0.17 &  0.07 & 0.12 & 0.09 &0.04 \\
s.e. & (0.0729) & (0.0333) & (0.0658) & (0.0616) & (0.0316)\\
\hline\\\vspace*{-0.9cm}\\
$N$ & 1322 &  1322 &40 & 40 &  74229
\\\hline\hline
\end{tabular}%
\label{tab:reg12010}\vspace{0.2cm}
 \begin{minipage}{1\textwidth}

 \footnotesize{
Note: This table shows the results from regression \eqref{reg1} for the period between 2010 and 2020 for professional forecasters (SPF) as well as for consumers. For the SPF, I use the \citet{blundell1998initial} (BB) estimator (first column), as well as pooled OLS (column 2). For the Survey of Consumer, I consider average expectations (column 3) and median expectations (columns 4). The standard errors, reported in parentheses, are robust with respect to heteroskedasticity and serial correlation. Additonally, column 5 shows the results for consumer inflation expectations from the New York Fed Survey of Consumer Expectations.
 }%
 \end{minipage}
\end{table}

\subsubsection{When is Attention High?}
Lemma \ref{lemma_optgain1} states that two key drivers of attention are inflation volatility and inflation persistence.
To examine these relationships empirically, I estimate regression \eqref{reg1}, using a rolling-window approach in which every window is 10 years long, and estimate one attention parameter for every window, $\widehat{\gamma}_t$, as well as the period-specific inflation volatility, $\widehat{\sigma}_{\pi,t}$ and the persistence parameter, $\widehat{\rho}_{\pi,t}$. In particular, I use the window-specific standard deviation of inflation as my measure of $\widehat{\sigma}_{\pi,t}$ and the first-order autocorrelation of inflation for $\widehat{\rho}_{\pi,t}$. Appendix B.1 shows that the following results also hold for different window lengths or when using the standard deviation and persistence of \textit{expected} inflation for $\widehat{\sigma}_{\pi,t}$ and $\widehat{\rho}_{\pi,t}$.

\begin{figure}[ht]
\caption{Attention, Inflation Volatility and Inflation Persistence}
\centering    
\begin{tabular}{cc}  
(a) Inflation Volatility & (b) Inflation Persistence\\
\includegraphics[scale=0.35]{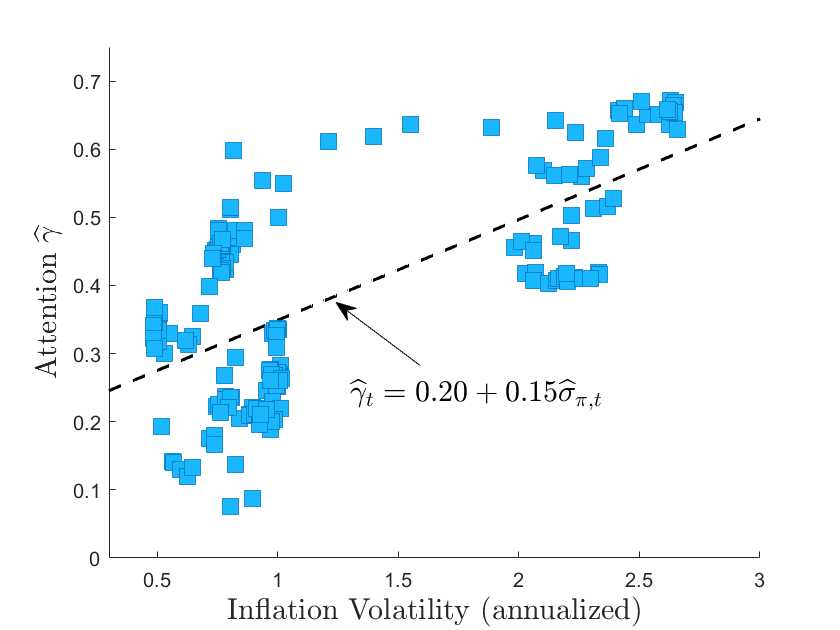}  &
\includegraphics[scale=0.35]{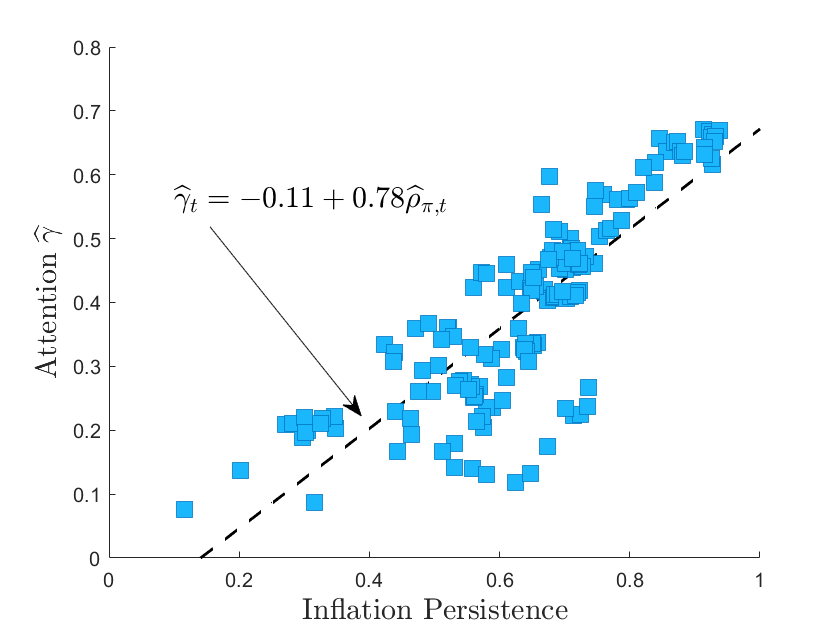}  
 \end{tabular}%
\\ \vspace{0.2cm}
 \begin{minipage}{\textwidth}
 \footnotesize{
 Notes: The right panel shows the relationship between inflation volatility and the estimated attention parameter, $\widehat{\gamma}$. The left panel shows the relationship between the persistence of inflation and the estimated attention parameter. Both panels report the results for individual professional forecasters where the attention parameter was estimated via pooled OLS.}
 \end{minipage}
\label{fig:attvola}
\end{figure}

Figure \ref{fig:attvola} summarizes the results graphically. The left scatterplot shows the inflation volatility on the horizontal axis and the estimated attention parameter, $\widehat{\gamma}$, on the vertical axis. The attention parameters shown in the figure are the ones for individual professional forecasters, obtained via pooled OLS. The right panel shows the relationship between attention and inflation persistence (on the horizontal axis). In both cases, we see that there is a clear positive relationship, just as the limited-attention model predicts.

To check if these findings are statistically significant, I regress attention on inflation volatility or on inflation persistence as follows
\begin{align}
\widehat{\gamma}_t &= \alpha_1+\beta \widehat{\sigma}_{\pi,t}+u_t \label{indgain} \\
\widehat{\gamma}_t &= \alpha_2+\zeta \widehat{\rho}_{\pi,t}+v_t. \label{indgain_rho}
\end{align}

\begin{table}[h!]

\caption{Attention, Inflation Volatility and Inflation Persistence}
\centering 
\vspace{-0.2cm}

\begin{tabular}{lcccc}
\hline\hline\vspace{-0.45cm}\\
 &  \multicolumn{2}{c}{Professional Forecasters} & &\multicolumn{1}{c}{Consumers}  \\ \cline{2-3} \cline{5-5} \\\vspace*{-0.9cm}\\
Estimator & Blundell-Bond & Pooled OLS && OLS \\
\hline\\\vspace*{-0.9cm}\\
$\widehat{\beta}$ & $0.13^{***}$ &  $0.15^{***}$ & &$0.09^{***}$ \\
s.e. & (0.0155) &  (0.0105) & &(0.0272) \\
$\widehat{\zeta}$ & $0.71^{***}$ & $0.78^{***}$ && $0.56^{***}$\\
s.e. & (0.0568) & (0.0349) & &(0.0714)\\
$N$ & 165 & 165 && 163
\\\hline\hline
\end{tabular}%
\label{tab:kg_vola}\vspace{0.2cm}
 \begin{minipage}{1\textwidth}

 \footnotesize{
Note: This table shows the results of regression \eqref{indgain} and \eqref{indgain_rho}. Standard errors are robust with respect to heteroskedasticity.  $ ^{***}:$ $p$-value $<$ 0.01, $ ^{**}:$ $p$-value $<$ 0.05, $ ^{*}:$ $p$-value $<$ 0.1.
 }%
 \end{minipage}
\end{table}

Table \ref{tab:kg_vola} reports the results. Standard errors are robust with respect to heteroskedasticity.
We see that the observed patterns in Figure \ref{fig:attvola} are indeed statistically significant. Attention to inflation is positively correlated with inflation volatility and inflation persistence, as Lemma \ref{lemma_optgain1} predicts. This is true for professional forecasters as well as for consumers. 
Overall, the magnitudes of the estimates indicate that the results are somewhat stronger for professional forecasters. Appendix B.1 shows that these results hold when regressing attention on inflation volatility and persistence jointly. I also show that the results are robust when controlling for the \textit{average level} of inflation and the average level of inflation has no significantly-positive effect on attention when controlling for volatility and persistence. 
Given the decline in inflation volatility and inflation persistence over the last fifty years (see Table 7 in Appendix B), the positive correlation with attention supports the findings in Table \ref{tab:reg1}: attention declined as the volatility and persistence of inflation declined.\footnote{\citet{benati2008investigating} documents a decline in inflation persistence in advanced economies, especially for countries that introduced inflation targeting regimes.}
\paragraph{Additional evidence and robustness.}
In Appendix B.1, I show that the presented results are robust to using different data sources, different sample splits, different specifications of the BB estimator, allowing for time-fixed effects to account, for example, for varying trend inflation, as well as constructing a quasi panel of consumers, based on their income. I further discuss the case in which the agent believes that inflation follows an AR(2) process. I again find that attention has declined substantially after the 1990s and that the additional coefficients showing up due to the AR(2) rather than the AR(1) tend to be insignificant.  In Appendix B.2, I document a decline in news coverage of inflation in popular news papers as well as in books, thus, providing additional, complementary evidence on the decline in attention after the 1980s.\footnote{\citet{carroll2003macroeconomic} proposes a micro-foundation of sticky information models (as in \citet{mankiw2002sticky}) that relies on news coverage of inflation. \citet{lamla2014role} and \citet{pfajfar2013news} test these predictions empirically.} Additionally, I show that another measure of attention (based on survey respondents answering "I don't know" when asked about their inflation expectations) is strongly correlated with my proposed measure of attention. Finally, I provide additional evidence favoring the proposed attention model compared to a setup in which the agent cannot distinguish between a trend and a cyclical component with time-varying volatilities of these two components. In that case, if the trend component's contribution to overall inflation increases, the agent's forecast would become more responsive to current inflation, too. I show that professional forecasters' \textit{nowcasts of inflation} are more accurate in times of high inflation volatility and persistence, which supports the proposed attention model rather than this alternative model.

\section{Monetary Policy Implications of Limited Attention}\label{sec:nkm}
How does the decline in attention to inflation affect the conduct of monetary policy? 
To answer this question, I augment the standard New Keynesian model with inflation expectations that are characterized by limited attention, and a lower-bound constraint on the nominal interest rate (model details and all derivations are in Appendix C). 
I build on the standard New Keynesian model without capital, with rigid prices in the spirit of \cite{Rotemberg1982} and  with a lower bound on the nominal interest rate. The government pays a subsidy to intermediate-goods producers to eliminate steady state distortions arising from market power. I focus on the case with zero-steady state inflation as my baseline case, but discuss the case of positive trend inflation in Section \ref{sec:nkpc_extensions}.

The linearized model consists of an aggregate supply equation, the \textit{New Keynesian Phillips Curve}, and an aggregate Euler (or IS) equation:
\begin{align}
\pi_t &= \beta  \pi^e_{t+1|t} + \kappa y^{gap}_t + u_t, \label{AS} \\
y^{gap}_t &= E_t y^{gap}_{t+1} - \varphi\left(i_t-\pi^e_{t+1|t}-r^n_t\right), \label{AD}
\end{align}
where $\kappa$ measures the sensitivity of aggregate inflation to changes in the output gap, $y^{gap}_t$, $\beta\in(0,1)$ denotes the time discount factor of the representative household, and $u_t$ are cost-push shocks, following an AR(1) process with persistence $\rho_u\in[0,1]$ and innovations $\varepsilon^u\sim i.i.N.(0,\sigma^2_u)$. The output gap is the log deviation of output from its efficient counterpart that would prevail under flexible prices. Altogether, equation \eqref{AS} summarizes the aggregate supply side of the economy.
Equation \eqref{AD}, together with monetary policy, determines aggregate demand in this model. Here, $\varphi > 0$ measures the  real rate elasticity of output, $i_t$ is the nominal interest rate which is set by the monetary authority, and $r^n_t$ is the natural interest rate. The natural interest rate is the real rate that prevails in the economy with fully flexible prices and is exogenous. It follows an AR(1) process with persistence $\rho_r\in[0,1]$ and innovations $\varepsilon^r\sim i.i.N.(0,\sigma^2_r),$ independent of $\varepsilon^u$. The nominal interest rate and the natural rate are both expressed in absolute deviations of their respective steady state values, $\bar{i}$ and $\bar{r}^n$.
$E_t$ denotes the full-information rational expectations operator. I relax the assumption that agents have rational expectations about the output gap in section \ref{sec:extensions} and in appendix E.1.

Inflation expectations are characterized by limited attention and are given by
\begin{equation}
\pi^e_{t+1|t} = (1-\rho_{\pi})\bar{\pi}+\rho_{\pi}\pi^e_{t|t-1}+\rho_{\pi}\gamma\left(\pi_t-\pi^e_{t|t-1}\right), \label{infexp}
\end{equation}
where the notation is the same as in Section \ref{sec:data}. Given the representative agent assumption, I abstract from noise shocks in \eqref{infexp}. The law of motion of inflation expectations, equation \eqref{infexp}, is driven by two main assumptions. First, the perceived law of motion follows an AR(1) process and second, paying attention is costly. In Section \ref{sec:full_attention}, I will disentangle the two and show that it is mainly the second assumption that drives the implications for optimal monetary policy.
For the most part, I will focus on $\rho_{\pi} = 1$ in which case average inflation expectations align with actual average inflation and long-run beliefs $\bar{\pi}$ are irrelevant. I discuss the case with $\rho_{\pi} < 1$ in Appendix E.4. For empirically-realistic values of $\rho_{\pi}$ the results are very similar to the case with $\rho_{\pi} = 1$.
This belief formation process is empirically plausible, in the sense that it is consistent with recent empirical findings, documented in \citet{angeletos2021imperfect}: after a shock, expectations initially underreact, followed by a delayed overreaction (see Appendix E.3).

As is standard in the rational inattention literature, I assume that the attention parameter $\gamma$ is constant.\footnote{See, e.g., \cite{mackowiak2009optimal}, \citet{mackowiak2018dynamic,mackowiak2020rational}.} The usual assumption to obtain this is that in period $t=0$ the agent chooses her level of attention and then obtains all future signals at this point. This leaves conditional second moments time-invariant and thus, the optimal level of attention constant. I will, however, compare economies with different levels of attention. In Section \ref{app:timevaryingattention}, I discuss the case in which attention to inflation is time varying.
\begin{table}[h]
\caption{Model Parameterization}
\label{tab_calib}\centering
\begin{tabular}{lll}
\hline \hline
Parameter & Value & Source/Target \\ \hline
\multicolumn{3}{c}{\rule{0pt}{0.4cm}\emph{Preferences and technology}} \\ 
$\beta $ & 0.9975 & Average natural rate of $1\%$ \\ 
$\varphi $ & $1$ & \citet{adam2006optimal} \\ 
$\kappa$ & $0.057$ & \citet{adam2006optimal} \\ 
\multicolumn{3}{c}{\rule{0pt}{0.4cm}\emph{Exogenous shock processes}} \\ 
$\rho_{r}$ & 0.8 & \citet{adam2006optimal} \\ 
$\sigma_{r}$ & 0.2940\%  & \citet{adam2006optimal} \\ 
$\rho_{u}$ & 0 & \citet{adam2006optimal} \\ 
$\sigma_{u}$ & 0.154\%  & \citet{adam2006optimal}
\\ 
\hline \hline
\end{tabular}%
\end{table}

\paragraph{Calibration.}
I calibrate the model to quarterly frequency. I assume an annualized steady state natural rate of 1\%.
The rest of the calibration is taken from \citet{adam2006optimal}. Table \ref{tab_calib} summarizes the calibration. The attention parameter $\gamma$ will be varied to understand its role for monetary policy.

\paragraph{Attention and the Phillips Curve.} 
Inflation expectations are a crucial driver of actual inflation through the supply side of the economy, and thus, changing levels of attention affect the Phillips Curve. The following Proposition summarizes these effects.
\begin{proposition}\label{prop:pc}
The New Keynesian Phillips Curve under limited attention is given by
\begin{equation}
\pi_t = \frac{\beta(1-\gamma)}{1-\beta\gamma}\pi^e_{t|t-1} + \frac{\kappa}{1-\beta\gamma}y^{gap}_t+\frac{1}{1-\beta\gamma}u_t. \notag
\end{equation}
\end{proposition}
\begin{proof}
See Appendix D.2.
\end{proof}

From Proposition \ref{prop:pc}, we see that for a given prior inflation expectation, $\pi^e_{t|t-1}$, and a given realization of the exogenous shock $u_t$, inflation becomes less sensitive to changes in the output gap at lower levels of attention. Put differently, a decrease in attention $\gamma$ resembles a flatter Phillips Curve. We also see that for a given output gap, inflation becomes less sensitive to cost-push shocks $u_t$ at lower levels of attention.

Thus, Proposition \ref{prop:pc} captures the stabilizing effects of lower attention. As attention declines, firms' inflation expectations react less to changes in actual inflation. Through the Phillips Curve, this muted reaction of expectations in turn stabilizes inflation itself.
What cannot be seen from Proposition \ref{prop:pc}, but what will be crucial in the subsequent analysis, is that lower attention not only affects the initial response of inflation and inflation expectations but the dynamics as well. Changes in inflation and inflation expectations become more persistent at low levels of attention, even though the initial response is muted. 
I now show in a numerical example that at low levels of attention the economy can get stuck in an \textit{inflation-attention trap}.\footnote{In Appendix D, I show analytically how lower levels of attention mute the effects of forward guidance.}

\subsection{Inflation-Attention Traps} 
To close the model from Section \ref{sec:nkm}, I assume for now that, away from the lower bound, the monetary authority sets the nominal interest rate according to a Taylor rule
\begin{equation}
\tilde{i}_t = \rho_i \tilde{i}_{t-1} + (1-\rho_i)\left(\phi_{\pi}\pi_t+\phi_y y^{gap}_t\right), \label{taylor}
\end{equation}
where $\rho_i\in [0,1)$ captures interest rate smoothing, $\phi_{\pi} > 1$ and $\phi_y \geq 0$ denote the reaction coefficients to inflation and the output gap, respectively. The actual interest rate, $i_t$, however is constrained by the lower bound
\begin{equation}
i_t = max\{\tilde{i}_t, -\bar{i}\},
\end{equation}
where I set the lower bound (in levels) to zero.

I set the persistence parameter of the nominal interest rate to 0.7, and the reaction coefficients $\phi_{\pi}=2$ and $\phi_y = 0.5$, as in \citet{Andrade_etal2019}. In Appendix E.2, I show that the exact specification of the Taylor rule is inconsequential for the following results.

\begin{figure}[ht]
\caption{Impulse Response Functions to a Negative Natural Rate Shock}
\centering 
\begin{tabular}{cccc}  
\multicolumn{2}{c}{\includegraphics[scale=0.35]{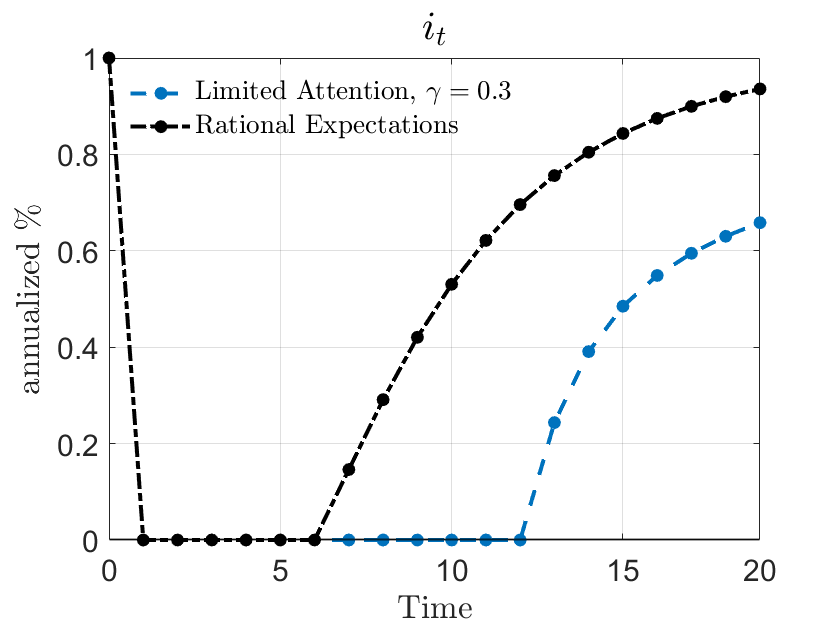}} & \multicolumn{2}{c}{\includegraphics[scale=0.35]{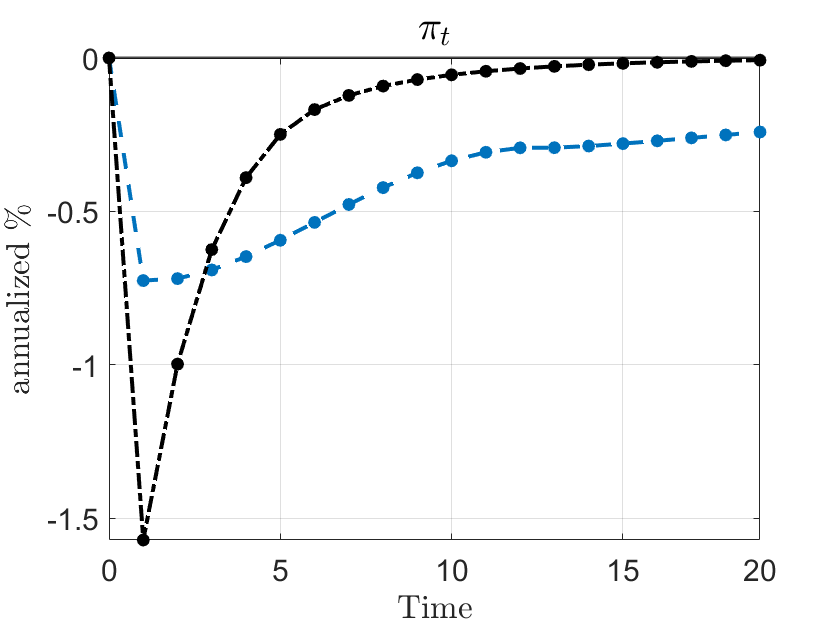}} 
\\ \multicolumn{2}{c}{\includegraphics[scale=0.35]{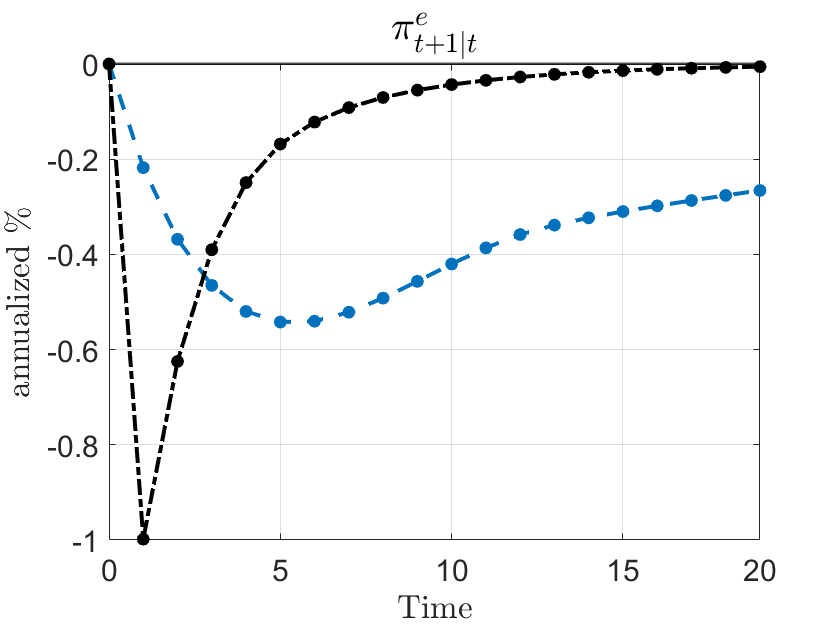}}
&  \multicolumn{2}{c}{ \includegraphics[scale=0.35]{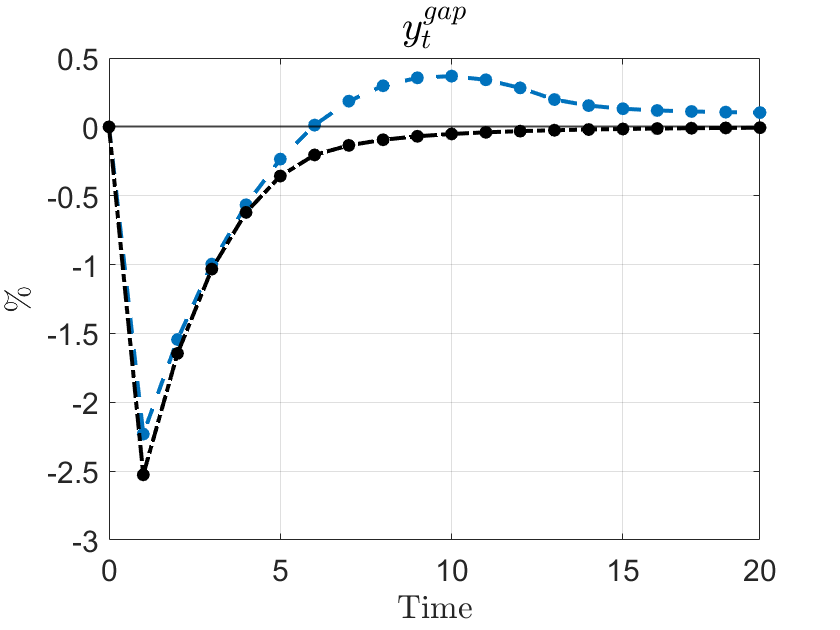}}
 \end{tabular}%
 \label{fig:irf_rn}
 \par
\vspace{0.3cm}
 \begin{minipage}{1\textwidth}

 \footnotesize{
Note: This figure shows the impulse-response functions of the nominal interest rate (upper-left panel), inflation (upper-right panel), inflation expectations (lower-left) and the output gap (lower-right) to a negative natural rate shock of three standard deviations. The blue-dashed lines show the case for the limited-attention model and the black-dashed-dotted lines for the rational expectations model. Everything is in terms of percentage deviations from the respective steady state levels, except the nominal rate is in levels.
 }%
 \end{minipage}

\end{figure}

Figure \ref{fig:irf_rn} plots the impulse response functions of the model's main variables to a negative natural rate shock of three standard deviations that pushes the nominal interest rate to the lower bound. The black-dashed-dotted lines are the IRFs in the model under FIRE and the blue-dashed lines are the ones under limited attention for the case $\gamma = 0.3$.

In both cases, the shock is large enough to push the economy to the lower bound. While the reaction of the output gap is very similar in both economies, the responses of inflation and inflation expectations are strikingly different. Initially, the muted response of inflation expectations to the adverse shock is reflected in a smaller downturn of inflation itself under limited attention. This captures the stabilizing effects that come with lower attention. The sluggish adjustment of inflation expectations in the following, however, leads to a very persistent undershooting of inflation. Even five years after the shock, inflation and inflation expectations are still substantially below their steady state levels of zero. The result is a prolonged period of a binding lower bound. While the economy under rational expectations escapes the ELB six periods after the shock, the economy under limited attention is stuck for twice as long. This is what I label \textit{inflation-attention trap}. A side-effect of these traps is that the long ELB period leads to an output boom.
As discussed earlier, this (expected) output boom in the future, however, has rather small effects on the economy today if people are inattentive.

Overall, limited attention to inflation offers a possible explanation for why several advanced economies were stuck at the ELB after the financial crisis, as well as inflation that undershot the central banks' inflation targets, even though the initial decrease was muted and output recovered quite strongly after declining severely initially (\citet{del2020s}).
In other words, the limited-attention model can explain the \textit{missing deflation puzzle} as well as the \textit{missing inflation puzzle} (\citet{coibion2015phillips}, \citet{constancio2015understanding}).

\section{Optimal Monetary Policy}\label{sec:omp}
To understand how monetary policy should optimally deal with declining attention, I now derive the Ramsey optimal monetary policy in this economy. I focus on the case of $\rho_{\pi} = 1$, in which average inflation expectations coincide with the actual inflation average. Appendix E.4 reports the results when relaxing this assumption.

The policymaker's objective is to maximize the representative household's utility, taking the household's and firms' optimal behavior, including their attention choice, as given. Thus, the policymaker cannot exploit the private agent's lack of information. Nevertheless, the policymaker can affect inflation expectations by influencing inflation itself and can set the average inflation expectations by setting the average inflation rate.  

The policymaker is paternalistic in the sense of \citet{benigno2014monetary} and evaluates the household's utility under rational expectations. A second-order approximation to the household's utility function yields the policymaker's objective
\begin{equation}
-\frac{1}{2}E_0\sum\limits_{t=0}^{\infty}\beta^t \left[\pi^2_t+\chi \left(y^{gap}_t\right)^2\right], \label{lossfct}
\end{equation}
where $\chi$ is the relative weight of the output gap, which I set to $\chi = 0.007$ as in \citet{adam2006optimal}. In the following, I refer to \eqref{lossfct} as \textit{welfare}.

In sum, the optimal policy problem is given by
\begin{align}
&\max_{\pi_t,y^{gap}_t,i_t} \quad -\frac{1}{2}E_0\sum\limits_{t=0}^{\infty}\beta^t\left[\pi^2_t+\chi \left(y^{gap}_t\right)^2\right]  \label{opp}
\end{align}
subject to
\begin{align}
&\pi_t = \beta\pi^e_{t+1|t}+\kappa y^{gap}_t+u_t \label{nkpc}\\
& y^{gap}_t = E_ty^{gap}_{t+1}-\varphi\left(i_t-\pi^e_{t+1|t}-r^n_t\right) \label{is}\\
& \pi^e_{t+1|t} = \pi^e_{t|t-1}+\gamma \left(\pi_{t}-\pi^e_{t|t-1}\right) \label{exp}\\
& u_t = \rho_u u_{t-1}+\varepsilon^u_t  \label{cps}\\
& r^n_t = \rho_{r} r^n_{t-1}+\varepsilon^r_t\label{rns}\\
& i_t \geq -\bar{i}, \label{elb}
\end{align}
with $\varepsilon^u_t \sim i.i.N.\left(0,\sigma^2_u\right)$ and $ \varepsilon^r_t \sim i.i.N.\left(0,\sigma^2_{r^n}\right)$ and \eqref{elb} is the lower-bound constraint.\footnote{I solve this numerically by recursifying the constrained optimization problem, as in \citet{marcet2019recursive}.} All variables are in percent deviations from their respective steady state, except the nominal interest rate and the natural rate which are in absolute deviations. 
\subsection{The Optimal Inflation Target}
What do low levels of attention imply for inflation volatility and the optimal inflation target? For this, I solve the Ramsey problem for different levels of attention, namely $\gamma\in \{0.05, 0.1, 0.2, 0.3\}$. An attention parameter of 0.3 is close to the estimates for consumers' attention after 1990, and the lower levels of 0.05 and 0.1 are close to the ones observed since 2010.

\begin{figure}[ht]
\caption{Optimal Inflation and Inflation Volatility}
\centering    
\begin{tabular}{cc}
(a) Optimal Inflation Target & (b) Inflation Volatility\\
\includegraphics[scale=0.35]{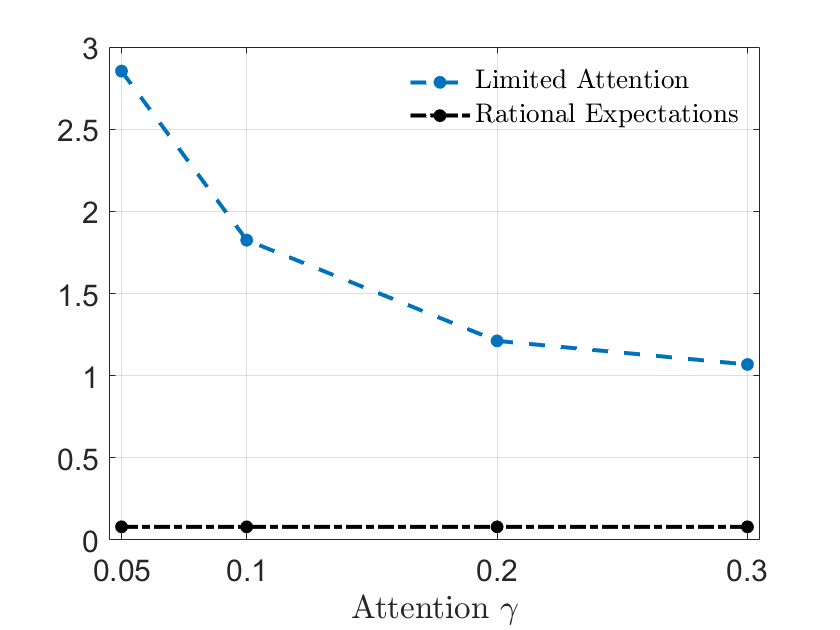}  & \includegraphics[scale=.35]{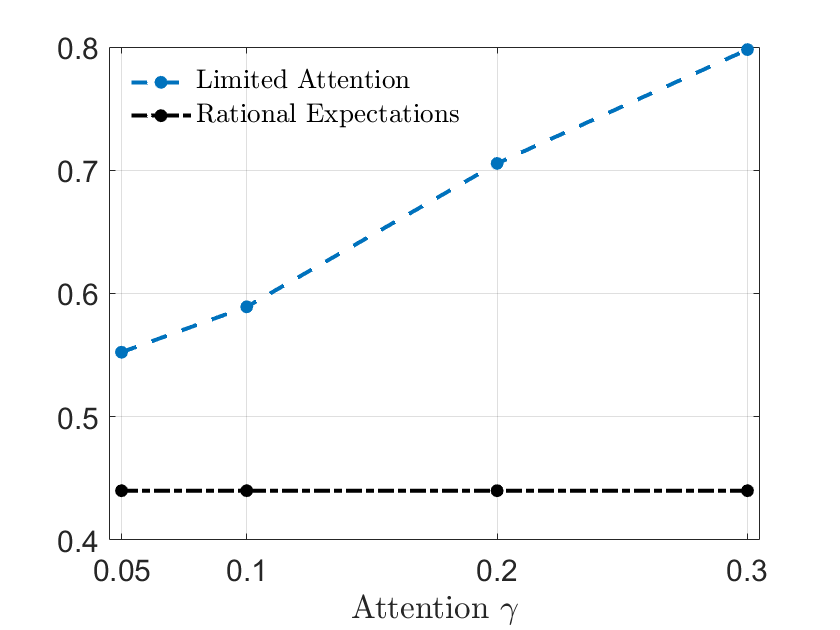}
\end{tabular}
\vspace{0cm}\\ 
 \begin{minipage}{\textwidth}
 \footnotesize{
 Notes: This figure shows the average inflation rate under Ramsey optimal policy (left panel) and inflation volatility for different attention levels. The blue-dashed lines show the results for the model under limited attention, and the black-dashed-dotted lines for the rational-expectations model.}%
 \end{minipage}
\label{fig:romp}
\end{figure} 

Figure \ref{fig:romp} shows the results. The average inflation rate under Ramsey optimal policy---what I refer to as the optimal inflation target---is plotted in the left panel and the inflation volatility in the right panel. The blue-dashed lines show the results for the model under limited attention, and the black-dashed-dotted lines for the rational-expectations model. We see that the optimal inflation target increases substantially as attention declines. At the levels of attention estimated just before the Covid crisis, $\gamma \in\{0.05, 0.1\}$, the inflation target is about 2-3 percentage points higher than under rational expectations due to the discussed ineffectiveness of forward-guidance policies. By increasing the average level of inflation the average nominal interest rate increases and thus makes it less likely that the ELB becomes binding. Indeed, the frequency of a binding ELB decreases substantially. For $\gamma = 0.3$, the ELB is binding 22\% of the time under optimal policy, whereas this value shrinks to 1.4\% for an attention level of 0.05.

While lower attention renders forward guidance, make-up policies and other policies that work (partly) through inflation expectations less effective, lower attention also stabilizes inflation, as can be seen from the right panel in Figure \ref{fig:romp}. First, because lower attention mutes the inflation response to shocks and output (see Proposition \ref{prop:pc}). Second, the lower ELB frequency further stabilizes the economy. Thus, lower attention to inflation can help stabilizing actual inflation and reduces the number of binding-ELB periods. The lower inflation volatility at lower levels of attention in fact justifies these low attention levels, as optimal attention depends positively on inflation volatility (see Section \ref{sec:data}). This low volatility, however, requires an increase in the inflation target, which is costly. Thus, it is not clear \textit{a priori} whether lower attention leads to welfare gains or not. 

\subsection{Welfare}
What are the effects of declining attention on overall welfare? Welfare is given by equation \eqref{lossfct} and from the previous discussion, we know that lower attention poses a trade off. On the one hand, inflation volatility decreases and the ELB binds less frequently, when attention is low. This raises welfare. On the other hand, lower attention complicates managing inflation expectations and thus, the optimal average \textit{level} of inflation increases, which is costly.  Which effect dominates?

\begin{figure}[ht]
\caption{Welfare and Attention}
\centering    
\centering    
\begin{tabular}{cc}
(a) ELB & (b) No ELB\\
\includegraphics[scale=0.35]{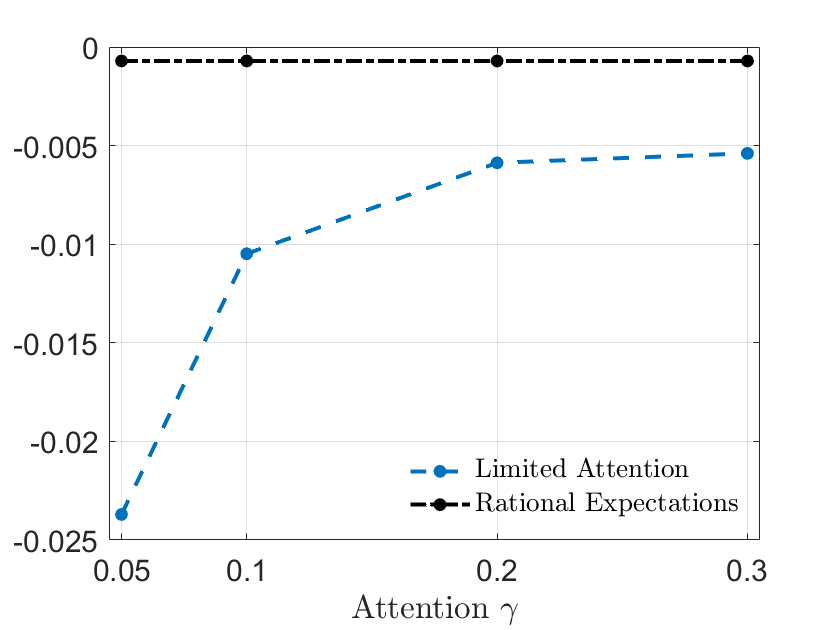}  & \includegraphics[scale=.35]{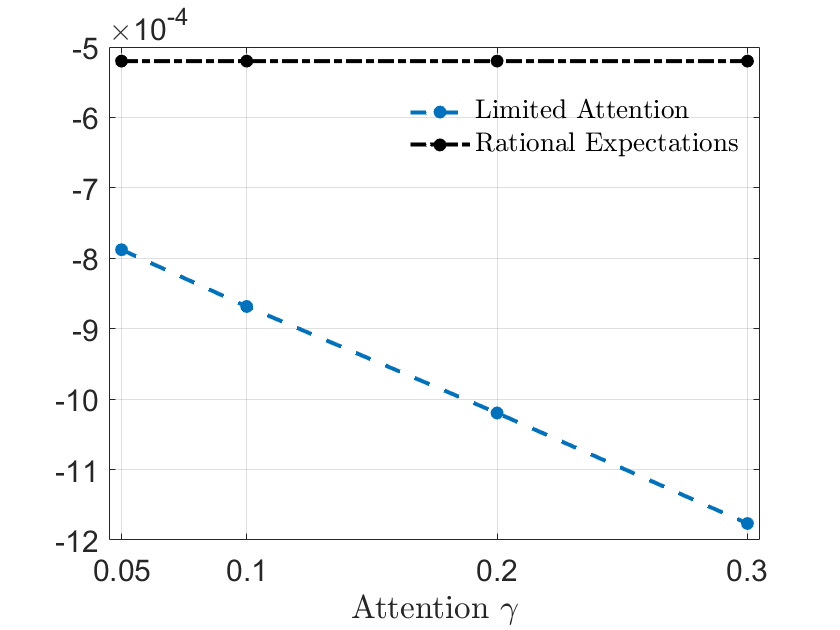}
\end{tabular}
\vspace{0.1cm}\\ 
 \begin{minipage}{\textwidth}
 \footnotesize{
 Notes: This figure shows welfare \eqref{lossfct} under Ramsey optimal policy for different levels of attention. The left panel shows the results for the case with an occasionally-binding ELB, and the right panel without an ELB. The blue-dashed lines show the results for the model under limited attention, and the black-dashed-dotted lines for the rational-expectations model.}%
 \end{minipage}
\label{fig:wf}
\end{figure}

Panel (a) in Figure \ref{fig:wf} shows that the cost of the level effect outweighs the stabilization benefits. As attention falls, welfare decreases. This is especially pronounced at low levels of attention, where the optimal inflation target increases substantially (see Figure \ref{fig:romp}). 

Absent the lower-bound constraint, the complications in managing inflation expectations due to limited attention are much less pronounced since managing expectations is particularly important at the lower bound. In fact, lower attention is welfare improving in the case without an ELB. Panel (b) in Figure \ref{fig:wf} shows this graphically. The stabilization benefits that arise from lower attention---which is reflected in more anchored expectations---lead to an increase in welfare. 

These findings show that accounting for the ELB is crucial for making a normative statement about costs and benefits of stabilizing inflation expectations. The ELB highlights the drawbacks that arise from the stabilization of expectations due to the fall in attention, as the management of expectations becomes particularly relevant when the ELB binds.

\subsection{Extensions}\label{sec:extensions}
In this section, I present several extensions of the baseline model and show that the overall implications derived so far remain robust, even when allowing (i) for a negative ELB, (ii) for non-rational output gap expectations, (iii) for a time-varying slope of the Phillips Curve or with non-zero trend inflation, and (iv) with time-varying attention to inflation. In \ref{sec:full_attention}, I further show that the results are largely driven by the agents' limited attention rather than by their perceived law of motion.

\subsubsection{Negative Interest Rate Policies}
In recent years, several central banks in advanced economies have implemented negative interest rate policies (NIRP).\footnote{See \citet{brandao2021negative} for a recent survey on negative interest rate policies and its effectiveness.} 
Could negative rates limit the negative consequences of declining attention? In order to answer this question, I solve the same Ramsey optimal policy problem as above, but set the effective lower bound to $-0.5\%$ (annualized).

\begin{figure}[ht]
\caption{Negative Interest Rate Policies and Attention}
\centering    
\centering    
\begin{tabular}{cc}
(a) Optimal Inflation Target & (b) Change in Inflation Target\\
\includegraphics[scale=0.35]{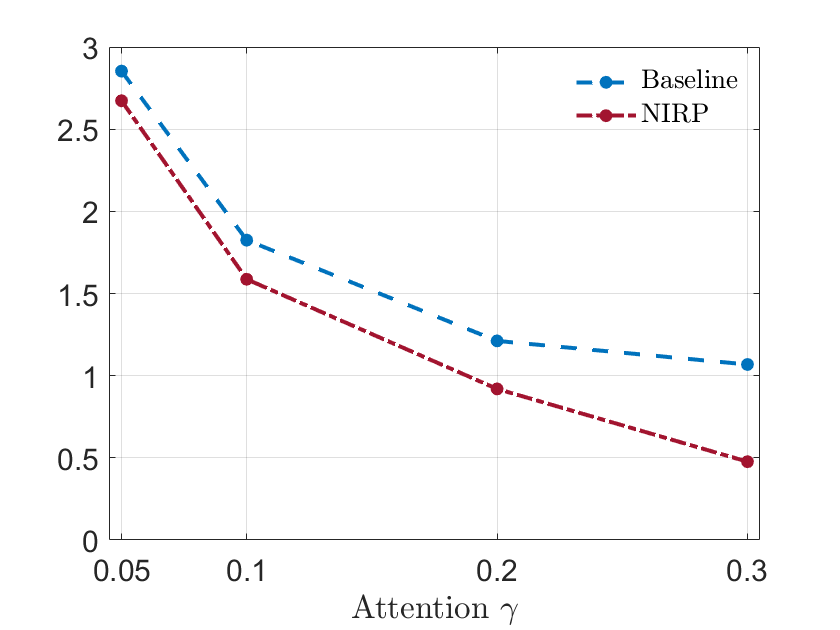}  & \includegraphics[scale=.35]{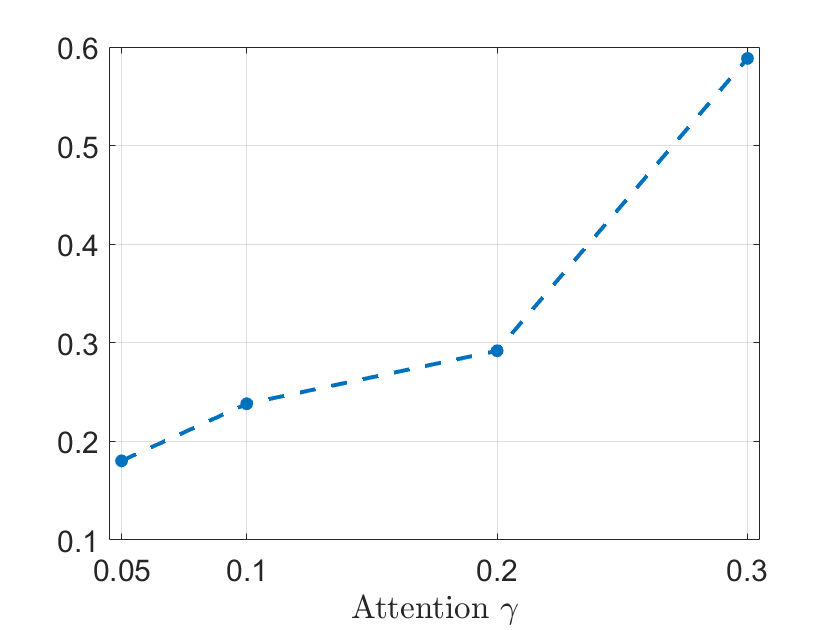}
\end{tabular}
\vspace{0.1cm}\\ 
 \begin{minipage}{\textwidth}
 \footnotesize{
 Notes: The left panel shows the average inflation rate under optimal policy for different degrees of attention $\gamma$. The blue-dashed lines show the results for the benchmark model where the lower bound is at 0, and the red-dashed-dotted lines show the results when allowing for negative interest rates up to $-0.5\%$ (annualized). The right panel shows the difference in the optimal inflation targets, defined as $\pi^{*,ZLB}-\pi^{*,NIRP}$, where $\pi^*$ denotes the optimal inflation target and the superscripts $ZLB$ and $NIRP$ denote the two cases where the ELB is at $0\%$ or $-0.5\%$, respectively.}%
 \end{minipage}
\label{fig:nirp}
\end{figure} 
Figure \ref{fig:nirp} reports the outcomes. Panel (a) shows the optimal inflation target (red-dashed-dotted line) and compares it to the case with an ELB at 0 (blue-dashed line). We see that the additional policy space due to the negative lower bound indeed calls for a lower inflation target. However, the decline in attention also weakens the effectiveness of NIRP. We see this by observing that the optimal inflation target under NIRP gets closer to the one without negative rates as attention declines. To see these gaps clearly, panel (b) shows the difference in the optimal inflation targets, defined as $\pi^{*,ZLB}-\pi^{*,NIRP}$, where $\pi^*$ denotes the optimal inflation target and the superscripts $ZLB$ and $NIRP$ denote the two cases where the ELB is at $0\%$ or $-0.5\%$, respectively. Overall, allowing for negative policy rates can help limiting the drawbacks of low attention but these policies itself become less effective as attention declines.

\subsubsection{Non-Rational Output Gap Expectations}

So far, I assumed that output gap expectations are fully rational. I now relax this assumption and show that in this case, (i) the optimal inflation target further increases and (ii) non-rational output gap expectations are welfare deteriorating. But before going into these results, I need to quantify people's attention to the output gap. As expectations about the output gap are not available, I use expectations about unemployment changes over the next year from the Michigan Survey and then estimate attention to unemployment $\gamma^y$ as in Section \ref{sec:data}. As I detail in Appendix E.1, I estimate attention to unemployment to have slightly increased from 0.09 to 0.1 from the period before 1990 to the period after 1990. This difference, however, is not statistically significant. Thus, I do not find any evidence for changes in people's attention to unemployment over the same period in which their attention to inflation decreased.

To understand the monetary policy implications of non-rational output gap expectations in the presence of an ELB, I extend the baseline model by allowing for limited attention to inflation and the output gap. Output gap expectations are given by
\begin{equation}
    y^{gap,e}_{t+1|t} = y^{gap,e}_{t|t-1} + \gamma^y\left(y^{gap}_t - y^{gap,e}_{t|t-1}  \right).  \label{eq:ey1}
\end{equation}
As I show in Appendix E.1, limited attention with respect to the output gap exacerbates attention traps. The economy remains stuck at the ELB for longer than in the case of rational expectations and, additionally, inflation, inflation expectations, and now also the output gap stay below their initial values very persistently. The reason for this is that make-up policies, such as forward guidance, not only work through inflation expectations but also through output gap expectations (via the IS equation). When households, however, form their expectations according to \eqref{eq:ey1}, output gap expectations become backward looking, which implies that these make-up policies do not stimulate output gap expectations any longer. Thus, the ELB becomes longer lasting.

To prevent these long periods at the ELB, I show in Appendix E.1 that the optimal inflation target becomes even higher than under rational output gap expectations and welfare deteriorates. Higher attention to the output gap serves a similar role as higher attention to inflation: the optimal inflation target is lower and welfare is higher at higher attention levels. These results hold independently of whether attention to inflation is higher than attention to the output gap (which was likely to be the case before the 1990s) or lower (which was the case before the Covid-19 pandemic).

\subsubsection{The Role of the Phillips Curve}\label{sec:nkpc_extensions}
\cite{ascari2022non} show that in high-inflation environments, the price level becomes more responsive to monetary policy shocks. Similarly, \cite{alvarez2019hyperinflation} find that prices are adjusted more frequently in such high-inflation environments (see also \cite{alexandrov2022effects}).
Thus, by inducing a higher average inflation rate, i.e., by raising the inflation target, the price-setting behavior of firms may change and thus, affect the optimal average inflation rate. To understand the implications of such changes in price-setting behavior, I now extend the analysis along three dimensions. First, I consider the case of a permanently steeper slope of the Phillips Curve, i.e., a higher $\kappa$ in equation \eqref{AS}, which may arise due to a lower price adjustment cost. Second, I let the slope of the Phillips Curve to be time-varying, i.e., I model $\kappa$ as a function of $\pi_t$. Third, I solve for the optimal policy for the case in which trend inflation is positive.\footnote{I use trend inflation and steady-state inflation interchangeably.} Table \ref{tab:nkpc} presents the results, which I now discuss in more detail.

\begin{table}[h]
 \caption{The Role of the Phillips Curve for Optimal Policy}\label{tab:nkpc}
 \centering
\begin{tabular}{lcccc}
\hline \hline
&   \multicolumn{2}{c}{Inflation Target} & \multicolumn{2}{c}{Welfare}  \\\cline{2-3} \cline{4-5} & $\gamma = 0.3$ & $\gamma = 0.1$ & $\gamma = 0.3$ & $\gamma = 0.1$ \\\hline\vspace{-0.4cm}\\
\underline{Baseline}  & 1.06\% & 1.82\% & -0.0049 & -0.0096 \\\vspace{-0.4cm}\\
\underline{Higher $\kappa$}  \\\vspace{-0.5cm}\\
Fixed $\chi$ & 0.95\% & 2.52\% & -0.0041 & -0.0173  \\\vspace{-0.4cm}\\
Higher $\chi$ & 1.02\% & 2.98\% & -0.0045 & -0.0239  \\\vspace{-0.4cm}\\
\underline{Time-varying $\kappa_t$} & 1.09\% & 1.83\% & -0.0050 & -0.0097 \\\vspace{-0.4cm}\\
\underline{Positive trend inflation} & 1.04\% & 2.02\% & -0.0029 & -0.0069
\\\vspace{-0.4cm}\\
\hline \hline
\end{tabular}\\\vspace{.1cm}
\begin{minipage}{1\textwidth}
\footnotesize{Notes: This table shows the implications of different Phillips Curve specifications for the optimal inflation target and welfare, for $\gamma = 0.3$ and $\gamma = 0.1$.
}%
 \end{minipage}
\end{table}

For the first case, I set the slope of the Phillips Curve $\kappa$ to a permanently higher level. This may reflect that in an economy with a higher inflation target, and a higher average inflation rate, the cost of adjusting prices is lower. I set $\kappa$ to 1.5 times the value I use in my baseline calibration (the baseline calibration sets $\kappa = 0.057$ as in \cite{adam2006optimal}). The row in Table \ref{tab:nkpc} labeled \textit{Fixed $\chi$} shows the implications of that change for the optimal inflation target and for welfare for the case in which attention to inflation is relatively high ($\gamma = 0.3$) and in which it is relatively low ($\gamma = 0.1$). The row labeled \textit{Higher $\chi$} shows the results for the case in which the welfare weight on the output gap is adjusted accordingly.\footnote{The weight on the output gap is given by $\chi = \frac{\kappa}{\epsilon}$, where $\epsilon$ denotes the price elasticity of demand for the intermediate goods \citep{adam2006optimal}.} When comparing it to the baseline calibration (row labeled \textit{Baseline}), we see that the optimal inflation target and welfare are not much affected in the case of $\gamma = 0.3$ by these changes. In the case of low attention, $\gamma = 0.1$, however, the optimal inflation target substantially increases by 0.7-1.16 percentage points and welfare decreases. The reason for this increase in the optimal inflation rate is that when the Phillips Curve is steep, fluctuations in the output gap translate into larger changes in inflation. When the economy is pushed to the ELB by adverse demand shocks, output gap decreases and hence, pushes inflation substantially down. Thus, the real rate is relatively high and hence, demand remains low. When attention is low, it takes longer for inflation expectations to recover after such a downturn and the economy remains at or close to the ELB for longer. To prevent this, it is optimal for the central bank to induce a higher inflation rate ex-ante which reduces the likelihood of reaching the ELB in the first place. 

In the second case, I allow $\kappa_t$ to be time-varying and to be a function of inflation $\pi_t$. In particular, I assume the following functional form:
\begin{equation*}
    \kappa_t = \bar{\kappa} + \kappa_1 \pi_t,
\end{equation*}
where I set $\bar{\kappa}$ to my baseline value of 0.057 and set $\kappa_1$ to 0.1. This captures the idea that the Phillips Curve steepens when inflation is high in a reduced form way. The row \textit{Time-varying $\kappa_t$} shows that this has barely an effect on the optimal inflation rate and welfare. The optimal inflation rate slightly increases, but the quantitative effects are very small.

In the third case, I allow for non-zero trend inflation (see Appendix C for details on the derivations, which follow \cite{ascari2012trend}). With Rotemberg price-adjustment costs \citep{Rotemberg1982}, this gives rise to the aggregate IS equation and Phillips Curve:
\begin{align*}
    \widehat{y}^{gap}_t &= E_t\widehat{y}^{gap}_{t+1} + \frac{\psi (\Bar{\Pi} - 1)\Bar{\Pi}}{1-\frac{\psi}{2}(\Bar{\Pi} -1)^2}\left[ \pi_t - \pi^e_{t+1|t}\right] -  \left(  i_t - \pi^e_{t+1|t} - r^n_t\right)
    \\
    \pi_t &= \zeta \left[ \frac{\epsilon(1+\nu)}{\psi}\widehat{y}^{gap}_t + \beta \Bar{\Pi}^2\pi^e_{t+1|t} + u_t + \Xi \pi^e_{t+1|t} \right]
\end{align*}
where
\begin{align*}
    \zeta &\equiv \frac{1}{\bar{\Pi}(2\bar{\Pi}-1)+\frac{\beta \psi \Bar{\Pi}(\bar{\Pi} - 1)^2}{1-\frac{\psi}{2}(\Bar{\Pi} -1)^2
}+\frac{\epsilon\Bar{\Pi}(\Bar{\Pi}-1)}{1-\frac{\psi}{2}(\Bar{\Pi} -1)^2}} \\
\Xi &\equiv \beta\Bar{\Pi}(\Bar{\Pi}-1)\left[1+\frac{\psi(\Bar{\Pi} -1)}{1-\frac{\psi}{2}(\Bar{\Pi} -1)^2} \right].
\end{align*}
Following \cite{ascari2012trend}, I set the demand elasticity $\epsilon = 10$. Given $\kappa = 0.057$, this implies a price-adjustment cost parameter $\psi = 350.87$.\footnote{In general, there would be an additional term with the expected change of the output gap in the Phillips Curve, but this term drops out in my case, because I set $\varphi = 1$ (see Appendix C for details).}
When trend inflation is zero, $\Bar{\Pi} = 1$, it follows that $\zeta = 1$ and $\Xi = 0$, so that we are back to the baseline case analyzed above. Note, that $\pi_t$ now denotes inflation in deviations from trend inflation (similarly, in the welfare objective \eqref{lossfct}).
The last row in Table \ref{tab:nkpc} labeled \textit{Positive trend inflation} shows the policy implications of trend inflation of 0.5\%. We see that the optimal inflation rate is barely affected for the case of high attention ($\gamma = 0.3$) but increases slightly  at low levels of attention ($\gamma = 0.1$). However, the increase in the average inflation rate is less than trend inflation. 

Overall these results highlight that the main policy implications of my analysis turn out to be robust to accounting for changing price-setting behavior in high-inflation environments: the optimal inflation target is substantially higher under limited attention to inflation and increases as attention falls.

\subsubsection{Time-Varying Attention}\label{app:timevaryingattention}
Up to now, I assumed that attention does not respond to short-term changes in the economy and therefore, compared economies with different degrees of attention but in which attention was time invariant within economy. To analyze how the results change, when attention responds to short-run changes, I impose that attention takes the form
\begin{equation*}
    \gamma_t = \bar{\gamma} + \gamma_1\pi_t. 
\end{equation*}
I set $\bar{\gamma}$ to  an intermediate value of 0.2 and solve for the optimal policy for the cases $\gamma_1 = 0.3$ and $\gamma_1 = -0.3$. Table \ref{tab:timevary} shows that when attention increases with inflation ($\gamma_1 = 0.3$), the optimal inflation target increases and welfare decreases. In contrast, when attention decreases with inflation ($\gamma_1 = -0.3$), the optimal inflation rate decreases and welfare increases. When the economy is pushed to the ELB due to an adverse demand shock, inflation decreases. In the case of $\gamma_1 = 0.3$, attention therefore then decreases as well which implies that inflation expectations tend to be lower for longer, keeping actual inflation low, too. Thus, the economy is likely to stay at the ELB for longer. In order to prevent this, it is therefore optimal to induce a higher average inflation rate which makes the ELB less likely to be binding. However, higher average inflation is costly from a welfare perspective, and hence, welfare is lower in that case. When $\gamma_1 = -0.3$, the opposite is the case, and therefore, the optimal inflation rate is lower and welfare is higher.

\begin{table}[h]
 \caption{Time-Varying Attention to Inflation}\label{tab:timevary}
 \centering
\begin{tabular}{lcc}
\hline \hline
&   \multicolumn{1}{c}{Inflation Target} & \multicolumn{1}{c}{Welfare}  \\\hline\vspace{-0.4cm}\\
{Baseline}  & 1.20\% & -0.0060 \\\vspace{-0.4cm}\\
{$\gamma_1 = 0.3$} & 1.32\% & -0.0069   \\\vspace{-0.4cm}\\
{$\gamma_1 = -0.3$} & 0.98\% & -0.0046
\\\vspace{-0.4cm}\\
\hline \hline
\end{tabular}\\\vspace{.1cm}
\begin{minipage}{1\textwidth}
\footnotesize{Notes: This table shows the implications of time-varying attention to inflation for the optimal inflation target and welfare.
}%
 \end{minipage}
\end{table}

\subsubsection{Full Attention}\label{sec:full_attention}
There are two main assumptions underlying the law of motion of inflation expectations \eqref{infexp}. First, the assumption that agents perceive inflation to follow an AR(1) process. Second, that paying attention to inflation is costly and thus, their attention is limited. To disentangle these two effects, Figure 15 in Appendix E.5 shows the implications of shutting down the second channel, i.e., if we set $\gamma = 1$. Put differently, how much of the results in previous sections is solely due to the misperception of the law of motion of inflation? Panel (a) shows that the optimal average inflation rate is practically identical to the one under rational expectations. Thus, as argued above, it is really the lack of attention that pushes up the optimal inflation rate, whereas the assumption of a misperceived law of motion is rather innocuous from this perspective. 

Panel (b) in Figure 15 reports the inflation volatility under Ramsey optimal policy for different levels of attention, including the full attention case, $\gamma = 1$. Inflation is substantially more volatile than under rational expectations. Thus, while the misperception of the law of motion of inflation is rather inconsequential for the optimal inflation target, it predicts a higher inflation volatility. Indeed, there is a trade-off. On the one hand, higher attention increases the response of inflation to shocks. On the other hand, if agents are fully attentive, inflation also reacts more strongly to changes in expected future output. Thus, to achieve a certain effect on today's inflation rate, the policymaker is required to make smaller promises about its policies in the future which in turn stabilizes inflation and output already today. While the first channel dominates at lower levels of $\gamma$, the second effect pushes inflation volatility down at higher levels of $\gamma$.

Finally, panel (c) in Figure 15 shows the welfare implications of the misperceived law of motion. Welfare is slightly more negative when agents have a misperceived law of motion of inflation compared to fully rational agents. Given the results in panels (a) and (b), we see that these additional welfare losses are mainly due to increased inflation volatility rather than its level.  

\section{Conclusion}\label{sec:conclusion}
With the stabilization of inflation in advanced economies since the Great Inflation period, inflation has become less important in people's lives. In this paper, I quantify this using a limited-attention model of inflation expectations. In line with this model, I show that attention to inflation decreased together with inflation volatility and inflation persistence since the 1970s. Especially in the period between 2010 and 2020, the general public's attention to inflation was close to zero.

For monetary policy the decline in attention was desirable at first, since lower attention stabilizes inflation expectations and hence, stabilizes actual inflation. With the outbreak of the Great Recession and nominal rates at their lower bound, however, managing inflation expectations became a central tool for monetary policy. But managing inflation expectations is difficult when people are inattentive. 

The optimal policy response is a substantial increase in the inflation target. This increases the average nominal rate and thus, binding ELB periods become less likely. The cost of this increase in inflation, however, outweighs the stabilization benefits of lower attention. Lower attention, therefore, decreases welfare if we account for the lower bound. This stands in stark contrast to the case without an ELB in which case lower attention leads to welfare gains through the stabilization of inflation expectations and inflation. My paper thus shows that accounting for the ELB is crucial when assessing the role of the public's attention to inflation.


\clearpage
\newpage

\singlespacing

\bibliography{bibPS}

\clearpage
\newpage

\onehalfspacing

\appendix
\title{\Huge{\textbf{Online Appendix}} }
\maketitle
\begin{large}
\begin{centering}

\end{centering}
\end{large}


\setcounter{equation}{20}
\setcounter{proposition}{1}
\setcounter{figure}{5}
\setcounter{table}{6}
\section{A Limited-Attention Model of Inflation Expectations}\label{sec:theory}

In this section, I derive the expectations-formation process under limited attention sketched in Section 2. 
%
%
The agent believes that (demeaned) inflation tomorrow, $\pi'$, depends on (demeaned) inflation today, $\pi$, as follows
\begin{equation}
\pi' =   \rho_{\pi} \pi + \nu, \notag
\end{equation}
where $\rho_{\pi}\in[0,1]$ denotes the perceived persistence of inflation and $\nu \sim i.i.N.(0,\sigma^2_{\nu})$. 
Inflation in the current period is unobservable, so before forming an expectation about future inflation, the agent needs to form an expectation about today's inflation. I denote this nowcast $\widetilde{\pi}$, and the resulting forecast about next period's inflation $\pi^e = \rho_{\pi}\widetilde{\pi}$.  Given her beliefs, the full-information forecast $\pi^{e*}$ is 
\begin{equation}
\pi^{e*} \equiv \rho_{\pi} \pi. \notag
\end{equation}
But since $\pi$ is not perfectly observable, the actual forecast will deviate from the full-information forecast. Deviating, however, is costly, as this causes the agent to make mistakes in her decisions.

The agent's choice is not only about how to form her expectations given certain information, but about how to choose this information optimally, while taking into account how this will later affect her forecast. That is, she chooses the form of the signal $s$ she receives about current inflation. Since acquiring information is costly, it cannot be optimal to acquire different signals that lead to an identical forecast. Due to this one-to-one relation of signal and forecast, we can directly work with the joint distribution of $\pi^e$ and $\pi$, $f(\pi^e, \pi)$, instead of working with the signal.

Let $U(\pi^e, \pi)$ denote the negative of the loss that is incurred when the agent's forecast deviates from the forecast under full information, and $C(f)$ the cost of information. Then, the agent's problem is given by
\begin{align}
&\max_{f} \int U(\pi^e, \pi)f(\pi^e, \pi)d\pi d\pi^e - C(f) \label{pr1}\\
\text{subject to } & \int f(\pi^e, \pi) d\pi^e = g(\pi), \text{ for all } \pi, \notag
\end{align}
where $g(\pi)$ is the agent's prior, which is assumed to be Gaussian; $\pi \sim N\left(\hat{\pi},\sigma^2_{\pi}\right)$.
$C(.)$ is the cost function that captures how costly information acquisition is. It is linear in \textit{mutual information} $I(\pi;\pi^e)$, i.e., the expected reduction in entropy of $\pi$ due to knowledge of $\pi^e$:
\begin{equation}
C(f) = \lambda I(\pi;\pi^e) = \lambda\left(H(\pi)-E\left[H(\pi|\pi^e)\right]\right), \notag
\end{equation}
where $H(x) = -\int f(x)log(f(x))dx$ is the entropy of $x$ and $\lambda$ is a parameter that measures the cost of information.

The objective function $U(.)$ is assumed to be quadratic:
\begin{equation}
U(\pi^e,\pi) = -r\left(\rho_{\pi}\pi-\pi^e\right)^2, \notag
\end{equation}
where $r$ measures the stakes of making a mistake.\footnote{A quadratic loss function is usually derived from a second-order approximation of the household's utility function or the firm's profit function (see, e.g., \citet{mackowiak2009optimal}).}$^,$\footnote{These stakes (or also the information cost parameter $\lambda$) can be interpreted as a way to incorporate other variables to which the agent might pay attention. For example, a household might not only want to forecast inflation but also her own income stream going forward. In this case, a smaller $r$ could capture an increase in her idiosyncratic income volatility. Thus, paying attention to inflation is relatively less beneficial, as the relative importance of her idiosyncratic income increases. Such an interpretation also explains why professional forecasters might not be fully informed about inflation, given that they usually forecast a whole array of variables.}

In this setup, Gaussian signals are optimal (and in fact the unique solution, see \citet{matvejka2015rational}). The optimal signal thus has the form
\begin{equation}
s = \pi+\varepsilon, \notag
\end{equation}
with $\varepsilon\sim i.i.N.(0,\sigma^2_{\varepsilon})$.\footnote{In this case, the entropy becomes $H(x) = \frac{1}{2}log(2\pi e \sigma^2_x)$, where $\sigma^2_x$ is the variance of $x$. Note, that here $\pi$ denotes the number ``pi'' and not inflation.} The problem \eqref{pr1} now reads
\begin{equation}
\max_{\sigma^2_{\pi|s}\leq \sigma^2_{\pi}}E_{\pi}\left[E_s\left[-r\rho_{\pi}^2\left(\pi-E[\pi|s]\right)^2\right]\right] - \lambda I(\pi;\pi^e)= \max_{\sigma^2_{\pi|s}\leq \sigma^2_{\pi}} \left(-r\rho_{\pi}^2\sigma^2_{\pi|s}-\frac{\lambda}{2} log\frac{\sigma^2_{\pi}}{\sigma^2_{\pi|s}}\right). \label{pr2}
\end{equation}
The optimal forecast is given by $\pi^e = \rho_{\pi} E\left[\pi|s\right]$, and Bayesian updating implies
\begin{equation}
\pi^e = \rho_{\pi}\left(1-\gamma\right)\hat{\pi}+\rho_{\pi}\gamma s,\label{updating_theo}
\end{equation}
where $\gamma = 1-\frac{\sigma^2_{\pi|s}}{\sigma^2_{\pi}}\in[0,1]$ measures how much attention the agent pays to inflation, and $\hat{\pi}$ denotes the prior mean of $\pi$.

An equivalent way of writing $\gamma$ is
\begin{equation}
\gamma = \frac{\sigma^2_{\pi}}{\sigma^2_{\pi}+\sigma^2_{\varepsilon}}. \label{gamma1}
\end{equation}
Now, since the agent \textit{chooses} the level of attention, we can re-formulate \eqref{pr2} as
\begin{equation}
\max_{\gamma\in[0,1]}\left(-r\rho_{\pi}^2(1-\gamma)\sigma^2_{\pi}-\frac{\lambda}{2}log\frac{1}{1-\gamma}\right).\label{maxi}
\end{equation}
Writing the cost of information relative to the stakes, $\tilde{\lambda}\equiv\frac{\lambda}{r}$, and solving the optimization problem \eqref{maxi} yields the \textit{optimal} level of attention, presented in Lemma 1.

\clearpage
\newpage

\section{Appendix to Empirical Results}\label{app:stats}

\begin{figure}[ht]
\centering
\caption{Inflation and Inflation Expectations}
\vspace{-0.4cm}

\includegraphics[scale=.5]{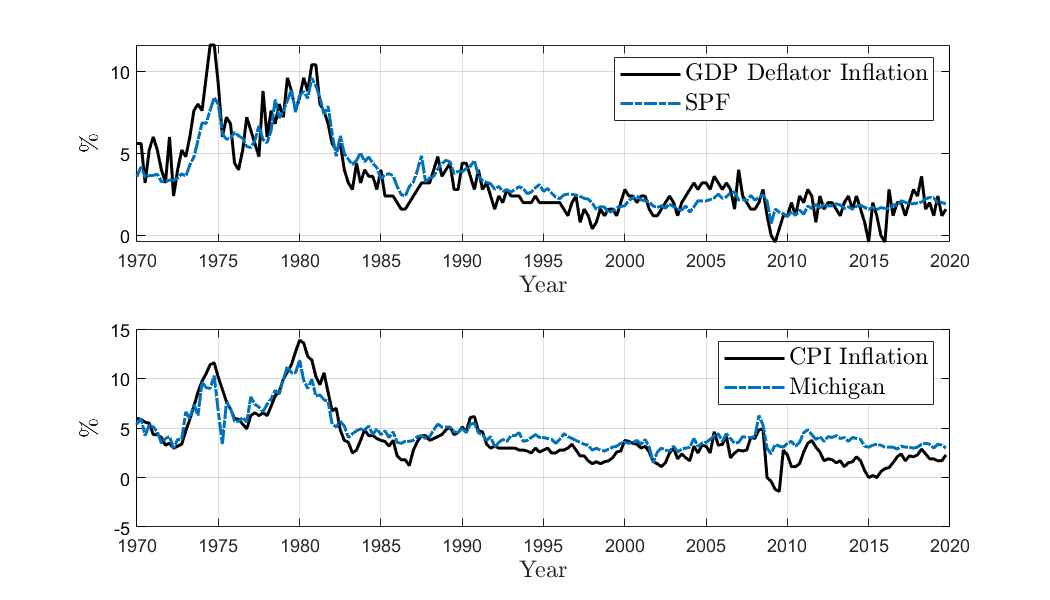}
\label{fig:inf}
 \begin{minipage}{\textwidth}

 \footnotesize{
Note: This figure shows the raw time series of inflation, as well as survey expectations about future inflation. Everything is in annualized percentages.
 }%
 \end{minipage}\\
\end{figure}

Figure \ref{fig:inf} shows the main time series that are used in the empirical analyses of Section 2. Apart from the apparent decrease in the level and volatility of inflation as well as inflation expectations, we see that expectations became more and more detached from actual inflation. First, consumer expectations seem to be biased on average in the most recent decades, as can be seen in the lower panel. While these expectations closely tracked inflation in the 70s and 80s, this is not the case anymore.\footnote{In the empirical analysis I account for this mean bias by including an intercept in the regressions.} Second, professional forecasters' expectations seem to perform quite well on average. In the last twenty years, however, they barely react to actual changes in inflation anymore. Overall, these observations suggest that attention decreased in the last decades.

Table \ref{tab:stats} shows the summary statistics, for the period before and after the 1990s, separately. 
For professional forecasters, the perceived persistence is higher than the actual one. This is especially the case when the actual persistence is relatively low, as was the case after 1990. \citet{afrouzi2020overreaction} document a similar finding in an experimental setting. This might point towards lower attention since the 1990s. Note, that in the empirical analysis I account for changes in the perceived persistence.

\begin{table}[ht]

\caption{Summary Statistics}
\centering 
\vspace{-0.2cm}

\begin{tabular}{@{\extracolsep{4pt}}lcccc@{}}
\hline\hline\vspace{-0.45cm}\\
 & \multicolumn{2}{c}{GDP Deflator Inflation} & \multicolumn{2}{c}{SPF Expectations}\\\cline{2-3}\cline{4-5}\\\vspace{-0.9cm}\\
 & 1968-1990 & 1990-2020 & 1968-1990 & 1990-2020 
\\\hline\vspace{-0.4cm}\\
Mean $(\%)$ & $5.44$ & $2.00$ & $5.18$ & $2.15$  \\
Std. Dev. $(\%)$ & $2.43$ & $0.90$ & $1.87$  & $0.63$ \\
Persistence & 0.84 & 0.55 & 0.93 & 0.92 \\\bottomrule\vspace{-0.4cm}\\
 & \multicolumn{2}{c}{CPI Inflation} & \multicolumn{2}{c}{Consumer Expectations}\\\cline{2-3}\cline{4-5}\\\vspace{-0.9cm}\\
 & 1968-1990 & 1990-2020 & 1968-1990 & 1990-2020 
\\\hline\vspace{-0.4cm}\\
Mean $(\%)$ & 6.09 & 2.42 & 6.00 & 3.62 \\
Std. Dev. $(\%)$ & 3.00 & 1.26 & 2.17 & 0.68  \\
Persistence & 0.96 & 0.77 & 0.85 & 0.70 \\
\hline\hline
\end{tabular} \\
\label{tab:stats}\vspace{0.2cm}
 \begin{minipage}{\textwidth}

 \footnotesize{
Note: This table shows the summary statistics of the data. The upper panel shows the statistics for the quarter-on-quarter GDP deflator inflation (left) and the corresponding inflation expectations from the Survey of Professional Forecasters (right). The lower panel shows the year-on-year CPI inflation (left) and the corresponding inflation expectations from the Survey of Consumers from the University of Michigan. All data are annualized.
 }%
 \end{minipage}\\

\end{table}

\clearpage
\newpage

\subsection{Robustness and Additional Evidence}\label{sec:robust}
In this section, I show that the empirical results are robust along several dimensions.

\subsubsection*{Additional Data Sources}
In Table \ref{tab:reg1_datas}, I show how attention changed over time for different data sources. The first two columns show the results for the Greenbook forecasts, columns 3-4 for the Livingston Survey, and columns 5-6 and 7-8 are for CPI forecasts from the SPF instead of forecasts about the GDP deflator. As in the main text, I use two different estimators. First, the Blundell-Bond estimator (columns 5-6) and pooled OLS (columns 7-8). All standard errors are robust with respect to heteroskedasticity and serial correlation. We see that the main finding of lower attention in inflation expectations in the period after 1990 compared to the period before is robust to these changes in the data source and/or exact variable.

\begin{table}[ht]

\caption{Regression Results of Equation (4)}
\vspace{-0.2cm}
\hspace{-1cm}
\begin{tabular}{lcccccccc}
\hline\hline\vspace{-0.45cm}\\
 & \multicolumn{2}{c}{Greenbook} & \multicolumn{2}{c}{Livingston} & \multicolumn{2}{c}{SPF CPI BB} & \multicolumn{2}{c}{SPF CPI OLS} \\
 \cline{2-3} \cline{4-5} \cline{6-7} \cline{8-9}\\\vspace*{-0.9cm}\\
 &  $<1990$\phantom{<} &  $\geq 1990$\phantom{<} &   $<1990$\phantom{<} &  $\geq 1990$\phantom{<}&   $<1990$\phantom{<} &  $\geq 1990$\phantom{<}&   $<1990$\phantom{<} &  $\geq 1990$\phantom{<}  \\
\hline\\\vspace*{-0.9cm}\\
$\widehat{\gamma}$ & 0.39 & 0.24 & 0.28 & 0.17 & 0.36 & 0.23& 0.17 & 0.13\\
s.e. & (0.0851) & (0.0715) & (0.0554) & (0.0624) &(0.1444) & (0.0328)& (0.0409) & (0.0142) \\
\hline\\\vspace*{-0.9cm}\\
$N$ & 84 & 100 & 83 & 61 &550 &3,577 & 550 & 3,577
\\\hline\hline
\end{tabular}%
\label{tab:reg1_datas}\vspace{0.2cm}
 \begin{minipage}{1\textwidth}

 \footnotesize{
Note: This table shows the results from regression (4) for different data sources. The standard errors, reported in parentheses, are robust with respect to heteroskedasticity and serial correlation.
 }%
 \end{minipage}
\end{table}

\clearpage
\newpage

\subsubsection*{Different Sample Splits}\label{app:splits}
Table 1 in the main body of the paper shows that attention to inflation declined by focusing on a sample split in 1990. To show that this is robust to the exact split point, Tables \ref{tab:reg1_85} and \ref{tab:reg1_95} show that the result holds when splitting the sample in 1985 or 1995, respectively. In fact, the decline in attention is even somewhat more pronounced when splitting the sample in 1985. This is in line with the theoretical prediction of the limited-attention model. Namely, the period between 1985 and 1990 was a period of relatively low and stable inflation compared to the period pre 1985 (see Figure \ref{fig:inf}), and thus, a period in which the model would predict a relatively low level of attention.

\begin{table}[ht]

\caption{Regression Results of Equation (4), pre 1985 vs. post 1985}
\vspace{-0.2cm}
\hspace{-1cm}
\begin{tabular}{lcccccccc}
\hline\hline\vspace{-0.45cm}\\
 &  \multicolumn{4}{c}{Professional Forecasters} &  \multicolumn{4}{c}{Consumers}  \\ \cline{2-5} \cline{6-9}\\\vspace*{-0.9cm}\\
 & \multicolumn{2}{c}{Blundell Bond} & \multicolumn{2}{c}{Pooled OLS} & \multicolumn{2}{c}{Averages} & \multicolumn{2}{c}{Median} \\
 \cline{2-3} \cline{4-5} \cline{6-7} \cline{8-9}\\\vspace*{-0.9cm}\\
 &  $<1985$\phantom{<} &  $\geq 1985$\phantom{<} &   $<1985$\phantom{<} &  $\geq 1985$\phantom{<} &  $<1985$\phantom{<} &  $\geq 1985$\phantom{<} &  $<1985$\phantom{<} &  $\geq 1985$\phantom{<} \\
\hline\\\vspace*{-0.9cm}\\
$\widehat{\gamma}$ & 0.75 &0.37 & 0.45& 0.25& 0.77 &0.31 & 0.50 & 0.26  \\
s.e. & (0.1247) & (0.0399) &(0.0403) &(0.0338) & (0.1688) & (0.0811) & (0.0955) & (0.0561) \\
\hline\\\vspace*{-0.9cm}\\
$N$ &1914  & 3887& 1914& 3887& 64 & 140 & 27 & 140
\\\hline\hline
\end{tabular}%
\label{tab:reg1_85}\vspace{0.2cm}
 \begin{minipage}{1\textwidth}

 \footnotesize{
Note: This table shows the results from regression (4) for professional forecasters (SPF) as well as for consumers. For the SPF, I use the \citet{blundell1998initial} (BB) estimator (first two columns), as well as pooled OLS (columns 3-4). For the Survey of Consumer, I consider average expectations (columns 5-6) and median expectations (columns 7-8). The standard errors, reported in parentheses, are robust with respect to heteroskedasticity and serial correlation.
 }%
 \end{minipage}
\end{table}
\begin{table}[ht]

\caption{Regression Results of Equation (4), pre 1995 vs. post 1995}
\vspace{-0.2cm}
\hspace{-1cm}
\begin{tabular}{lcccccccc}
\hline\hline\vspace{-0.45cm}\\
 &  \multicolumn{4}{c}{Survey of Professional Forecasters} &  \multicolumn{4}{c}{Survey of Consumers}  \\ \cline{2-5} \cline{6-9}\\\vspace*{-0.9cm}\\
 & \multicolumn{2}{c}{Blundell Bond} & \multicolumn{2}{c}{Pooled OLS} & \multicolumn{2}{c}{Averages} & \multicolumn{2}{c}{Median} \\
 \cline{2-3} \cline{4-5} \cline{6-7} \cline{8-9}\\\vspace*{-0.9cm}\\
 &  $<1995$\phantom{<} &  $\geq 1995$\phantom{<} &   $<1995$\phantom{<} &  $\geq 1995$\phantom{<} &  $<1995$\phantom{<} &  $\geq 1995$\phantom{<} &  $<1995$\phantom{<} &  $\geq 1995$\phantom{<} \\
\hline\\\vspace*{-0.9cm}\\
$\widehat{\gamma}$ &0.70 &0.41 & 0.44 & 0.21 & 0.72 &0.27 & 0.43 & 0.22  \\
s.e. &(0.0907) & (0.0654)& (0.0379) & (0.0344) & (0.1473) & (0.0962) & (0.0819) & (0.0654) \\
\hline\\\vspace*{-0.9cm}\\
$N$ &2708 &3093 &2708 &3093 & 104 & 100 & 67 & 100
\\\hline\hline
\end{tabular}%
\label{tab:reg1_95}\vspace{0.2cm}
 \begin{minipage}{1\textwidth}

 \footnotesize{
Note: This table shows the results from regression (4) for professional forecasters (SPF) as well as for consumers. For the SPF, I use the \citet{blundell1998initial} (BB) estimator (first two columns), as well as pooled OLS (columns 3-4). For the Survey of Consumer, I consider average expectations (columns 5-6) and median expectations (columns 7-8). The standard errors, reported in parentheses, are robust with respect to heteroskedasticity and serial correlation.
 }%
 \end{minipage}
\end{table}

\clearpage
\newpage

\subsubsection*{Different Specifications of the BB Estimator}
In the baseline estimation, reported in Table 1, I included all potential lags for the Blundell-Bond estimation. To show that the results are robust to this specification, I show in Table \ref{tab:reg1_bb} that for maximum lag lengths of 20 and 10 periods, the estimated attention parameter $\widehat{\gamma}$ is in all cases higher before 1990 compared to the period after 1990.

\begin{table}[ht]

\caption{Different Maximum Lag Lengths}
\centering 
\vspace{-0.2cm}

\begin{tabular}{lcccccc}
\hline\hline\vspace{-0.45cm}\\
 & \multicolumn{2}{c}{All Lags} & \multicolumn{2}{c}{20 Lags} & \multicolumn{2}{c}{10 Lags}  \\
 \cline{2-3} \cline{4-5} \cline{6-7} \\\vspace*{-0.9cm}\\
 &  $<1990$\phantom{<} &  $\geq 1990$\phantom{<} &   $<1990$\phantom{<} &  $\geq 1990$\phantom{<}&   $<1990$\phantom{<} &  $\geq 1990$\phantom{<} \\
\hline\\\vspace*{-0.9cm}\\
$\widehat{\gamma}$ & 0.70 & 0.41 & 0.74 & 0.51 & 0.84 & 0.69   \\
s.e. & (0.1005) & (0.0522) & (0.1086) & (0.0632) & (0.1247) &  (0.1127)   \\
\hline\\\vspace*{-0.9cm}\\
$N$ & 2235 & 3566 & 2235 & 3566 & 2235 & 3566
\\\hline\hline
\end{tabular}%
\label{tab:reg1_bb}\vspace{0.2cm}
 \begin{minipage}{1\textwidth}

 \footnotesize{
Note: This table shows the results from regression (4) for different numbers of lags included in the BB estimation. The standard errors, reported in parentheses, are robust with respect to heteroskedasticity and serial correlation.
 }%
 \end{minipage}
\end{table}

\clearpage
\newpage

\subsubsection*{Time Fixed Effects}
To account for potential changes in trend inflation, I include time-fixed effects in regression (4). To do so, recall that (4) is given by
\begin{equation}
\pi^e_{t+1|t,i} = \beta_i + \beta_1 \pi^e_{t|t-1,i} +\beta_2 \left(\pi_t - \pi^e_{t|t-1,i}\right)+u_{i,t}. \label{reg1_2}
\end{equation}
To include time fixed effects, I first compute a period-specific persistence parameter, $\rho_{\pi}$. Note, that in \eqref{reg1_2}, $\beta_1$ measures this persistence. Therefore, I subtract $\widehat{\rho}_{\pi}\pi^e_{t|t-1,i} $ from both sides and then to directly estimate $\gamma$, I further divide both sides by $\widehat{\rho}_{\pi}$:
\begin{equation}
\frac{\pi^e_{t+1|t,i} - \widehat{\rho}_{\pi}\pi^e_{t|t-1,i}}{\widehat{\rho}_{\pi}}  = \delta_i +d_t +\gamma \left(\pi_t - \pi^e_{t|t-1,i}\right)+v_{i,t}, \label{reg1_tfe}
\end{equation}
where $d_t$ captures time-fixed effects, $\delta_i = \frac{\beta_i}{\widehat{\rho}_{\pi}}$ and $v_{i,t} = \frac{u_{i,t}}{\widehat{\rho}_{\pi}}$. I do this transformation for the period before and after 1990 separately. 
Note, that this transformation also deals with the endogeneity problem explained in Section 2.

The estimated attention levels are 0.75 (s.e. 0.0327) for the period before 1990 and 0.61 (s.e. 0.0295) after 1990 if I use the first-order autocorrelation of expected inflation as my measure of $\rho_{\pi}$. If I use the estimate of $\beta_1$ from equation \eqref{reg1_tfe} as my measure of $\rho_{\pi}$, the estimated attention before the 1990s is $0.68$ (s.e. 0.0252) and the one after the 1990s is $0.46$ (s.e. 0.0242). Thus, we see that the decrease in attention is robust to controlling for time-fixed effects, even though the decline is somewhat muted. 

When using the first-order autocorrelation of expected inflation as my measure of $\rho_{\pi}$, estimating equation (5) in this way, delivers a point estimate of 0.06 (s.e. 0.0111) that is statistically significant on all conventional significance levels. The estimate for $\zeta$ in regression (6) is 0.23 (s.e. 0.0220), statistically significant on all conventional significance levels. When using $\hat{\beta}_1$ from \eqref{reg1_tfe} as the measure of $\rho_{\pi}$, the point estimate of $\beta$ in equation (5)  is 0.06 (s.e. 0.0074) and the estimate of $\zeta$ in (6) is 0.29 (s.e. 0.0306), both statistically significant on all conventional levels of significance.  Thus, the positive relationships between attention and volatility, as well as between attention and inflation persistence, are robust to controlling for time fixed effects.

\clearpage
\newpage

\subsubsection*{Professional Forecasters in the Aggregate}
When estimating attention of professional forecastors' average expectations instead of individual ones, we obtain a value of 0.24 (s.e. 0.0481) for the period before 1990 and of 0.09 (s.e. 0.0353) after 1990. Consistent with the main results, attention substantially decreased in recent decades and is about half after 1990 compared to before.

Estimating regression (5) on aggregate SPF data delivers a coefficient of 0.15 ($p$-value of 0.000) and the estimate of $\zeta$ in regression (6) is 0.69 ($p$-value of 0.000). Thus, the results reported in the main text are robust.

\subsubsection*{Joint Regressions}
Instead of running regressions (5) and (6) separately, I estimate
\begin{equation}
\widehat{\gamma}_t = \alpha + \beta\widehat{\sigma}_{\pi,t} + \zeta\widehat{\rho}_{\pi,t}+u_t.  \label{joint}
\end{equation}
Table \ref{tab:kg_vola_joint} shows that the results are robust to this change in specification.

\begin{table}[h!]

\caption{Attention, Inflation Volatility and Inflation Persistence}
\centering 
\vspace{-0.2cm}

\begin{tabular}{lcccc}
\hline\hline\vspace{-0.45cm}\\
 &  \multicolumn{2}{c}{Survey of Professional Forecasters} & &\multicolumn{1}{c}{Michigan Survey}  \\ \cline{2-3} \cline{5-5} \\\vspace*{-0.9cm}\\
Estimator & Blundell-Bond & Pooled OLS && OLS \\
\hline\\\vspace*{-0.9cm}\\
$\widehat{\beta}$ & $0.04^{***}$&$0.05^{***}$ & &$0.06^{***}$ \\
s.e. & (0.0153) & (0.0128) &  &(0.0150) \\
$\widehat{\zeta}$ &$0.59^{***}$ & $0.65^{***}$  && $0.31^{***}$\\
s.e. & (0.0597) & (0.0499) & &(0.0772)\\
$N$ & 165 & 165 && 163
\\\hline\hline
\end{tabular}%
\label{tab:kg_vola_joint}\vspace{0.2cm}
 \begin{minipage}{1\textwidth}

 \footnotesize{
Note: This table shows the results of regression \eqref{joint}. Standard errors are robust with respect to heteroskedasticity.  $ ^{***}:$ $p$-value $<$ 0.01, $ ^{**}:$ $p$-value $<$ 0.05, $ ^{*}:$ $p$-value $<$ 0.1.
 }%
 \end{minipage}
\end{table}

\clearpage
\newpage

\subsubsection*{Controlling for Average Inflation}
A potential confounder in regression \eqref{joint} above is the average level of inflation. Thus, I now control for the average level of inflation, computed as the average inflation rate in the respective 10-year window. In particular, I run the following regression
\begin{equation}
\widehat{\gamma}_t = \alpha + \beta\widehat{\sigma}_{\pi,t} + \zeta\widehat{\rho}_{\pi,t} + \omega \widehat{\bar{\pi}}_t+u_t,  \label{avg}
\end{equation}
where $\widehat{\bar{\pi}}_t$ is the estimated average inflation rate. Table \ref{tab:kg_mean} reports the results for the professional forecasters. We see that the volatility and the persistence of inflation are positively related with attention and that these relationships are statistically significant even when controlling for the average level of inflation. The average level of inflation, on the other hand, does not have a positive, statistically-significant, effect on the estimated attention when we control for the volatility and persistence of inflation. These results are consistent with the underlying theoretical model. 

\begin{table}[h!]

\caption{Controlling for Average Inflation}
\centering 
\vspace{-0.2cm}

\begin{tabular}{lcccc}
\hline\hline\vspace{-0.45cm}\\
 &  \multicolumn{2}{c}{Survey of Professional Forecasters}   \\ \cline{2-3}  \\\vspace*{-0.9cm}\\
Estimator & Blundell-Bond & Pooled OLS  \\
\hline\\\vspace*{-0.9cm}\\
$\widehat{\beta}$ & $0.13^{***}$ & $0.06^{***}$  \\
s.e. & (0.0264) & (0.0200)&  \\
$\widehat{\zeta}$ &  $0.69^{***}$ & $0.82^{***}$  \\
s.e. & (0.1153)& (0.0824) \\
$\widehat{\omega}$ & $-0.02^{**}$ & 0.01 \\
s.e. &  (0.0095) & (0.0082)\\
$N$ & 165 & 165 
\\\hline\hline
\end{tabular}%
\label{tab:kg_mean}\vspace{0.2cm}
 \begin{minipage}{1\textwidth}

 \footnotesize{
Note: This table shows the results of regression \eqref{avg}. Standard errors are robust with respect to heteroskedasticity.  $ ^{***}:$ $p$-value $<$ 0.01, $ ^{**}:$ $p$-value $<$ 0.05, $ ^{*}:$ $p$-value $<$ 0.1.
 }%
 \end{minipage}
\end{table}

\clearpage
\newpage

\subsubsection*{Quasi-Panel of Consumers}
The Survey of Consumers does not follow consumers over time. Therefore, I could not allow for individual-specific fixed effects but rather consider average and/or median inflation expectations. I now group the survey respondents into four groups, based on their income. The SoC provides data on this starting in the last quarter of 1979. 

Table \ref{tab:reg1_qp} shows the results. The first two columns report the results for the split point in 1990, and the third and fourth column for the split point in 1995. We see that the estimated attention levels using this quasi panel are similar to the ones obtained using average expectations (Table 1).

\begin{table}[ht]

\caption{Regression Results of Equation (4), Quasi-Panel}
\centering 
\vspace{-0.2cm}

\begin{tabular}{lcccc}
\hline\hline\vspace{-0.45cm}\\
 &   \multicolumn{4}{c}{Survey of Consumers}  \\ \cline{2-5} \\\vspace*{-0.9cm}\\
 &  $<1990$\phantom{<} &  $\geq 1990$\phantom{<} &  $<1995$\phantom{<} &  $\geq 1995$\phantom{<}  \\
\hline\\\vspace*{-0.9cm}\\
$\widehat{\gamma}$ & 0.77 & 0.33 & 0.70 & 0.29  \\
s.e. & (0.0933) & (0.0263) & (0.1078) & (0.0289) \\
\hline\\\vspace*{-0.9cm}\\
$N$ & 160 & 480 & 240 & 400
\\\hline\hline
\end{tabular}%
\label{tab:reg1_qp}\vspace{0.2cm}
 \begin{minipage}{1\textwidth}

 \footnotesize{
Note: This table shows the results from regression (4), estimated using the \citet{blundell1998initial} estimator, for consumers grouped into four groups, based on their income. The standard errors, reported in parentheses, are robust with respect to heteroskedasticity and serial correlation.
 }%
 \end{minipage}
\end{table}

Table \ref{tab:kg_vola_qp} shows the results of regressions (5) and (6) (first column), as well as of the joint regression \eqref{joint}, using this quasi panel of consumers. We see that the results are robust and that there is indeed a significantly positive relation between attention and inflation volatility, as well as between attention and inflation persistence.

\begin{table}[h!]

\caption{Attention, Inflation Volatility and Inflation Persistence}
\centering 
\vspace{-0.2cm}

\begin{tabular}{lcc}
\hline\hline\vspace{-0.45cm}\\
 &  \multicolumn{2}{c}{Survey of Consumers}  \\ \cline{2-3} \\\vspace*{-0.9cm}\\
Estimator & Separate & Joint \\
\hline\\\vspace*{-0.9cm}\\
$\widehat{\beta}$ &$0.13^{***}$ & $0.13^{***}$  \\
s.e. & (0.0106) & (0.0126) \\
$\widehat{\zeta}$ & $0.20^{***}$ & $0.12^{***}$ \\
s.e. & (0.0787) & (0.0620) \\
$N$ & 121 & 121
\\\hline\hline
\end{tabular}%
\label{tab:kg_vola_qp}\vspace{0.2cm}
 \begin{minipage}{1\textwidth}

 \footnotesize{
Note: This table shows the results of regressions (5), (6) (first column) and \eqref{joint} (second column) using a quasi panel of consumers. The attention parameters have been estimated using the BB-estimator. Standard errors are robust with respect to heteroskedasticity.  $ ^{***}:$ $p$-value $<$ 0.01, $ ^{**}:$ $p$-value $<$ 0.05, $ ^{*}:$ $p$-value $<$ 0.1.
 }%
 \end{minipage}
\end{table}
\clearpage
\newpage

\subsubsection*{Volatility and Persistence of Inflation Expectations}
Table \ref{tab:kg_vola_perc_5} shows the results of regressions (5) and (6) using the volatility and persistence of inflation expectations instead of actual inflation as independent variables. Standard errors are robust with respect to heteroskedasticity.

\begin{table}[h!]

\caption{Attention, Inflation Volatility and Inflation Persistence}
\centering 
\vspace{-0.2cm}

\begin{tabular}{lcccc}
\hline\hline\vspace{-0.45cm}\\
 &  \multicolumn{2}{c}{Survey of Professional Forecasters} & &\multicolumn{1}{c}{Michigan Survey}  \\ \cline{2-3} \cline{5-5} \\\vspace*{-0.9cm}\\
Estimator & Blundell-Bond & Pooled OLS && OLS \\
\hline\\\vspace*{-0.9cm}\\
$\widehat{\beta}$ &$0.14^{***}$ &$0.16^{***}$ &  &$0.13^{***}$  \\
s.e. &(0.0153)&(0.0098)&& (0.0172)  \\
$\widehat{\zeta}$ & $1.06^{***}$ &$1.21^{***}$ &&$0.23^{***}$ \\
s.e. &(0.1272) & (0.0838)& &(0.0685)\\
$N$ &165 &165 &&163 
\\\hline\hline
\end{tabular}%
\label{tab:kg_vola_perc_5} \vspace{0.2cm}
 \begin{minipage}{1\textwidth}

 \footnotesize{
Note: This table shows the results of regressions (5) and (6) using the volatility and persistence of inflation expectations instead of actual inflation as dependent variables. Standard errors are robust with respect to heteroskedasticity. $ ^{***}:$ $p$-value $<$ 0.01, $ ^{**}:$ $p$-value $<$ 0.05, $ ^{*}:$ $p$-value $<$ 0.1.
 }%
 \end{minipage}
\end{table}
\clearpage
\newpage

\subsubsection*{Window Length}
As predicted by the underlying model of optimal information acquisition, I showed that there is indeed a positive relationship between attention to inflation and inflation volatility, as well as between attention and inflation persistence. 
In the baseline specification, I relied on a rolling-window approach in which every window was 10 years. Tables \ref{tab:kg_vola_joint_5} and \ref{tab:kg_vola_joint_15} show that these results are robust to using different window lengths, namely 5 and 15 years.

\begin{table}[h!]

\caption{Attention, Inflation Volatility and Inflation Persistence}
\centering 
\vspace{-0.2cm}

\begin{tabular}{lcccc}
\hline\hline\vspace{-0.45cm}\\
 &  \multicolumn{2}{c}{Survey of Professional Forecasters} & &\multicolumn{1}{c}{Michigan Survey}  \\ \cline{2-3} \cline{5-5} \\\vspace*{-0.9cm}\\
Estimator & Blundell-Bond & Pooled OLS && OLS \\
\hline\\\vspace*{-0.9cm}\\
$\widehat{\beta}$ &-0.01 &$0.06^{***}$ & &$0.13^{***}$ \\
s.e. &(0.1643) &(0.0185) &  &(0.0411) \\
$\widehat{\zeta}$ & $0.73$ &$0.44^{***}$  && $0.40^{***}$\\
s.e. &(0.6731) & (0.0551)& &(0.1547)\\
$N$ & 185&185 && 183
\\\hline\hline
\end{tabular}%
\label{tab:kg_vola_joint_5}\vspace{0.2cm}
 \begin{minipage}{1\textwidth}

 \footnotesize{
Note: This table shows the results of regression \eqref{joint} using windows of 5 years each. Standard errors are robust with respect to heteroskedasticity. $ ^{***}:$ $p$-value $<$ 0.01, $ ^{**}:$ $p$-value $<$ 0.05, $ ^{*}:$ $p$-value $<$ 0.1.
 }%
 \end{minipage}
\end{table}

\begin{table}[h!]

\caption{Attention, Inflation Volatility and Inflation Persistence}
\centering 
\vspace{-0.2cm}

\begin{tabular}{lcccc}
\hline\hline\vspace{-0.45cm}\\
 &  \multicolumn{2}{c}{Survey of Professional Forecasters} & &\multicolumn{1}{c}{Michigan Survey}  \\ \cline{2-3} \cline{5-5} \\\vspace*{-0.9cm}\\
Estimator & Blundell-Bond & Pooled OLS && OLS \\
\hline\\\vspace*{-0.9cm}\\
$\widehat{\beta}$ &$0.01$ &0.01 & &$0.07^{***}$ \\
s.e. & (0.0124)&(0.0115) &  &(0.0136) \\
$\widehat{\zeta}$ &$0.90^{***}$& $1.00^{***}$ && $0.43^{***}$\\
s.e. & (0.0603)& (0.0552)& &(0.0706)\\
$N$ &145 & 145&& 143
\\\hline\hline
\end{tabular}%
\label{tab:kg_vola_joint_15}\vspace{0.2cm}
 \begin{minipage}{1\textwidth}

 \footnotesize{
Note: This table shows the results of regression \eqref{joint} using windows of 15 years each. Standard errors are robust with respect to heteroskedasticity. $ ^{***}:$ $p$-value $<$ 0.01, $ ^{**}:$ $p$-value $<$ 0.05, $ ^{*}:$ $p$-value $<$ 0.1.
 }%
 \end{minipage}
\end{table}

\clearpage
\newpage

\subsubsection*{Attention over Time}
\begin{figure}[ht]
\caption{Attention and Inflation Volatility over Time}
\centering    
\includegraphics[scale=0.4]{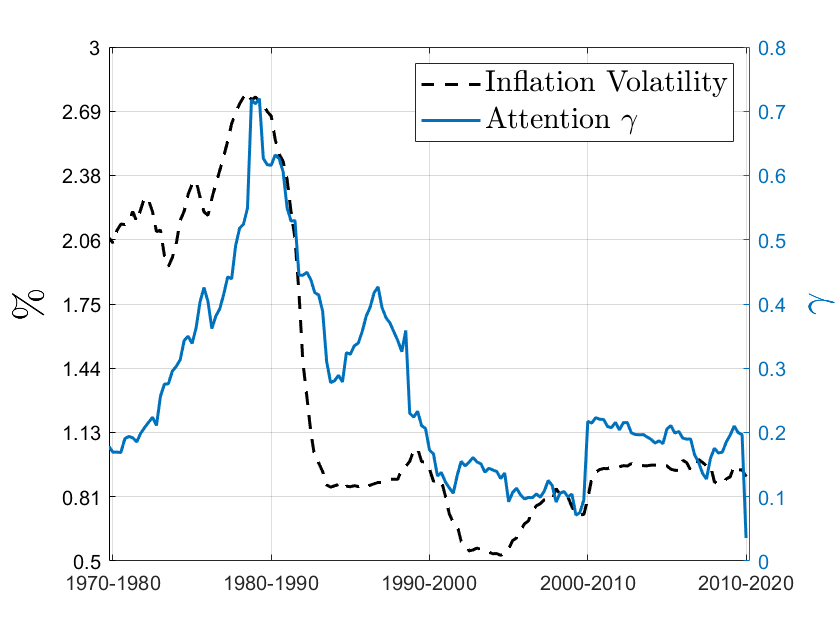}
\vspace{0cm}\\ 
 \begin{minipage}{\textwidth}
 \footnotesize{
 Notes: This figure shows the estimated attention levels, $\gamma$, (black-solid line) from the SPF consensus forecasts, together with the volatility of GDP deflator inflation (blue-dashed lines).}%
 \end{minipage}
\label{fig:overtime}
\end{figure}

Figure \ref{fig:overtime} shows the estimated attention levels, $\gamma$, (black-solid line) from the SPF consensus forecasts, together with the volatility of GDP deflator inflation (blue-dashed lines). We clearly see the aforementioned decrease in attention over time, as well as the positive correlation of attention and inflation volatility.

\subsubsection*{AR(2) Beliefs}

In the main part of the paper, I assume that agents believe that inflation follows an AR(1). I now show that the main results are unchanged when instead assuming that agents believe that inflation follows an AR(2). 

Assume agents have a law of motion of demeaned inflation given by
\begin{equation*}
    \pi_t = \phi_1 \pi_{t-1} + \phi_2 \pi_{t-2} + \nu_t,
\end{equation*}
and that they receive signals of the form $s_t = \pi_t + \epsilon_t$, and the disturbances $\nu$ and $\epsilon$ are i.i.d.\ zero-mean normally-distributed random variables with time-invariant volatilities.\footnote{Assuming a signal of the form $s_t = \varphi_1 \pi_t + \varphi_2 \pi_{t-1} + \epsilon_t$ (which would be the optimal signal in the AR(2) case, see \cite{mackowiak2018dynamic} and \cite{jurado2023rational}) would lead to an identification issue because the weights in the signal $\varphi_1$ and $\varphi_2$ could not be disentangled from the updating weights, $k_1$ and $k_2$.} 
Following the arguments in \cite{hamilton1994time}, the steady state Kalman filter then yields:
\begin{equation*}
    \begin{pmatrix}
        \pi^e_{t+1|t} \\ \pi^e_{t|t}
    \end{pmatrix} = \begin{pmatrix}
        \phi_1 & \phi_2 \\ 1 & 0
    \end{pmatrix} \begin{pmatrix}
        \pi^e_{t|t-1} \\ \pi^e_{t-1|t-1}
    \end{pmatrix} + \begin{pmatrix}
        k_1 \\ k_2
    \end{pmatrix}\left(\pi_t + \epsilon_t - \pi^e_{t|t-1} \right). 
\end{equation*}
The second equation, shifted one period backwards, yields
\begin{equation*}
    \pi^e_{t-1|t-1} = \pi^e_{t-1|t-2} + k_2\left( \pi_{t-1} + \epsilon_{t-1} - \pi^e_{t-1|t-2}\right),
\end{equation*}
which we can then plug into the first equation to obtain an expression for the one-period ahead expectations $\pi^e_{t+1|t}$:
\begin{equation}
           \pi^e_{t+1|t} = \phi_1 \pi^e_{t|t-1} + \phi_2\pi^e_{t-1|t-2} + k_1 \left( \pi_t - \pi^e_{t|t-1}\right) + \underbrace{K_2}_{=\phi_2k_2} \left(\pi_{t-1} - \pi^e_{t-1|t-2} \right) + u_t, \label{eq:ar2}
\end{equation}
where $k_1$ and $k_2$ are the coefficients in the Kalman gain matrix (denoted by $K$ in \cite{hamilton1994time}), and $u_t = k_1\epsilon_t + K_2 \epsilon_{t-1}$. I consider household average and median expectations when estimating regression \eqref{eq:ar2}, and I include an intercept. To account for serial correlation in the error term, I apply the Newey-West estimator using four lags (\citet{newey1987simple}). Table \ref{tab:ar2} shows the results.

\begin{table}[h!]

\caption{AR(2) Perceived Law of Motion}
\centering 
\vspace{-0.2cm}

\begin{tabular}{lcccc}
\hline\hline\vspace{-0.45cm}\\
 &   \multicolumn{4}{c}{Michigan Survey}  \\ \cline{2-5}  \\\vspace*{-0.9cm}\\
&\multicolumn{2}{c}{Average Expectations} & \multicolumn{2}{c}{Median Expectations} \\ \cline{2-3} \cline{4-5}
& pre 1990 & post 1990 & pre 1990 & post 1990\\
\hline\\\vspace*{-0.9cm}\\
$\widehat{k}_1$ & $0.86$ & 0.30 & 0.446 & 0.21 \\
s.e. & (0.167) & (0.084) & (0.107) & (0.0599)   \\
$\widehat{K}_2$ & -0.35 & -0.14 & -0.105 & -0.0889 \\
s.e. & (0.156) & (0.082) & (0.112) & (0.0529) \\
$\widehat{\phi}_1$ & $1.11$ & 0.88 & 1.09 & 0.82 \\
s.e. & (0.178) & (0.100) & (0.226) & (0.127) \\
$\widehat{\phi}_2$ & -0.413 & -0.136 & -0.276 & -0.166 \\
s.e. & (0.190) & (0.083) &  (0.2203) & (0.085)
\\\hline\hline
\end{tabular}%
\label{tab:ar2}\vspace{0.2cm}
 \begin{minipage}{1\textwidth}

 \footnotesize{
Note: This table shows the results of regression \eqref{eq:ar2} for household average and median expectations. Standard errors are robust with respect to heteroskedasticity and serial correlation (Newey-West with 4 lags). 
 }%
 \end{minipage}
\end{table}

When using the sum of the two updating gains, $k_1$ + $K_2$, as the measure of attention, we see that attention clearly decreased from the period before the 1990s to the period after the 1990s. Before 1990, the sum of the two updating gains when focusing on average expectations is 0.51 (with s.e.\ of 0.13, so statistically significantly different from 0 at the 1\% significance level). After 1990, this measure of attention decreased to 0.16 (again, statistically significantly different from 0 at the 1\% significance level). When focusing on median expectations, the sum of the two before the 1990s is 0.341 and it decreased in the period after 1990 to 0.12. This decrease in attention---measured as how strongly households update their expectations---is consistent with the findings in the main text in Section 2.

Furthermore, the estimates in Table \ref{tab:ar2} show that none of the estimated coefficients that are arise due to the AR(2) assumption---the estimates for $K_2$ and $\phi_2$---are statistically significant at the 1\% level (often, not even at the 5\% or 10\% level). When estimating attention in exactly the same way as in Section 2, i.e., computing attention as $\frac{k_1}{\phi_1}$, I obtain estimates that are very close to the ones in the main text where I ignore the effects arising from the second lag of inflation in the perceived law of motion. For average expectations, I estimate for the period before 1990 an attention parameter of 0.78 (it was 0.75 in the main text), and for the period after 1990 a value of 0.34 (0.31 in the main text). For median expectations, I obtain attention estimates of 0.41 for the period before 1990 (it is 0.43 in the main text) and 0.255 for the period after 1990 (0.24 in the main text).
These findings give empirical support to the assumption that the perceived law of motion for inflation follows an AR(1).

\subsection{Other Measures of Attention}\label{app:news}
In this section, I provide complementary evidence to the one presented in Section 2 based on news coverage, based on the share of survey respondents that answer "I don't know" when asked about their inflation expectations, and based on assessing the accuracy of nowcasts of inflation.

\subsubsection*{News Coverage of Inflation}
\begin{figure}[ht]
\caption{News Coverage of Inflation}
\centering    
\begin{tabular}{cc}  
(a) New York Times & (b) Washington Post\\
\includegraphics[scale=0.3]{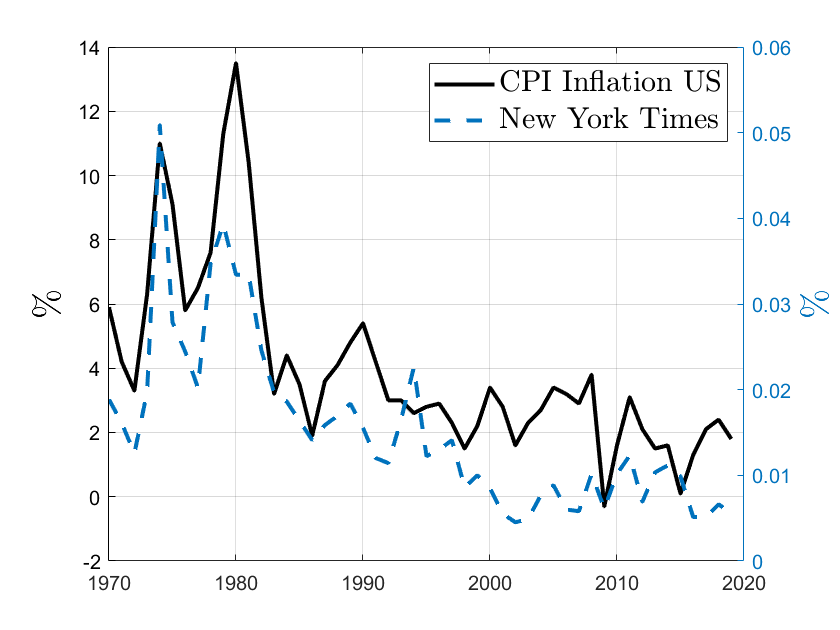}  &
\includegraphics[scale=0.3]{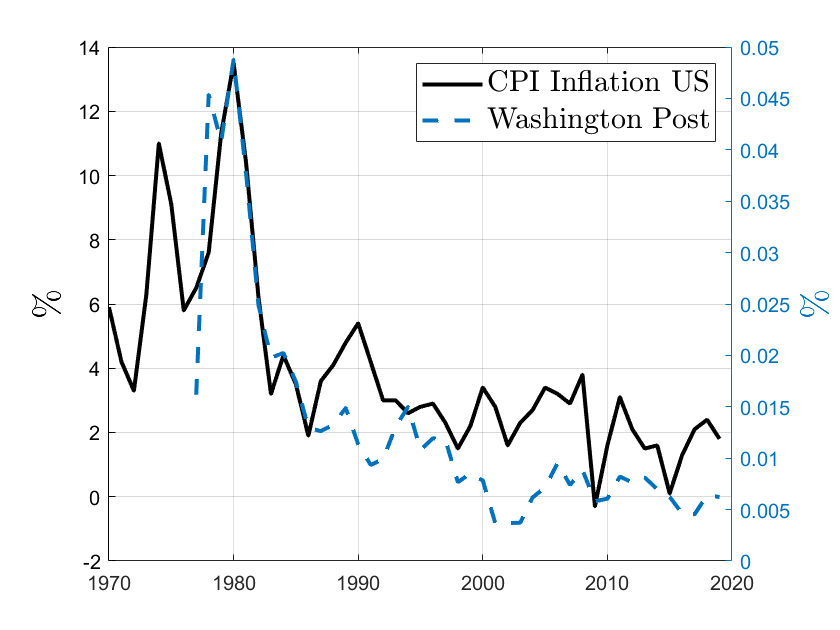}  

 \end{tabular}%
\vspace{0cm}\\ 
 \begin{minipage}{\textwidth}
 \footnotesize{
 Notes: This figure shows the relative frequency (blue dashed lines, right axis) of the word ``inflation'' in the New York Times (left) and the Washington Post (right). The black solid line shows annual U.S. CPI inflation (left axis).}%
 \end{minipage}
\label{fig:news}
\end{figure}

Figure \ref{fig:news} shows the relative frequency of the word "inflation" among all words in two major U.S. newspapers (blue-dashed lines), the New York Times (left panel) and the Washington Post (right panel), together with the annual U.S. CPI inflation (black-solid lines). It is evident that news coverage is higher in times of high and volatile inflation as was the case during the 1970s and early 1980s. Moreover, the figure suggests that the public's attention to inflation---proxied here by news coverage---has not always been as low as in recent years, but declined over time.

\begin{figure}[ht]
\caption{Book Coverage of Inflation}
\vspace{1em}
\centering    
\begin{tabular}{cc}  
(a) Google Books English & (b) Google Books Spanish\\
\includegraphics[scale=0.3]{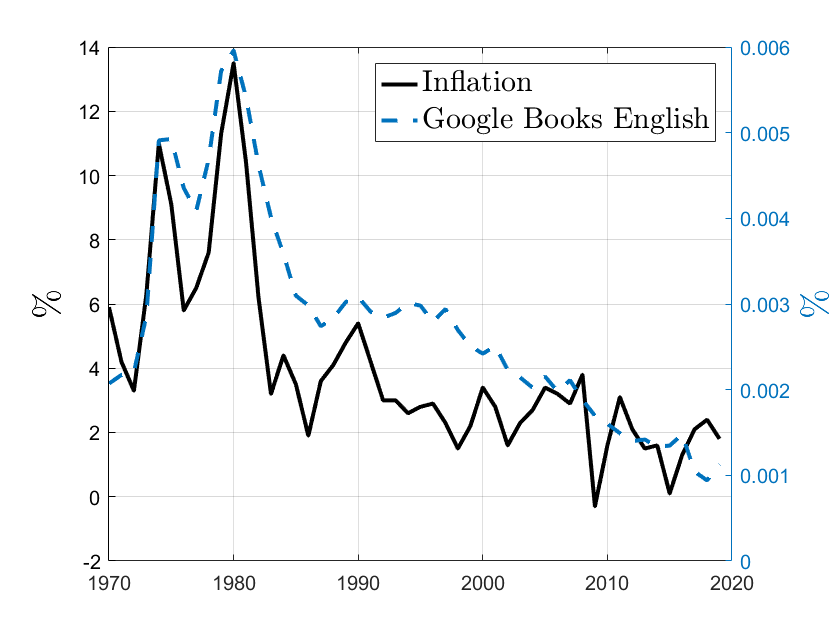}  &
\includegraphics[scale=0.3]{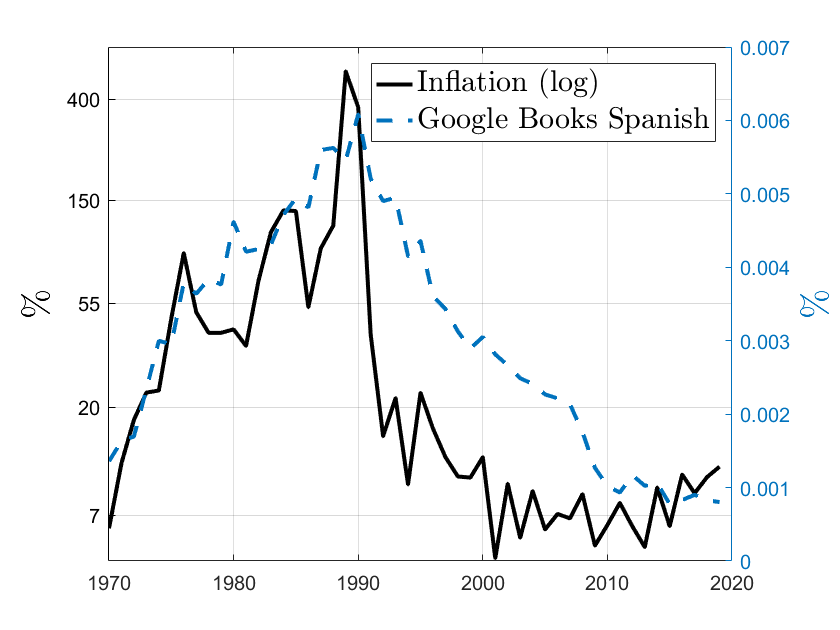}  

 \end{tabular}%
\vspace{0.5cm}\\ 
 \begin{minipage}{\textwidth}
 \footnotesize{
 Notes: The blue dashed lines show the frequency of the words ``inflation'' and ``inflaci\'on'', respectively, in English and Spanish books, according to Google Books Ngram Viewer. The black solid line shows the corresponding inflation rates.}%
 \end{minipage}
\label{fig:news2}
\end{figure}

In Figure \ref{fig:news2}, we see that a similar picture emerges when looking at the coverage of "inflation" in books, according to \textit{Google Books Ngram Viewer}. In the left panel, we see that "inflation" is covered more frequently in English books written in times of high inflation. But this is not simply a U.S. phenomenon. To see this, I show the same statistic for books written in Spanish for the word "inflaci\'on". To contrast this with inflation, the black solid line shows the average inflation (in logs) of the four largest Spanish-speaking countries, weighted by their 2020 population size. These are Argentina, Colombia, Mexico and Spain. Again, we observe that attention to inflation---measured by book coverage---is higher in times of high and volatile inflation.

\clearpage

\subsubsection*{Households answering "I don't know"}
Another potential measure of people's inattention is to see how the share of survey respondents that answer the question about their inflation expectations with "I don't know" changes over time. The Michigan survey provides these shares. Following a rolling-windows approach, I compute for each 10-year window the estimated attention parameter $\widehat{\gamma}_t$ as well as the average share of households within these 10 years that say "I don't know". When using average expectations to estimate $\gamma$, I find that the two are strongly negatively correlated with a correlation coefficient of -0.4. When regressing the share of "don't know" respondents on the estimated attention parameter, I obtain a regression coefficient of -7.23 (p-value of 0.010). When controlling for the window-specific inflation volatility, autocorrelation and average inflation rate, the regression coefficient is -5.64 with a p-value of 0.016. When using median expectations to compute $\gamma$, the results are even slightly stronger. The raw correlation is -0.47, the regression coefficient without controls is -4.29 with a p-value of 0.000, and when adding controls it equals -3.16 with a p-value of 0.013.
These results indicate that inattention (as measured by the share of respondents answering "I don't know") is lower in times my measure of attention, $\gamma$, is higher. Thus, these findings support the view that my measure of attention indeed captures people's attention to inflation.

\subsubsection*{Accuracy of Nowcasts}
In a setup in which the agent cannot distinguish between a trend and a cyclical component of inflation with time-varying volatilities of these two components, if the trend component's contribution to overall inflation increases, the agent's forecast would become more responsive to current inflation, too, similarly to an increase in the attention parameter $\gamma$. To differentiate these two models, I therefore now also consider \textit{nowcasts} of inflation and their accuracy.
The optimal attention choice problem presented in Section 2 says that more attentive agents receive more precise signals about current inflation and should therefore make smaller nowcast errors in times of high attention. This prediction is exclusive to the proposed model of attention and does not apply to the alternative model of the trend and cycle component of inflation. Using the nowcasts from the Survey of Professional Forecasters, I now test this prediction of the model. To do this, I take the absolute value of forecast errors of current inflation or the squared forecast errors as my two measures of the accuracy of the forecasters' nowcasts. I then compute the average across all forecasters and estimate a time series of these average forecast errors using a rolling-windows approach where each window is 10 years long. Similarly, I estimate the window-specific volatility and persistence of perceived inflation. Consistent with my theory of attention, I find strong negative correlations between inflation volatility and forecast errors, as well as between inflation persistence and forecast errors. This holds for both measures of forecast errors, i.e., for the absolute values and the squared values of forecast errors. The correlations are indeed quite strong. For the squared forecast errors, I find a correlation with inflation persistence of -0.58 and with inflation volatility of -0.25. For the absolute values of forecast errors, the correlation with persistence is -0.64 and with inflation volatility -0.40.
These results are consistent with the recent findings in \cite{weber2023tell} who find that households that report to pay more attention to inflation have inflation expectations that are much closer to the actual level of inflation.

\clearpage
\newpage

\section{Model Details and Derivations}\label{app:derivations}
In this Appendix, I derive the New Keynesian Phillips Curve and the aggregate IS equation under limited attention (equations (7) and (8)). To nest the case of positive trend inflation (see Section 4.3.3), I do this for the general case that allows for an arbitrary steady state inflation rate, following \cite{ascari2012trend} who derive the New Keynesian Phillips Curve with positive trend inflation with Rotemberg price adjustment costs \citep{Rotemberg1982}. A key assumption that I make throughout the following derivations is that firms set their prices optimally for a given inflation expectation, and these inflation expectations are provided by forecaster, as in \cite{adam2011inflation}.

\paragraph{Households.}
There is a representative household obtaining utility from consumption and disutility from working, with lifetime utility
\begin{equation}
\Tilde{E}_{0}\sum_{t=0}^{\infty }\beta^{t}Z_t\left[ \frac{C_t^{1-\sigma}}{1-\sigma}-\Psi\frac{H_t^{1+\nu}}{1+\nu}\right] ,  \label{utilorig}
\end{equation}
where $C_{t}$ is consumption of the final good, $H_{t}$ is hours worked, $\beta$ is the household's time discount factor, and $\tilde{E}_t$ denotes the household's subjective expectations operator based on information available in period $t$. $Z_t$ are exogenous preference shocks. The parameters $\sigma$ and $\nu$ pin down the relative risk aversion and the inverse Frisch labor elasticity, respectively. $\Psi$ is the utility weight on hours worked.

Households maximize their lifetime utility subject to the flow budget constraints 
\begin{align}
& C_{t}+B_{t}= w_tH_t+\frac{1+i_{t-1}}{1+\pi _{t}}B_{t-1}+\frac{T_{t}^H}{P_{t}}, \quad\text{ for all } t, \label{BC_final}
\end{align}%
where $B_t$ is the real value of government bonds, $w_t$ the real wage, $\pi_t$ is the net inflation rate, and $i_t$ the nominal interest rate. $T_t^H$ denotes lump-sum taxes and transfers from the government.

Maximizing \eqref{utilorig} subject to \eqref{BC_final} yields the Euler equation
\begin{equation}
    Z_tC_t^{-\sigma} = \beta(1+i_t)\Tilde{E}_t\left[Z_{t+1}C_{t+1}^{-\sigma}\frac{1}{1+\pi_{t+1}}
\right], \label{eq:euler_model}
\end{equation}
and the labor-leisure condition
\begin{equation}
w_{t}C_t^{-\sigma}= \Psi H_t^{\nu}. \label{eq:ll}
\end{equation}%

\paragraph{Final goods producer.} There is a representative final good producer that aggregates the intermediate goods $Y_t(j)$ to a final good $Y_t$, according to
\begin{equation}
    Y_t = \left( \int_0^1 Y_t(j)^{\frac{\epsilon-1}{\epsilon}} dj\right)^{\frac{\epsilon}{\epsilon -1}},
\end{equation}
with $\epsilon > 1$. Nominal profits are given by
$
    P_t \left( \int_0^1 Y_t(j)^{\frac{\epsilon-1}{\epsilon}} dj\right)^{\frac{\epsilon}{\epsilon -1}} - \int_0^1P_t(j)Y_t(j)dj,
$
and profit maximization gives rise to the demand for each variety $j$:
\begin{equation}
    Y_t(j) = \left( \frac{P_t(j)}{P_t} \right)^{-\epsilon} Y_t.
\end{equation}
Thus, demand for variety $j$ is a function of its relative price, the price elasticity of demand $\epsilon$ and aggregate output $Y_t$. The aggregate price level is given by
\begin{equation}
    P_t = \left( \int_0 ^1 P_t(j)^{1-\epsilon}dj \right)^{\frac{1}{1-\epsilon}}.
\end{equation}

\paragraph{Intermediate goods producers.} Intermediate producer of variety $j$ produces output $Y_t(j)$ using labor $H_t(j)$ as its only input
\begin{equation}
    Y_t(j) = H_t(j).
\end{equation}
All intermediate producers pay the same wage $w_t$ and a sales tax (or subsidy) $\tau_t$, which in steady state is set such that profits in steady state are 0. These taxes are given back to firms in a lump-sum fashion, denoted $t^F_t(j)$. Taxes are assumed to be constant in the efficient economy, i.e., absent price rigidities, but fluctuate around their steady state in the economy with price rigidities in order to give rise to exogenous cost-push shocks. 

Each intermediate firm has two managers: one is responsible for the firm's forecasts and the other manager sets the price of firm $j$ given these forecasts, similar to the setup in, e.g., \cite{adam2011inflation}.

When adjusting the price, the firm is subject to a \cite{Rotemberg1982} price-adjustment friction. 
Their per-period profits (in real terms) are given by
\begin{equation}
    (1-\tau_t)P_t(j)\left( \frac{P_t(j)}{P_t}\right)^{-\epsilon} \frac{Y_t}{P_t} - w_t H_t(j) - \frac{\psi}{2}\left(\frac{P_t(j)}{P_{t-1}(j)} - 1\right)^2Y_t + t^F_t(j),
\end{equation}
where $\psi \geq 0$ captures the price-adjustment cost parameter. They set prices to maximize
\begin{equation}
    Profits_0(j) = \Tilde{E}^j_0\sum\limits_{t=0}^{\infty}D_{0,t}\left[(1-\tau_t)P_t(j)\left( \frac{P_t(j)}{P_t}\right)^{-\epsilon} \frac{Y_t}{P_t} - mc_t H_t(j) - \frac{\psi}{2}\left(\frac{P_t(j)}{P_{t-1}(j)} - 1\right)^2Y_t + t^F_t(j)\right], \notag
\end{equation}
where $D_{0,t}\equiv \beta^t\left(\frac{C_{t}}{C_0}\right)^{-\sigma} $ is the stochastic discount factor (for simplicity, I assume that firm managers are not subject to preference shocks), $mc_t = w_t$ denotes the real marginal cost which is the same for every firm. Using the production function to substitute for $H_t(j)$ and the demand for firm $j$'s product from the final goods producer, the corresponding first order condition is then given by
\begin{align*}
        T_t (\epsilon-1)\left(\frac{P_t(j)}{P_t} \right)^{-\epsilon}  = &\epsilon mc_t \left(\frac{P_t(j)}{P_t} \right)^{-\epsilon -1}  - \psi\left(\frac{P_t(j)}{P_{t-1}(j)} -1\right)\frac{P_t}{P_{t-1}(j)}\\
&+ \beta \psi \tilde{E}^j_t\left[\left(\frac{C_{t+1}}{C_t}\right)^{-\sigma}\left(\frac{P_{t+1}(j)}{P_t(j)}-1\right)\frac{P_{t+1}(j)}{P_t(j)}\frac{P_{t}}{P_t(j)}\frac{Y_{t+1}}{Y_t} \right],
\end{align*}
where $T_t \equiv 1-\tau_t$.

\paragraph{Government.} The government imposes a sales tax $\tau_t$ on sales of intermediate goods, issues nominal bonds, and pays lump-sum
taxes and transfers $T_{t}^H$ to households and $t^F_t(j)$ to firms. The real government budget constraint is given
by 
\begin{equation*}
B_{t}=B_{t-1}\frac{1+i_{t-1}}{\Pi_t}+\frac{T_{t}^H}{P_{t}}-\tau Y_{t} + t^f_t.
\end{equation*}%
Lump-sum taxes and transfers are
set such that they keep real government debt constant at the initial level $%
B_{-1}/P_{-1}$, which I set to zero. 

\paragraph{Steady State.} The resource constraint is given by $Y_t = C_t + \frac{\psi}{2}\left( \frac{P_t}{P_{t-1}} -1\right)^2 Y_t$. With steady state inflation denoted by $\bar{\Pi}$, it follows that in steady state we have
\begin{equation}
    C = \left(1-\frac{\psi}{2}(\bar{\Pi}-1)^2\right)Y.
\end{equation}
From the production function, we have $Y = H$ and marginal costs are equal to the real wage, $mc = w$. Given the assumption that intermediate producers receive the subsidy $\tau$ which is set to induce the efficient steady state, it follows that $mc = w =1$. Since all firms set the same price in steady state, it follows from the intermediate producers' first-order conditions, that 
\begin{equation*}
    mc = \frac{(1-\tau)(\epsilon -1)}{\epsilon} + \psi \bar{\Pi}(\bar{\Pi} -1)\frac{1}{\epsilon}(1-\beta),
\end{equation*}
which implies that the steady state subsidy is equal to
\begin{equation*}
    T = 1- \tau = \frac{\epsilon -\psi(\bar{\Pi} -1)\bar{\Pi}(1-\beta)}{\epsilon -1}.
\end{equation*}
From the labor-leisure equation and the resource constraint, we obtain
\begin{equation*}
    Y = \left(\frac{1}{\Psi\left(1-\frac{\psi}{2}(\bar{\Pi} -1)^2 \right)^{\sigma}}\right)^{\frac{1}{\nu + \sigma}}.
\end{equation*}

\paragraph{Linearization.} Linearizing the Euler equation \eqref{eq:euler_model} yields
\begin{equation}
    \widehat{c}_t = \tilde{E}_t\widehat{c}_{t+1} - \varphi\left(\Tilde{i}_t - \Tilde{E}_t\pi_{t+1} - (\widehat{z_t} - \Tilde{E}_tz_{t+1})\right), \label{eq:euler_lin}
\end{equation}
where $\varphi \equiv \frac{1}{\sigma}$.
Linearizing the resource constraint, we obtain
\begin{equation*}
    \widehat{y}_t = \widehat{c}_t + \frac{\psi(\bar{\Pi}-1)\bar{\Pi}}{1-\frac{\psi}{2}(\bar{\Pi}-1)^2}\pi_t,
\end{equation*}
where $\pi_t$ now denotes inflation in deviations from its steady state value. Plugging this into \eqref{eq:euler_lin}, we get
\begin{equation*}
    \widehat{y}_t = \tilde{E}_t\widehat{y}_{t+1} + \frac{\psi(\bar{\Pi}-1)\bar{\Pi}}{1-\frac{\psi}{2}(\bar{\Pi}-1)^2}\left[ \pi_t - \tilde{E}_t{\pi}_{t+1}\right] - \varphi \left(\Tilde{i}_t - \Tilde{E}_t\pi_{t+1} - (\widehat{z_t} - \Tilde{E}_tz_{t+1})\right).
\end{equation*}
In order to express this in terms of the output gap, rather than output, we have to solve for the efficient output that prevails in the economy absent price rigidities (denoted by a "$^*$"). From the production function, we have $Y^*_t = H^*_t$. The real wage is constant $w^*_t = 1$. From the labor-leisure equation \eqref{eq:ll}, we get that potential output is therefore also constant and equal to
\begin{equation}
    Y^*_t = \Psi^{-\frac{1}{\nu+\sigma}}.
\end{equation}
Thus, potential output in log-deviations is 0. The Euler equation in the flexible-price economy is therefore given by
\begin{equation}
    0 =  - \varphi\left(r_t - (\widehat{z_t} - \Tilde{E}_tz_{t+1})\right). \notag
\end{equation}
Since the natural rate is defined as the real rate that prevails under flexible prices, $r_t$, it follows that
\begin{equation}
    r^n_t = \widehat{z}_t - \Tilde{E}_tz_{t+1}.
\end{equation}
Substituting $\widehat{z}_t - \Tilde{E}_tz_{t+1}$ with $r^n_t$ in \eqref{eq:euler_lin} and using that $\widehat{y}_t = y^{gap}_t $, since potential output in deviations from steady state is 0, yields the aggregate IS equation
\begin{equation*}
    y^{gap}_t = \tilde{E}_ty^{gap}_{t+1} + \frac{\psi(\bar{\Pi}-1)\bar{\Pi}}{1-\frac{\psi}{2}(\bar{\Pi}-1)^2}\left[ \pi_t - \tilde{E}_t{\pi}_{t+1}\right] - \varphi \left(i_t - \Tilde{E}_t\pi_{t+1} - (\widehat{z_t} - \Tilde{E}_tz_{t+1})\right).
\end{equation*}
I assume for the most part of the analysis that output gap expectations are rational, $\tilde{E}_ty^{gap}_{t+1} = E_ty^{gap}_{t+1}$, and that households believe that inflation follows an AR(1) process and that they receive signals of the form $s_t = \pi_t + \varepsilon_t$ with normally distributed noise $\varepsilon_t$ (see Section \ref{sec:theory} for details). For tractability, I abstract from noise shocks and therefore assume that $\varepsilon_t = 0$ for all $t$ but that the household behaves as if there was noise. This then gives rise to the law of motion for inflation expectations stated in equation (9). Taking everything together, we can therefore write 
\begin{equation*}
    y^{gap}_t = E_ty^{gap}_{t+1} + \frac{\psi(\bar{\Pi}-1)\bar{\Pi}}{1-\frac{\psi}{2}(\bar{\Pi}-1)^2}\left[ \pi_t - \pi^e_{t+1|t}\right] - \varphi \left(i_t - \pi^e_{t+1|t} - r^n_t\right).
\end{equation*}
In the case of zero trend inflation, $\bar{\Pi} = 1$, this collapses to 
\begin{equation*}
    y^{gap}_t = E_ty^{gap}_{t+1} - \varphi \left(i_t - \pi^e_{t+1|t} - r^n_t\right),
\end{equation*}
as stated in equation (8).

In order to derive the Phillips Curve, we need to linearize the intermediate producers' first-order condition. This condition is given by
\begin{align*}
        \underbrace{T_t (\epsilon-1)\left(\frac{P_t(j)}{P_t} \right)^{-\epsilon}}_{I}  = &\underbrace{\epsilon mc_t \left(\frac{P_t(j)}{P_t} \right)^{-\epsilon -1}}_{II}  \underbrace{- \psi\left(\frac{P_t(j)}{P_{t-1}(j)} -1\right)\frac{P_t}{P_{t-1}(j)}}_{III}\\
&+ \underbrace{\beta \psi \tilde{E}^j_t\left[\left(\frac{C_{t+1}}{C_t}\right)^{-\sigma}\left(\frac{P_{t+1}(j)}{P_t(j)}-1\right)\frac{P_{t+1}(j)}{P_t(j)}\frac{P_{t}}{P_t(j)}\frac{Y_{t+1}}{Y_t} \right]}_{IV}
\end{align*}
The linearization of the terms $I$ to $IV$ yields
\begin{align*}
    I: \quad \quad & (\epsilon -1 )T \widehat{T}_t - \epsilon(\epsilon -1) T \widehat{p}^j_t + (\epsilon -1)\epsilon T \widehat{p}_t \\
    II: \quad \quad & \epsilon(\widehat{mc}_t - (1+\epsilon)\widehat{p}^j_t + (1+\epsilon)\widehat{p}_t) \\
    III: \quad \quad & -\psi\bar{\Pi}(\bar{\Pi}-1)\widehat{p}_t - \psi\bar{\Pi}^2\widehat{p}^j_t + \psi\bar{\Pi}(2\bar{\Pi}-1)\widehat{p}^j_{t-1} \\
    IV: \quad \quad & \beta \psi \bar{\Pi}^2 \Tilde{E}^j_t\pi^j_{t+1} + \beta\psi\bar{\Pi}(\bar{\Pi} -1 )\Bigg[ (1-\sigma)\tilde{E}^j_t \widehat{y}_{t+1} + (\sigma -1 )\widehat{y}_t \\ &+  \frac{\sigma \psi (\bar{\Pi} -1 )}{1- \frac{\psi}{2}(\bar{\Pi} -1 )^2}\left(\Tilde{E}^j_t\pi_{t+1} - \pi_t \right) + \Tilde{E}^j_t\pi^j_{t+1} + \widehat{p}_t - \widehat{p}^j_t \Bigg],
\end{align*}
where I used the linearized resource constraint to arrive at the expression $IV$. When trend inflation is zero, $\bar{\Pi} = 1$, the expression $IV$ becomes simply $\beta\psi \Tilde{E}^j_t\pi^j_{t+1}$. Since, I focus on the case $\sigma = 1$, the terms relating to output in expression $IV$ drop out. Thus, the only reason why prices may differ across firms $j$ is due to different forecasts of future inflation (either of aggregate or of firm-specific inflation). Following the assumption in \cite{adam2011inflation}, I assume that these forecasts are provided by forecasters that are different from the price setting managers. Given that there are no idiosyncratic shocks, I assume that the forecaster of firm $j$ expects firm-specific inflation to be equal to aggregate inflation, $\Tilde{E}^j_t\pi^j_{t+1} = \Tilde{E}^j_t\pi_{t+1}$. These forecasters then form their inflation expectations in the same way as households and as detailed in the limited attention problem in Section 2. Given that I abstract from noise (but signals are perceived as noisy), all forecasters then receive exactly the same signals and therefore also form their expectations equally. Note, that a weaker assumption would be sufficient to arrive at the following results: namely, that all forecasters receive the same signal.\footnote{If all forecasters receive the same signal, they form identical inflation expectations. However, the noise term would give rise to a "belief shock" in the Phillips Curve (and also the aggregate IS equation). I ignore these belief shocks.} Thus, $\Tilde{E}^j_t\pi^j_{t+1} = \Tilde{E}^j_t\pi_{t+1} = \Tilde{E}_t\pi_{t+1}$. Therefore, all price setters set the same price, $\widehat{p}^j_t = \widehat{p}_t$. Thus, the belief that firm-specific inflation coincides with aggregate inflation is satisfied in equilibrium, which confirms the forecasters' belief and she therefore does not have an incentive to update that belief.

Using this, and 
\begin{equation*}
    \widehat{mc}_t = (1+\nu)\widehat{y}_t - \frac{\psi(\bar{\Pi}-1)\bar{\Pi}}{1-\frac{\psi}{2}(\bar{\Pi}-1)^2}\pi_t,
\end{equation*}
which follows from the labor-leisure equation, the production function and the resource constraint, we then obtain the following Phillips Curve with trend inflation and limited attention:
\begin{align*}
    \pi_t &= \zeta \left[ \frac{\epsilon(1+\nu)}{\psi}\widehat{y}^{gap}_t + \beta \Bar{\Pi}^2\pi^e_{t+1|t} + u_t + \Xi \pi^e_{t+1|t} \right]
\end{align*}
where
\begin{align*}
    \zeta &\equiv \frac{1}{\bar{\Pi}(2\bar{\Pi}-1)+\frac{\beta \psi \Bar{\Pi}(\bar{\Pi} - 1)^2}{1-\frac{\psi}{2}(\Bar{\Pi} -1)^2
}+\frac{\epsilon\Bar{\Pi}(\Bar{\Pi}-1)}{1-\frac{\psi}{2}(\Bar{\Pi} -1)^2}} \\
\Xi &\equiv \beta\Bar{\Pi}(\Bar{\Pi}-1)\left[1+\frac{\psi(\Bar{\Pi} -1)}{1-\frac{\psi}{2}(\Bar{\Pi} -1)^2} \right],
\end{align*}
as stated in Section 4.3.3, and with the cost-push shock defined as $u_t \equiv -\frac{(\epsilon -1)T}{\psi}\widehat{T}_t$. For the case of zero trend inflation, we get $\zeta = 1$ and $\Xi = 0$, so that the Phillips Curve reduces to 
\begin{align}
\pi_t &= \beta  \pi^e_{t+1|} + \kappa y^{gap}_t + u_t, \notag
\end{align}
which is equation (7).

\clearpage
\newpage

\section{Analytical Results and Proofs}\label{app:fg}
To see how lower attention weakens the effectiveness of forward guidance, consider the following stylized experiment.\footnote{For clarity, I focus on the case $\rho_{\pi} = 1$. I discuss the general case with $\rho_{\pi} < 1$ in section \ref{app:extensions_fg}.} 
The economy is hit by a negative natural rate shock in period $t=0$ that pushes the nominal interest rate to the effective lower bound, i.e.,  $r^n_0 < 0$  and $i_0 = -\underline{i}$. In $t=1$, the natural rate returns to its steady state value and stays there indefinitely, $r^n_t = 0$ for all $t\geq 1$. From period $t=2$ onwards, the output gap, and the real rate are back at their steady states, $y^{gap}_t = 0$ and $i_t -\pi^e_{t+1|t} = 0$ for all $t\geq 2$.

To model forward guidance, the real rate is assumed to be below the natural rate in $t=1$. To make it comparable across different degrees of attention, I impose that 
\begin{equation}
r_1 \equiv i_1-\pi^e_{2|1} < 0
\notag
\end{equation} 
is the same for all $\gamma$ and known in advance.\footnote{This is different to \citet{angeletos2018forward}, where private agents are uncertain about future policies.} Hence, forward guidance here means to announce a certain value for the {real} rate. I discuss the implications of forward guidance via the {nominal} rate in section \ref{app:extensions_fg}.
In the following, I assume that $\left(-\underline{i}-r^n_0+r_1\right)$ is negative, which means that the announced policy, captured by $r_1 < 0$, \textit{makes up} for the binding lower bound in $t=0$, captured by $-\underline{i}-r^n_0 > 0$.

Given the real rate $r_1$ and the fact that $y^{gap}_2 = 0$, the Euler equation in $t=1$ determines the output gap in period 1 as
\begin{equation}
y_1^{gap} = -\varphi\left(r_1\right) > 0. \label{compensation1}
\end{equation}
Equation \eqref{compensation1} captures the \textit{make-up policy:} by keeping the real rate below the natural rate, output is above potential after the lower-bound constraint stops to be binding.

In $t=0$, the ELB binds and the natural rate is negative. Thus, the Euler equation in $t=0$ yields
\begin{equation}
y_0^{gap} = \underbrace{-\varphi\left(r_1\right)}_{=E_0 y^{gap}_1}- \varphi\left(-\underline{i}-\pi^e_{1|0}-r^n_0 \right).  \notag
\end{equation}
Substituting the law of motion for inflation expectations
\begin{equation}
\pi^e_{1|0} = (1-\gamma)\pi^e_{0|-1}+\gamma \pi_0, \notag
\end{equation}
into the Phillips Curve
\begin{equation}
\pi_0 =  \frac{\beta}{1-\beta\gamma}(1-\gamma)\pi^e_{0|-1}+\frac{\kappa}{1-\beta\gamma}y^{gap}_0 \notag
\end{equation}
yields an expression for inflation expectations:
\begin{align*}
\pi^e_{1|0} &=   \frac{1-\gamma}{1-\beta\gamma}\pi^e_{0|-1}+\frac{\kappa\gamma}{1-\beta\gamma}y^{gap}_0.
\end{align*}
Putting everything together, we arrive at the following result.
\begin{proposition}\label{prop:fg1}
The output gap in the period when the shock hits, $t=0$, is given by
\begin{equation}
\begin{aligned}
y_0^{gap} 
&= -\frac{\varphi\left(1-\beta\gamma\right)}{1-\gamma(\beta+\varphi\kappa)} \left[-\underline{i}-r^n_0+r_1\right] +\frac{\varphi(1-\gamma)}{1-\gamma(\beta+\varphi\kappa)}\pi^e_{0|-1} \label{ygap_fg1}
\end{aligned}
\end{equation}
and inflation in $t=0$ is given by
\begin{equation}
\begin{aligned}
\pi_0 &=-\frac{\kappa\varphi}{1-\gamma(\beta+\varphi\kappa)}
\left[-\underline{i}-r^n_0+r_1\right]
+(1-\gamma)\left[\frac{\beta}{1-\beta\gamma}+\frac{\varphi}{1-\gamma(\beta+\varphi\kappa)}\right]\pi^e_{0|-1}. \label{pi_fg1}
\end{aligned}
\end{equation}
\end{proposition}

Proposition \ref{prop:fg1} captures the effectiveness of forward guidance on the output gap and inflation in the period when the shock hits.
Assuming $\left(1-\gamma\left(\beta+\varphi\kappa\right)\right)$ is positive makes sure that forward guidance has a stimulating effect on output and inflation in $t=0$.
Proposition \ref{prop:fg1} captures several channels how a change in attention affects the economy's response to forward guidance, which I discuss in the following two corollaries.
\begin{corollary}\label{lemma_fg11}
Lower attention weakens
\begin{itemize}
\item[(i)] the negative effects of the shock, 
\item[(ii)] the positive effects of forward guidance,
\item[(iii)] the positive effects of a decrease in the lower bound $-\underline{i}$ 
\end{itemize}
on the output gap and inflation.
\end{corollary}
Corollary \ref{lemma_fg11} follows from the fact that the terms $\frac{\varphi\left(1-\beta\gamma\right)}{1-\gamma(\beta+\varphi\kappa)}$ and $\frac{\kappa\varphi}{1-\gamma(\beta+\varphi\kappa)}$ in front of $\left[-\underline{i}-r^n_0+r_1\right]$ are both increasing in $\gamma$.
Points $(i)$ and $(ii)$ capture the main trade off of lower attention. While lower attention has a stabilizing effect via more anchored inflation expectations (point $(i)$), it renders forward guidance less effective (point $(ii)$). The reason why forward guidance becomes less effective as attention declines is because inflation expectations increase less in response to the announced policy, and thus, the real rate remains higher.
Point $(iii)$ illustrates an additional drawback of lower attention. A reduction of the effective lower bound, $-\underline{i}$, is less stimulating if agents in the economy are less attentive. Thus, going from a zero lower bound to a lower bound in negative territory, as conducted in several advanced economies over the last ten years, becomes less effective in terms of stimulating output and inflation if the public is inattentive (consistent with the exercise in figure 5). Away from the lower bound, point $(iii)$ implies that the effectiveness of conventional monetary policy via the nominal interest rate becomes less effective as attention declines.

%
How attention matters for the transmission of prior inflation expectations on the output gap and inflation is ambiguous, as the following Corollary shows.
\begin{corollary}\label{corollary_fg31}
Lower attention
\begin{itemize}
\item[(i)] weakens the positive effect of higher prior inflation beliefs, $\pi^e_{0|-1}$, on the output gap if and only if,
\begin{equation}
\left(\beta+\varphi\kappa\right) > 1, \label{cond_prior1}
\end{equation} 
\item[(ii)] weakens the positive effect of higher prior inflation beliefs on inflation if and only if
\begin{equation}
\frac{\beta\left(\beta-1\right)}{\left(1-\beta\gamma\right)^2}+\frac{\varphi\left((\beta+\varphi\kappa)-1\right)}{\left(1-\gamma(\beta+\varphi\kappa)\right)^2} > 0. \label{cond_prior_inf1}
\end{equation}
\end{itemize}
\end{corollary}
Overall, the role of attention for the effects of higher prior beliefs on output and inflation is ambiguous. This is mainly the case because, on the one hand, lower attention implies that agents put more weight on their prior beliefs. On the other hand, as discussed previously, lower attention leads to more stable inflation overall, thus, weakening the effects of prior beliefs. 

Given the calibration in Table 4, conditions \eqref{cond_prior1} and \eqref{cond_prior_inf1} both hold for all $\gamma < 0.99$.
The effects of changes in $\gamma$, however, are numerically small. Thus, an increase in the average inflation rate---which increases average prior beliefs---is a promising monetary instrument to combat the loss of control via forward guidance as attention declines. By \textit{ex-ante} increasing the average inflation rate, the policymaker not only supports higher inflation expectations and thus, lower real rates for a given nominal rate, but also gains additional policy space through the increase in the average nominal rate. 

\subsection{Extensions}\label{app:extensions_fg} 
I now show that all the results go through when relaxing the assumption that $\rho_{\pi} = 1$ and also discuss how forward guidance via the nominal (instead of the real) interest rate changes the results and I also allow for attention heterogeneity across firms and households.
We consider the same stylized experiment but now the law of motion for inflation expectations is given by
\begin{equation}
\pi^e_{1|0} = (1-\rho_{\pi})\bar{\pi}+\rho_{\pi}(1-\gamma)\pi^e_{0|-1}+\rho_{\pi}\gamma \pi_0, \notag
\end{equation}
which can be substituted into the Phillips Curve:
\begin{equation}
\pi_0 =  \frac{\beta}{1-\beta\rho_{\pi}\gamma}\left((1-\rho_{\pi})\bar{\pi}+\rho_{\pi}(1-\gamma)\pi^e_{0|-1}\right)+\frac{\kappa}{1-\beta\rho_{\pi}\gamma}y^{gap}_0. \notag
\end{equation}
Thus, inflation expectations are given by
\begin{align*}
\pi^e_{1|0} &= \frac{1-\rho_{\pi}}{1-\beta\rho_{\pi}\gamma} \bar{\pi} +  \frac{\rho_{\pi}(1-\gamma)}{1-\beta\rho_{\pi}\gamma}\pi^e_{0|-1}+\frac{\kappa\rho_{\pi}\gamma}{1-\beta\rho_{\pi}\gamma}y^{gap}_0.
\end{align*}
Putting everything together, we arrive at the following Proposition.
\begin{proposition}\label{prop:fg}
The output gap in the period when the shock hits, $t=0$, is given by
\begin{equation}
\begin{aligned}
y_0^{gap} 
&= -\frac{\varphi\left(1-\beta\rho_{\pi}\gamma\right)}{1-\rho_{\pi}\gamma(\beta+\varphi\kappa)} \left[-\underline{i}-r^n_0+r_1\right] \notag\\
&+\frac{\varphi}{1-\rho_{\pi}\gamma(\beta+\varphi\kappa)}\left[(1-\rho_{\pi})\bar{\pi}+\rho_{\pi}(1-\gamma)\pi^e_{0|-1}\right] \label{ygap_fg}
\end{aligned}
\end{equation}
and inflation in $t=0$ is given by
\begin{equation}
\begin{aligned}
\pi_0 &=-\frac{\kappa\varphi}{1-\rho_{\pi}\gamma(\beta+\varphi\kappa)}
\left[-\underline{i}-r^n_0+r_1\right]
+ \left(1-\rho_{\pi}\right)\left[\frac{\beta}{1-\beta\rho_{\pi}\gamma}+\frac{\varphi}{1-\rho_{\pi}\gamma(\beta+\varphi\kappa)}\right]\bar{\pi}\\
&+
\rho_{\pi}(1-\gamma)\left[\frac{\beta}{1-\beta\rho_{\pi}\gamma}+\frac{\varphi}{1-\rho_{\pi}\gamma(\beta+\varphi\kappa)}\right]\pi^e_{0|-1}. \label{pi_fg}
\end{aligned}
\end{equation}
\end{proposition}

Proposition \ref{prop:fg} captures the effectiveness of forward guidance on the output gap and inflation in the period when the shock hits.
The assumption that $\left(1-\rho_{\pi}\gamma\left(\beta+\varphi\kappa\right)\right)$ is positive, makes sure that forward guidance, i.e, a lower $r_1$ has a stimulating effect on output and inflation in $t=0$.
Proposition \ref{prop:fg} captures several channels how a change in attention affects the economy's response to forward guidance, which I now collect in a series of corollaries.
\begin{corollary}\label{lemma_fg1}
Lower attention 
\begin{itemize}
\item[(i)] weakens the negative effect of the shock on impact, 
\item[(ii)] weakens the effects of forward guidance on the output gap and inflation,
\item[(iii)] weakens the stimulative effects of a decrease in the lower bound $-\underline{i}$. 
\end{itemize}
\end{corollary}
Corollary \ref{lemma_fg1} follows from the fact that the terms $\frac{\varphi\left(1-\beta\rho_{\pi}\gamma\right)}{1-\rho_{\pi}\gamma(\beta+\varphi\kappa)}$ and $\frac{\kappa\varphi}{1-\rho_{\pi}\gamma(\beta+\varphi\kappa)}$ in front of $\left[-\underline{i}-r^n_0+r_1\right]$ are both increasing in $\gamma$.
Points $(i)$ and $(ii)$ capture the main trade off of lower attention. While lower attention has a stabilizing effect via more anchored inflation expectations (point $(i)$), it renders forward guidance less effective (point $(ii)$). The reason why forward guidance becomes less effective as attention declines is because inflation expectations increase less in response to the announced policy, and thus, the real rate remains higher.
Point $(iii)$ illustrates an additional drawback of lower attention. A reduction of the effective lower bound, $-\underline{i}$, is less stimulating if agents in the economy are less attentive. Thus, going from a zero lower bound to a lower bound in negative territory, as conducted in several advanced economies over the last ten years, becomes less effective in terms of stimulating output and inflation if the public is inattentive. 
Note, that a decrease in the perceived inflation persistence, $\rho_{\pi}$, has the exact same implications as a decrease in $\gamma$.

The next corollary discusses how changes in attention affect the role of long-run inflation beliefs on the output gap and inflation.
\begin{corollary}\label{corollary_fg2}
Lower attention weakens the positive effects of higher long-run inflation beliefs $\bar{\pi}$ on output and inflation,
\end{corollary}
Corollary \ref{corollary_fg2} says that higher long-run beliefs have a positive effect on inflation and the output gap, but lower attention weakens these effects. However, as long as $\gamma\left(\beta+\varphi\kappa\right)<1$, a higher $\rho_{\pi}$ mutes the effects of $\bar{\pi}$ on the output gap. Since this condition is usually satisfied and because $\rho_{\pi}$ is in general close to 1, the role of high long-run inflation beliefs is quite weak. In the limit case $\rho_{\pi}\to 1$, long-run beliefs become irrelevant.

How attention matters for the transmission of prior inflation expectations on the output gap and inflation is ambiguous, as the following Corollary shows.
\begin{corollary}\label{corollary_fg3}
Lower attention
\begin{itemize}
\item[(i)] weakens the positive effect of higher prior inflation beliefs, $\pi^e_{0|-1}$, on the output gap if and only if,
\begin{equation}
\rho_{\pi}\left(\beta+\varphi\kappa\right) > 1, \label{cond_prior}
\end{equation} 
\item[(ii)] weakens the positive effect of higher prior inflation beliefs on inflation if and only if
\begin{equation}
\frac{\rho_{\pi}\beta\left(\rho_{\pi}\beta-1\right)}{\left(1-\beta\rho_{\pi}\gamma\right)^2}+\frac{\rho_{\pi}\varphi\left(\rho_{\pi}(\beta+\varphi\kappa)-1\right)}{\left(1-\rho_{\pi}\gamma(\beta+\varphi\kappa)\right)^2} > 0. \label{cond_prior_inf}
\end{equation}
\end{itemize}
\end{corollary}
Overall, the role of attention for the effects of higher prior beliefs on output and inflation is ambiguous. This is mainly the case because, on the one hand, lower attention implies that agents put more weight on their prior beliefs. On the other hand, as discussed previously, lower attention leads to more stable inflation overall, thus, weakening the effects of prior beliefs. This can also be seen in the discussion of the Phillips Curve, see Proposition 1.

Given the calibration in Table 4, conditions \eqref{cond_prior} and \eqref{cond_prior_inf} both hold.
The effects of changes in $\gamma$, however, are numerically small. Thus, an increase in the average inflation rate---which also increases average prior beliefs---is a promising monetary instrument to combat the loss of control via forward guidance as attention declines. By \textit{ex-ante} increasing the average inflation rate, the policymaker not only supports higher inflation expectations and thus, lower real rates for a given nominal rate, but also gains additional policy space through the increase in the average nominal rate. Higher average inflation, however, is also costly. In the analysis of optimal policy, later on, I will explore this trade off and characterize the optimal inflation target for different levels of attention.

\subsubsection{Forward Guidance via Nominal Interest Rates} So far, forward guidance was characterized as a promise to keep the \textit{real} rate low. Now, assume that forward guidance is conducted via promising lower \textit{nominal} rates instead. Thus, $i_1$ will be fixed across different $\gamma$. For simplicity, I focus on the case with $\rho_{\pi} = 1$ and $\pi^e_{0|-1} = 0$.  It follows from the Euler equation in $t=1$ that
\begin{align*}
y^{gap}_1  &= -\varphi\left(i_1 - \left(1-\gamma\right)\gamma\pi_0-\gamma\pi_1 \right).
\end{align*}
The Phillips Curve in $t=1$ yields
\begin{align*}
\pi_1&= \frac{(1-\gamma)\gamma	}{1-\beta\gamma}\pi_0 + \frac{\kappa}{1-\beta\gamma}y^{gap}_1,
\end{align*}
so that we get an expression for $y^{gap}_1$ in terms of $\pi_0$:
\begin{equation}
y^{gap}_1 = -\frac{\varphi\left(1-\beta\gamma\right)	}{1-\gamma\left(\beta+\varphi\kappa\right)}i_1 + \varphi(1-\gamma)\gamma\frac{1+\gamma(1-\beta)}{1-\gamma(\beta+\varphi\kappa)}\pi_0. \label{y1_fg}
\end{equation}
Given $\pi^e_{1|0} = \gamma\pi_0$, the Phillips Curve in $t=0$ yields
\begin{equation}
\pi_0 = \frac{\kappa}{1-\beta\gamma}y^{gap}_0, \notag
\end{equation}
and hence, $\pi^e_{1|0} = \frac{\kappa\gamma}{1-\beta\gamma}y^{gap}_0$. Plugging this into the Euler equation in $t=0$ gives
\begin{equation}
y^{gap}_0 = \mathbb{E}_0 y^{gap}_1-\varphi\left(-\underline{i}-\frac{\kappa\gamma}{1-\beta\gamma}y^{gap}_0-r^n_0\right). \notag
\end{equation}
Solving for $y^{gap}_0$ leads to the following Lemma.
\begin{lemma}\label{lemma_fg_nominal}
Forward guidance via the nominal interest rate yields the following output gap
\begin{equation}
y^{gap}_0 = A_1\left[-\frac{\varphi\left(1-\beta\gamma\right)	}{1-\gamma\left(\beta+\varphi\kappa\right)}i_1-\varphi\left(-\underline{i}-r^n_0\right)\right], \label{a1_y}
\end{equation}
 and inflation
 \begin{equation}
 \pi_0 = \frac{\kappa}{1-\beta\gamma}A_1\left[-\frac{\varphi\left(1-\beta\gamma\right)	}{1-\gamma\left(\beta+\varphi\kappa\right)}i_1-\varphi\left(-\underline{i}-r^n_0\right)\right], \label{a1_pi}
 \end{equation}
 where
 \begin{equation}
 A_1 \equiv \frac{1}{1- \varphi(1-\gamma)\gamma\frac{1+\gamma(1-\beta)}{1-\gamma(\beta+\varphi\kappa)}\frac{\kappa}{1-\beta\gamma}-\frac{\varphi\kappa\gamma}{1-\beta\gamma}}. \label{a1_defn}
 \end{equation}
\end{lemma}
Given the calibration in Table 4, $A_1$ is positive and increasing in $\gamma$. 
Thus, promising lower future nominal interest rates can indeed stimulate the economy. But similar to the case in which the policy maker commits to a certain future \textit{real} rate, forward guidance becomes less effective when agents are less attentive.
In fact, all three results from Corollary \ref{lemma_fg1} go through.

Recall equation \eqref{y1_fg}:
\begin{equation}
y^{gap}_1 = -\frac{\varphi\left(1-\beta\gamma\right)	}{1-\gamma\left(\beta+\varphi\kappa\right)}i_1 + \varphi(1-\gamma)\gamma\frac{1+\gamma(1-\beta)}{1-\gamma(\beta+\varphi\kappa)}\pi_0. \notag
\end{equation}
Note, that the first term becomes less negative as $\gamma$ declines. Given the calibration in Table 4, also the second term decreases as attention declines. Thus, for a given $\pi_0$, a particular $i_1$ has weaker effects on the output gap in $t=1$ at lower levels of attention. Since lower attention also weakens the positive effects of forward guidance on $\pi_0$, the output gap (and inflation) stay lower also in $t=1$.

Since inflation in $t=0$ and $t=1$ is lower at smaller values of $\gamma$, also $\pi^e_{2|1}$ will be lower and thus, for a given nominal rate $i_1$, the \textit{real} rate, $r_1 \equiv i_1-\pi^e_{2|1}$, will be higher. Hence, to achieve a certain forward guidance in terms of the real interest rate, the promise in terms of the nominal rate needs to be larger when firms and households are inattentive. Combining this with the findings on the effectiveness of forward guidance via the \textit{real} rate (Proposition \ref{prop:fg}) shows how lower attention renders forward guidance less powerful \textit{even though} the promise in terms of the \textit{nominal rate} is stronger.

\subsubsection{Heterogeneous Attention} So far, I assumed that firms and households are equally attentive. But what if firms and households differ in their attention to inflation? Let us denote firms' attention by $\gamma_F$ and households' attention by $\gamma_H$ with $\gamma_F \neq \gamma_H$. For clarity, I focus on the case with $\rho_{\pi} = 1$ and $\pi^{e,j}_{0|-1}=0$ for $j\in\{F,H\}$. 

\begin{lemma}\label{lemma_heterog}
With heterogeneous attention to inflation, the output gap in $t=0$ is given by
\begin{equation}
y^{gap}_0 = \frac{-\varphi\left(1-\beta\gamma_F\right)}{1-\beta\gamma_F-\kappa\varphi\gamma_H}\left[-\underline{i}+r_1-r^n_0\right], \label{gap_heterog}
\end{equation}
and inflation by
\begin{equation}
\pi_0 = \frac{-\varphi\kappa}{1-\beta\gamma_F-\kappa\varphi\gamma_H}\left[-\underline{i}+r_1-r^n_0\right], \label{pi_heterog}
\end{equation}
where $r_1 \equiv i_1-\pi^{e,H}_{2|1}$ is the real rate given the households' expectations.
\end{lemma}
Lemma \ref{lemma_heterog} shows that a similar result as in Corollary \ref{lemma_fg1} holds under heterogeneous attention levels. 
\begin{corollary}\label{coro_heterog1}
Lower attention of either firms or households
\begin{itemize}
\item[(i)] weakens the negative effect of the shock on the output gap and inflation on impact, 
\item[(ii)] weakens the effects of forward guidance on the output gap and inflation,
\item[(iii)] weakens the stimulative effects of a decrease in the lower bound $-\underline{i}$ on the output gap and inflation. 
\end{itemize}
\end{corollary}
The parts concerning the output gap in Corollary \ref{coro_heterog1} follow because the term in front of the brackets in equation \eqref{gap_heterog} becomes more negative as either of $\{\gamma_F,\gamma_H\}$ increases:
\begin{align*}
\frac{\partial \left[\frac{-\varphi\left(1-\beta\gamma_F\right)}{1-\beta\gamma_F-\kappa\varphi\gamma_H}\right]}{\partial \gamma_F} &=
-\frac{\beta\kappa\varphi^2\gamma_H}{(1-\beta\gamma_F-\kappa\varphi\gamma_H)^2} < 0 \\
\frac{\partial \left[\frac{-\varphi\left(1-\beta\gamma_F\right)}{1-\beta\gamma_F-\kappa\varphi\gamma_H}\right]}{\partial \gamma_H} &=
-\frac{\varphi^2(1-\beta\gamma_F)}{(1-\beta\gamma_F-\kappa\varphi\gamma_H)^2} < 0,
\end{align*}
and the parts concerning inflation because the term $\frac{-\varphi\kappa}{1-\beta\gamma_F-\kappa\varphi\gamma_H}$ in equation \eqref{pi_heterog} becomes more negative as either of $\{\gamma_F,\gamma_H\}$ increases, too.

Thus, if either firms or households (or both) become less attentive, forward guidance becomes less effective. 
In fact, the two degrees of attention reinforce each other, as the following Corollary shows.
\begin{corollary}
Lower levels of households' attention to inflation weaken the effectiveness of forward guidance, especially when firms' attention to inflation is low, and vice-versa.
\end{corollary}
To see this, note that
\begin{align*}
\frac{\partial^2\left[-\frac{\varphi\kappa}{1-\beta\gamma_F-\kappa\varphi\gamma_H}\right]}{\partial\gamma_F\partial\gamma_H } &= \frac{-2\varphi^2\kappa^2\beta}{\left(1-\beta\gamma_F-\kappa\varphi\gamma_H\right)^3} < 0,
\\
\frac{\partial^2 \left[\frac{-\varphi\left(1-\beta\gamma_F\right)}{1-\beta\gamma_F-\kappa\varphi\gamma_H} \right]}{\partial\gamma_F\partial\gamma_H} &= \frac{-\beta\kappa\varphi^2\left[1-\beta\gamma_F+\kappa\varphi\gamma_H\right]}{\left(1-\beta\gamma_F-\kappa\varphi\gamma_H\right)^3}<0.
\end{align*}

\clearpage
\newpage
\subsection{Proof of Proposition 1}\label{app:proofs}
\begin{proof}
The New Keynesian Phillips Curve is given by
\begin{align*}
\pi_t &= \beta \pi^e_{t+1|t}+\kappa y^{gap}_t+u_t.
\end{align*}
Substituting
\begin{equation}
\pi^e_{t+1|t} = \pi^e_{t|t-1}+\gamma\left(\pi_t - \pi^e_{t|t-1}\right) \notag
\end{equation}
for $\pi^e_{t+1|t}$ yields
\begin{align*}
\pi_t &= \beta\left(\pi^e_{t|t-1}+\gamma\left(\pi_t-\pi^e_{t|t-1}\right)\right)+\kappa y^{gap}_t+u_t\\
\Leftrightarrow & \pi_t(1-\beta\gamma)  = \beta\pi^e_{t|t-1}\left(1-\gamma\right)+\kappa y^{gap}_t+u_t\\
\Leftrightarrow & \pi_t  =\frac{ \beta\pi^e_{t|t-1}\left(1-\gamma\right)+\kappa y^{gap}_t+u_t}{(1-\beta\gamma)}\\
\Leftrightarrow & \pi_t =\frac{ \beta\left(1-\gamma\right)}{(1-\beta\gamma)}\pi^e_{t|t-1}+\frac{\kappa }{(1-\beta\gamma)}y^{gap}_t+\frac{u_t}{(1-\beta\gamma)}.
\end{align*}
Now, taking derivatives with respect to $y^{gap}_t$, $u_t$, and $\pi^e_{t|t-1}$, respectively, yields the results (i), (ii), and (iii).
\end{proof}
\clearpage
\newpage
\section{Additional Numerical Results}\label{app:extensions}

\subsection{Non-Rational Output Gap Expectations}\label{app:ey}
In this section, I estimate households' attention to the output gap (using expected unemployment changes as a proxy) and derive the policy implications of non-rational output gap expectations.
I use the Survey of Consumers from the University of Michigan which asks respondents about what they think will happen to unemployment over the next 12 months. A drawback of that question is that respondents give a qualitative answer, saying that they expect unemployment to either "go up", "stay about the same", or "go down". Following \cite{bhandari2019survey}, I translate these qualitative answers into quantitative answers (see \cite{bhandari2019survey}, or \cite{pfauti2022behavioral} for details). One assumption I need for this is that I have to impose what "about the same" means. I assume that survey respondents answer "about the same" when they believe that unemployment will change less than 0.15\% which is half a standard deviation of unemployment changes over the period 1978-2019.

I then estimate attention to unemployment, $\gamma^y$, in the same way I estimate attention to inflation in Section 2, and I do so separately for the period before 1990 and the period after 1990. 
Table \ref{tab:unemp} shows the results. Attention to unemployment slightly increased from 0.088 before the 1990s ($\widehat{\gamma}_{y,<1990}$ ) to 0.100 after the 1990s ($\widehat{\gamma}_{y,\geq 1990}$). These differences, however, are not statistically significant, as the last column indicates. Similarly, when I set the break point at 2000, I estimate attention levels of 0.098 before 2000, and 0.099 after 2000. Again, the difference between the two is not statistically significantly different from 0. These results therefore indicate that while there was a strong decline in people's attention to inflation, their attention to unemployment did not change.
\begin{table}[h]
 \caption{Attention to Unemployment}\label{tab:unemp}
 \centering
\begin{tabular}{lccc}
\hline \hline
&   $\widehat{\gamma}_{y,<1990}$ & $\widehat{\gamma}_{y,\geq 1990}$  & $p$-val.\ $\widehat{\gamma}_{y,<1990}  = \widehat{\gamma}_{y,\geq 1990} $\\\hline\vspace{-0.4cm}\\
Estimate & 0.088 & 0.100 & 0.554 \\\vspace{-0.5cm}\\
s.e.\ &   (0.0379) & (0.0263)
\\\vspace{-0.4cm}\\
\hline \hline
\end{tabular}\\\vspace{.1cm}
\begin{minipage}{1\textwidth}
\footnotesize{Notes: This table shows the estimated attention parameters with respect to unemployment, separately for the period before the 1990s ($\widehat{\gamma}_{y,<1990}$) and the period after 1990 ($\widehat{\gamma}_{y,\geq 1990}$). The last column shows the $p$-value for the null-hypothesis that the two coefficients are the same. Standard errors are robust with respect to serial correlation and heteroskedasticity (Newey-West with four lags).
}%
 \end{minipage}
\end{table}

Now, to understand the policy implications of limited attention to the output gap, I impose that output gap expectations are given by
\begin{equation}
    y^{gap,e}_{t+1|t} = y^{gap,e}_{t|t-1} + \gamma^y\left(y^{gap}_t - y^{gap,e}_{t|t-1}  \right).  \label{eq:ey}
\end{equation}
With these expectations, the aggregate IS equation is given by
\begin{equation}
y^{gap}_t = y^{gap,e}_{t+1|t} - \varphi\left(i_t - \pi^e_{t+1|t} - r^n_t \right),
\end{equation}
whereas the Phillips Curve and the Taylor rule remain unchanged.

\begin{figure}[ht]
\caption{Impulse Response Functions with Non-Rational Output Gap Expectations}
\centering 
\begin{tabular}{cccc}  
\multicolumn{2}{c}{\includegraphics[scale=0.35]{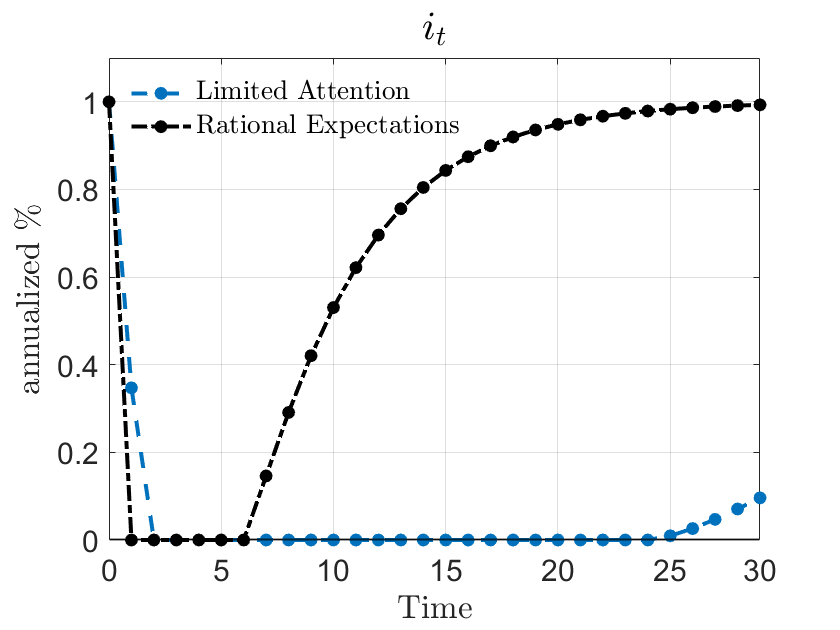}} & \multicolumn{2}{c}{\includegraphics[scale=0.35]{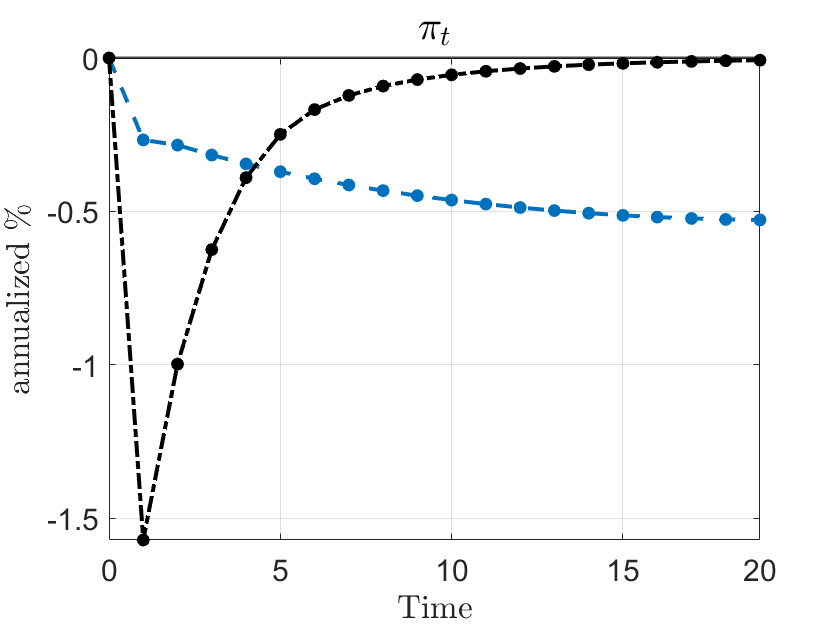}} 
\\ \multicolumn{2}{c}{\includegraphics[scale=0.35]{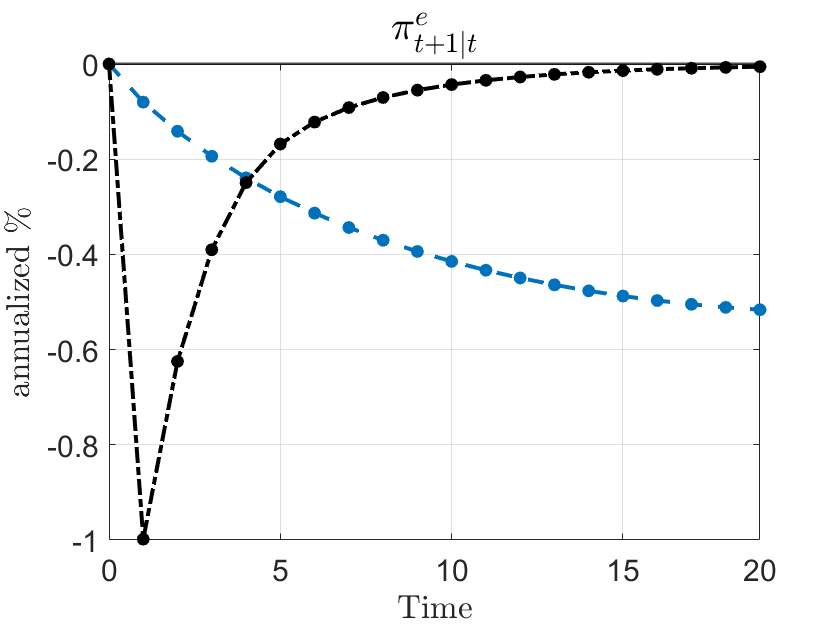}}
&  \multicolumn{2}{c}{ \includegraphics[scale=0.35]{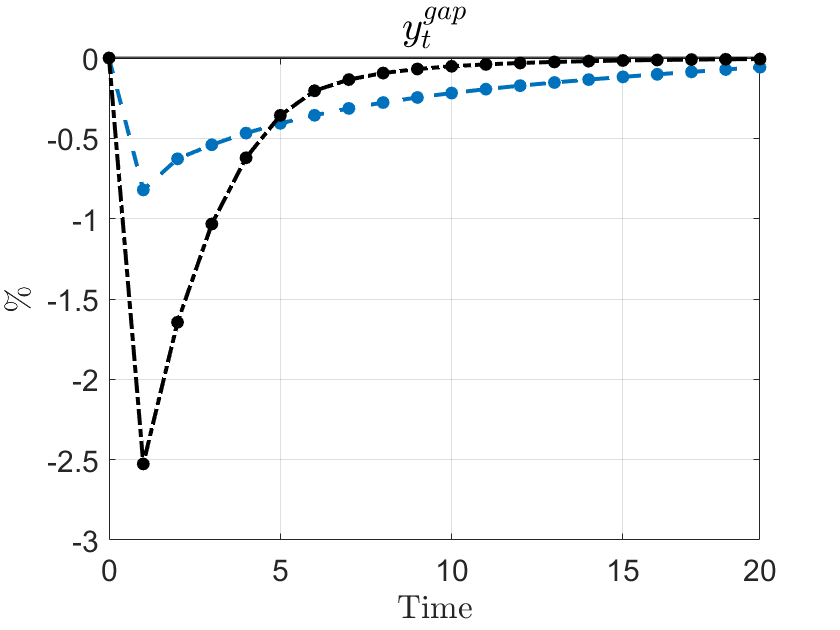}}
 \end{tabular}%
 \label{fig:ey}
 \par
\vspace{0.3cm}
 \begin{minipage}{1\textwidth}

 \footnotesize{
Note: This figure shows the impulse-response functions of the nominal interest rate (upper-left panel), inflation (upper-right panel), inflation expectations (lower-left) and the output gap (lower-right) to a negative natural rate shock of three standard deviations in the case where output gap expectations are given by equation \eqref{eq:ey}. The blue-dashed lines show the case for the limited-attention model and the black-dashed-dotted lines for the rational expectations model. Everything is in terms of percentage deviations from the respective steady state levels, except the nominal rate is in levels.
 }%
 \end{minipage}

\end{figure}

Figure \ref{fig:ey} shows the impulse response functions of the main variables in this economy after a negative three-standard deviation natural rate shock and with $\gamma^y = 0.1$ (I set $\gamma^{\pi}$ to 0.3, as in Section 3 and keep the rest of the calibraton also unchanged). We see that the inflation-attention traps get exacerbated. The reason is that now make-up policies are even less effective because not only inflation expectations are backward looking but also output gap expectations. Thus, even though there is interest-rate smoothing in the Taylor rule which features some form of make-up policy, this is not effective in stimulating expectations and thus, the economy remains stuck at the ELB even longer. Furthermore, inflation, inflation expectations, and now also the output gap stay below their initial values very persistently. 

Table \ref{tab:yexp} shows the implications of limited attention to the output gap for optimal policy. The upper part of the table shows the optimal inflation target and welfare when output gap expectations are rational, and the lower part shows the results when output gap expectations are non-rational. The table highlights the following two main results: (i) limited attention to the output gap increases the optimal inflation target quite substantially and decreases welfare (independent of the level of $\gamma^y$ or $\gamma^{\pi}$), and (ii) higher attention to the output gap reduces the optimal inflation target and increases welfare. The second result mirrors the main result regarding attention to inflation: lower attention (to inflation or the output gap) is welfare deteriorating in the presence of an effective lower bound constraint on nominal interest rates.

\begin{table}[h]
 \caption{Non-rational output gap expectations}\label{tab:yexp}
 \centering
\begin{tabular}{lccc}
\hline \hline
&   Inflation Target & Welfare \\\hline\vspace{-0.4cm}\\
\underline{Rational $E_ty^{gap}_{t+1}$} \\\vspace{-0.5cm}\\
$\gamma^{\pi} = 0.25$ & 1.18\% & -0.0053 \\\vspace{-0.5cm}\\
$\gamma^{\pi} = 0.2$ & 1.20\% & -0.0054 \\\vspace{-0.5cm}\\
$\gamma^{\pi} = 0.1$ & 1.82\% & -0.0105 \\\vspace{-0.3cm}\\
\underline{Limited attention $y^{gap,e}_{t+1|t}$} \\\vspace{-0.5cm}\\
$\gamma^{\pi} = 0.25$, $\gamma^y = 0.075$ & 2.46\% & -0.009 \\\vspace{-0.5cm}\\
$\gamma^{\pi} = 0.2$, $\gamma^y = 0.075$ & 3.22\% & -0.014 \\\vspace{-0.5cm}\\
$\gamma^{\pi} = 0.2$, $\gamma^y = 0.125$ & 2.99\% & -0.012
\\\vspace{-0.5cm}\\
$\gamma^{\pi} = 0.1$, $\gamma^y = 0.125$ & 3.04\% & -0.013
\\\vspace{-0.5cm}\\
$\gamma^{\pi} = 0.1$, $\gamma^y = 0.15$ & 2.86\% & -0.011 
\\\vspace{-0.4cm}\\
\hline \hline
\end{tabular}\\\vspace{.1cm}
\begin{minipage}{1\textwidth}
\footnotesize{Notes: This table shows the implications of limited attention to the output gap for the optimal inflation target and welfare, for different combinations of $\gamma^y$ (attention to the output gap) and $\gamma^{\pi}$ (attention to inflation).
}%
 \end{minipage}
\end{table}

\subsection{Different Taylor Rule}\label{app:taylor}
\begin{figure}[ht]
\caption{IRFs to Natural Rate Shock for Taylor rule \eqref{tay2}}
\centering 
\begin{tabular}{cccc}  
\multicolumn{2}{c}{\includegraphics[scale=0.3]{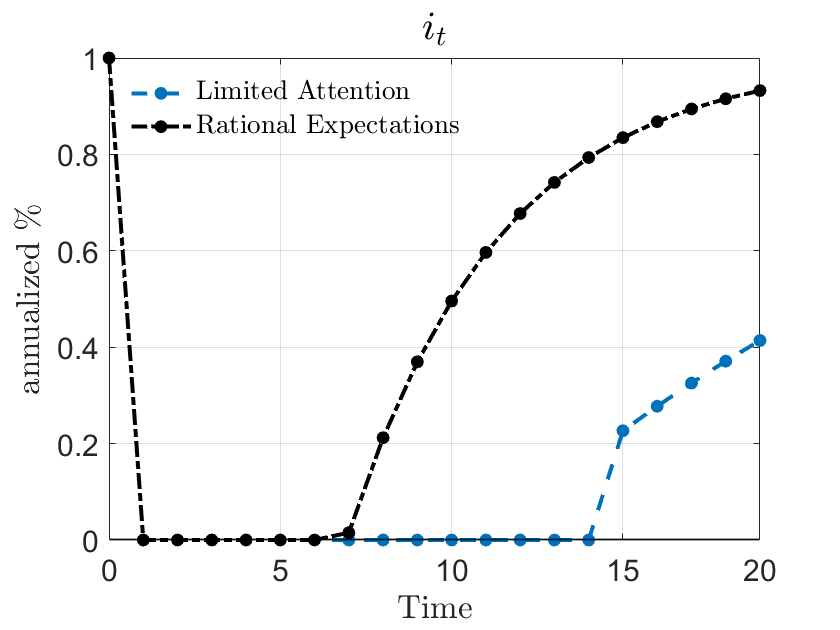}} & \multicolumn{2}{c}{\includegraphics[scale=0.3]{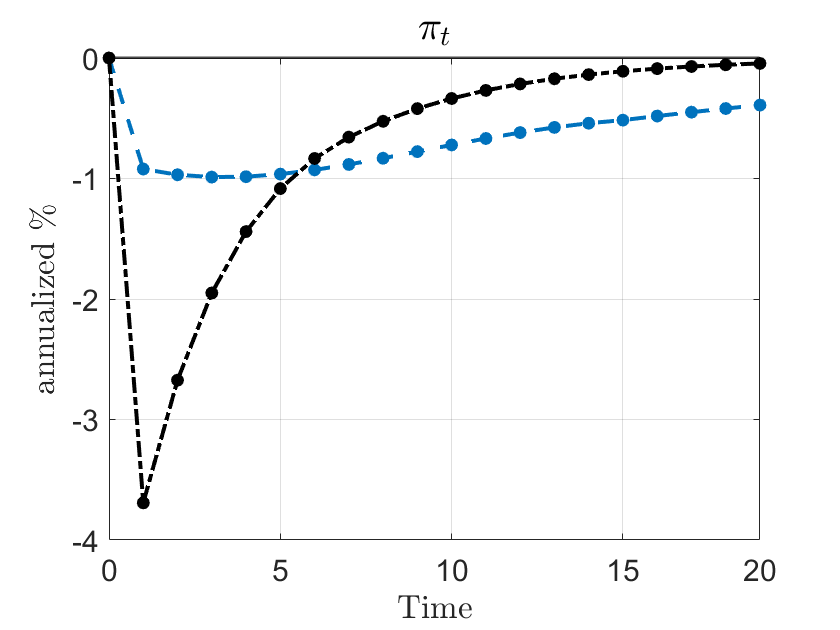}} 
\\ \multicolumn{2}{c}{\includegraphics[scale=0.3]{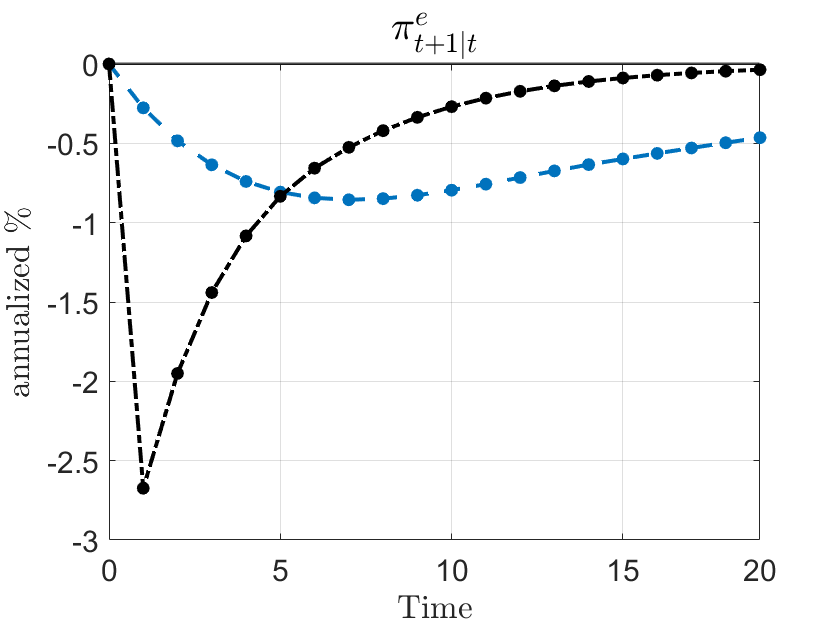}}
&  \multicolumn{2}{c}{ \includegraphics[scale=0.3]{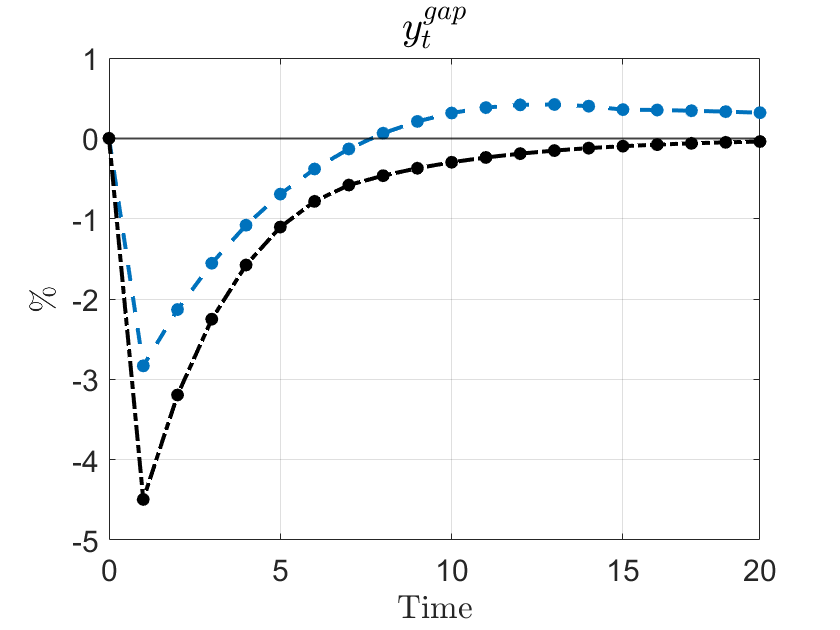}}
 \end{tabular}%
 \label{fig:irf_rn_tay}
 \par
\vspace{0.3cm}
 \begin{minipage}{1\textwidth}

 \footnotesize{
Note: This figure shows the impulse-response functions of the nominal interest rate (upper-left panel), inflation (upper-right panel), inflation expectations (lower-left) and the output gap (lower-right) to a negative natural rate shock of three standard deviations. The blue-dashed lines show the case for the limited-attention model and the black-dashed-dotted lines for the rational expectations model. Everything is in terms of percentage deviations from the respective steady state levels, except the nominal rate is in levels.
 }%
 \end{minipage}

\end{figure}

To show that the exact specification of the Taylor rule is not essential for the occurence of inflation-attention traps, Figure \ref{fig:irf_rn_tay} shows the impulse-response functions of the nominal interest rate, inflation, inflation expectations and the output gap for the model in which the Taylor rule absent the ELB is given by
\begin{equation}
i_t = 1.5 \pi_t. \label{tay2}
\end{equation}

\clearpage
\newpage

\subsection{Forecast Errors}\label{app:fe}

\begin{figure}[ht]
\caption{Impulse Response Functions of Forecast Errors}
\vspace{1em}
\centering 
\begin{tabular}{cc}  
\includegraphics[scale=0.3]{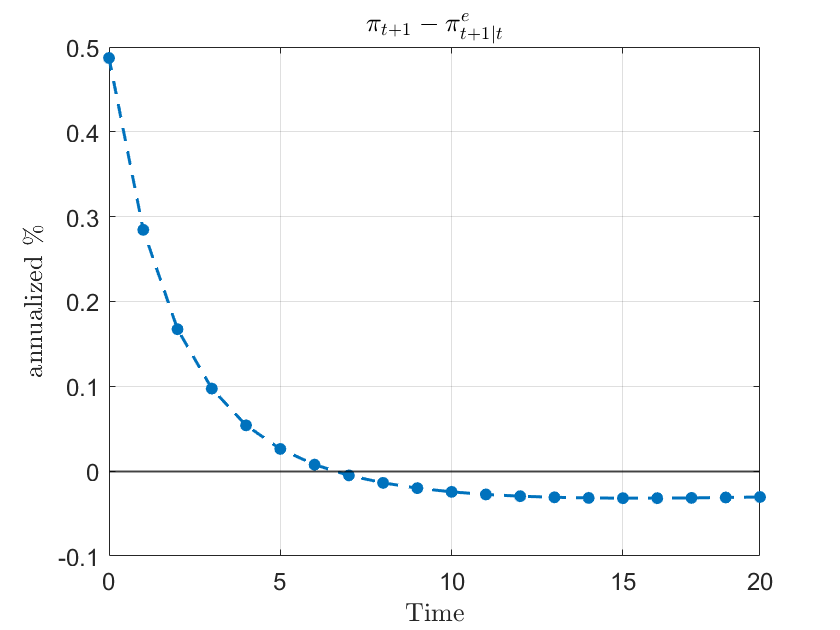} & \includegraphics[scale=0.3]{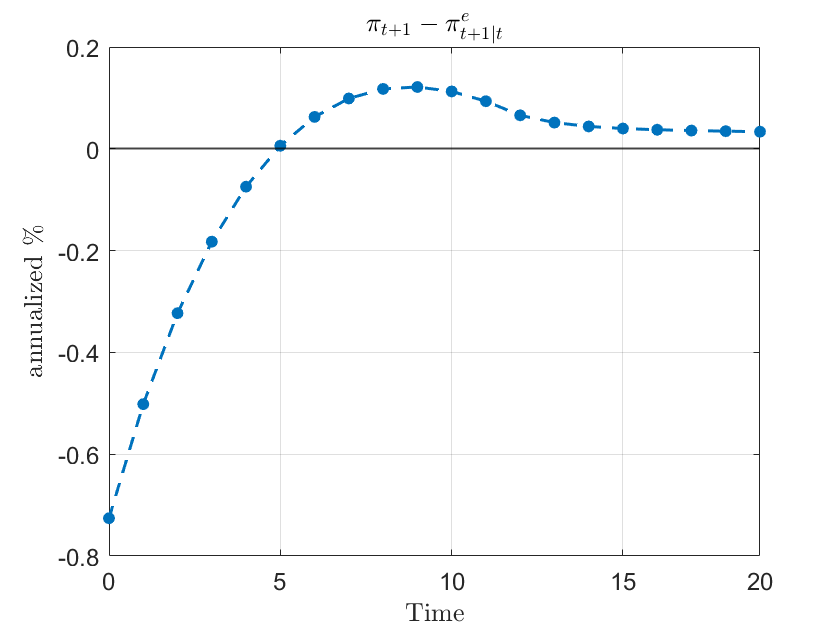}
 \end{tabular}
 \label{fig:irf_fe} 
 \vspace{0.3cm}
 \begin{minipage}{1\textwidth}

 \footnotesize{
Note: This figure shows the impulse-response functions of inflation forecast errors after a three-standard deviation positive (left) and negative (right) natural rate shock. 
 }%
 \end{minipage}

\end{figure}

\citet{angeletos2021imperfect} propose a new test of models that deviate from FIRE. Namely, that expectations should initially underreact but overshoot eventually. A straightforward way to test this is to look at the model-implied impulse response functions of the forecast error, $\pi_{t+1} - \pi^e_{t+1|t}$, to an exogenous shock. Figure \ref{fig:irf_fe} shows these IRFs. The left panel shows the IRF of the forecast error after a positive natural rate shock and the right panel shows the corresponding IRF to a negative natural rate shock. In both cases, we see an underreaction in expectations, which manifests itself in a positive forecast error after a shock that increases the forecasted variable, and vice-versa following a negative shock. After about 5-6 periods, the forecast error response, however, flips sign. This is exactly the eventual overreaction, mentioned above and documented in \citet{angeletos2021imperfect}. Thus, my model of inflation expectations matches these empirical findings.

\clearpage
\newpage
\subsection{No Random Walk}\label{app:095}
A potential concern with the results stated in Section 3, in particular the \textit{inflation-attention trap} in Figure 2, is that these findings are driven by the random walk assumption in the belief process of the agents. Relaxing the random-walk assumption requires to take a stand on the perceived average inflation. In this case, where I solve the model around the zero inflation steady state, this is quite innocuous. But later on, when I focus on Ramsey optimal policy, this cannot be done anymore without distorting the results, in the sense that agents might have a mean bias. 

\begin{figure}[ht]
\caption{Impulse Response Functions to a Negative Natural Rate Shock}
\centering 
\begin{tabular}{cccc}  
\multicolumn{2}{c}{\includegraphics[scale=0.3]{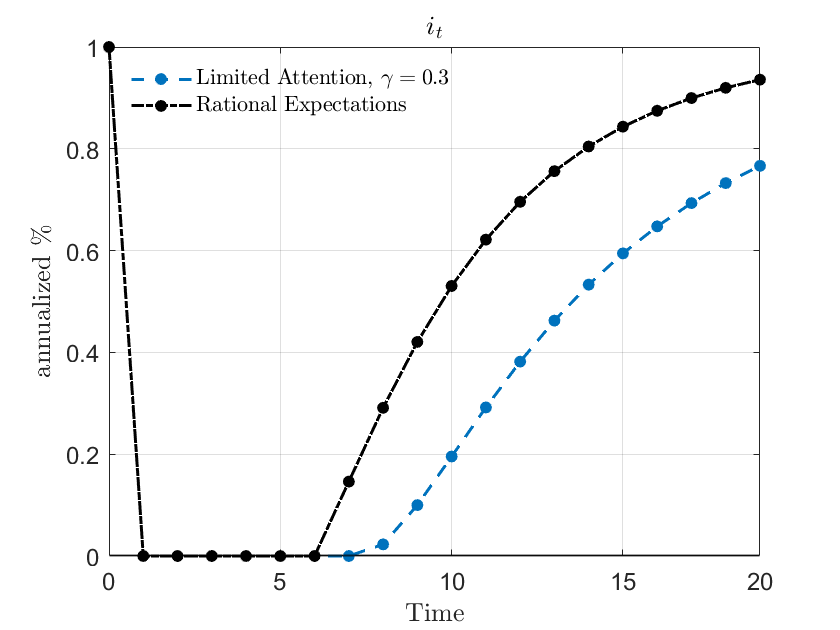}} & \multicolumn{2}{c}{\includegraphics[scale=0.3]{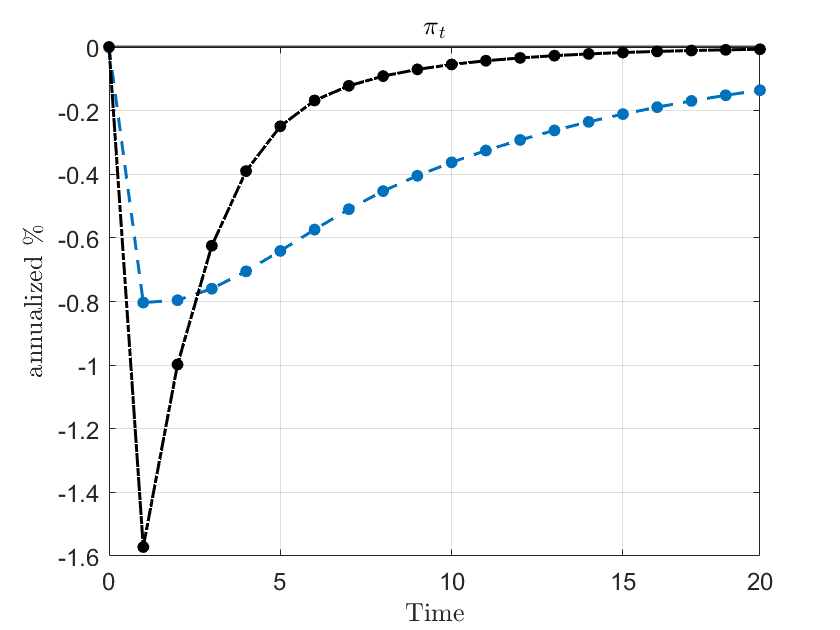}} 
\\ \multicolumn{2}{c}{\includegraphics[scale=0.3]{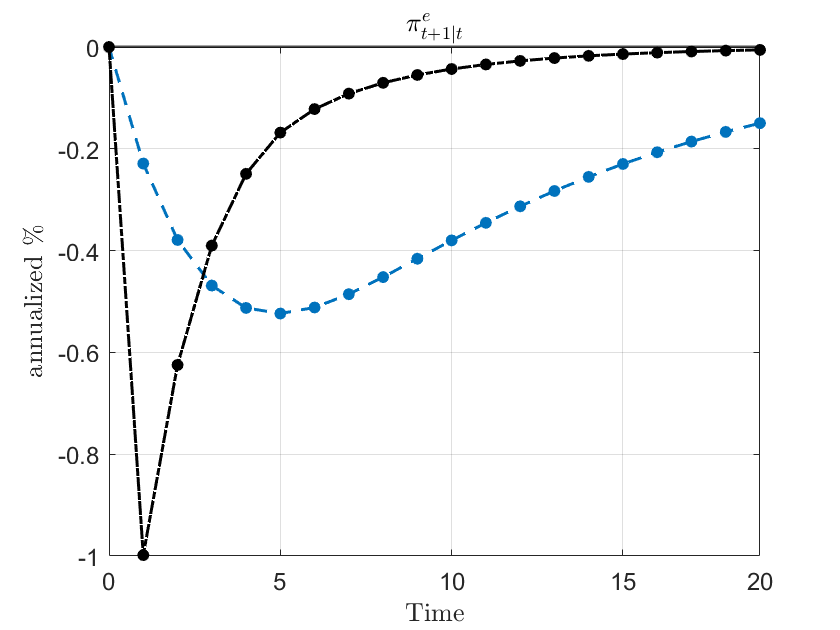}}
&  \multicolumn{2}{c}{ \includegraphics[scale=0.3]{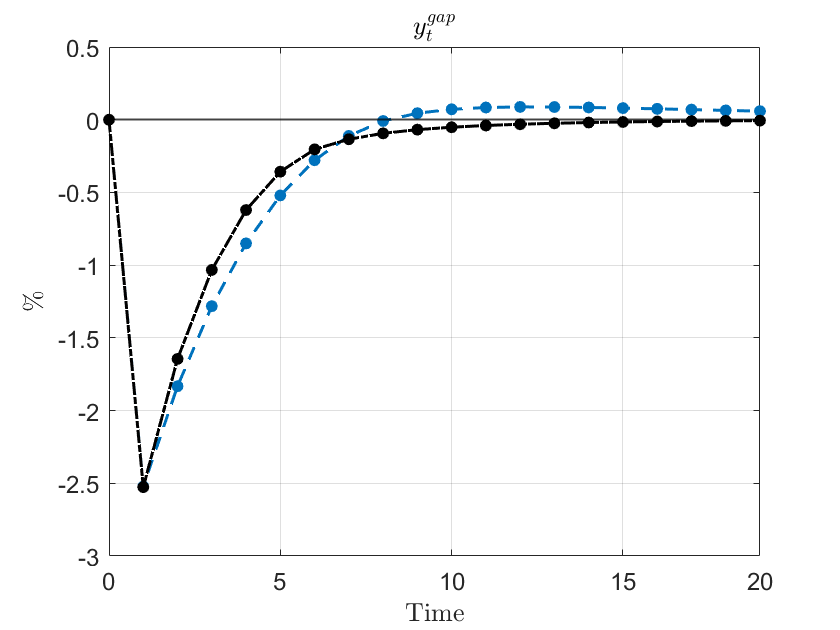}}
 \end{tabular}%
 \label{fig:irf_rn095}
 \par
\vspace{0.3cm}
 \begin{minipage}{1\textwidth}

 \footnotesize{
Note: This figure shows the impulse-response functions of the nominal interest rate (upper-right panel), inflation (upper-left panel), inflation expectations (lower-right) and the output gap (lower-left) to a negative natural rate shock of three standard deviations. The blue-dashed lines show the case for the limited-attention model and the black-dashed-dotted lines for the rational expectations model. Everything is in terms of percentage deviations from the respective steady state levels, expect the nominal rate is in levels.
 }%
 \end{minipage}

\end{figure}

Figure \ref{fig:irf_rn095} shows the same impulse response functions as reported in Figure 2 for the case of $\rho_{\pi} = 0.95$ and an average inflation of 0. We see a similar pattern, even though somewhat less pronounced. Inflation is persistently lower under limited attention due to slowly-adjusting inflation expectations. Expectations are updated even more sluggishly when $\rho_{\pi} < 1$. Further, this also dampens the initial response in inflation expectations, and thus, of inflation itself. Therefore, the attention trap is somewhat mitigated and the economy escapes the lower bound faster than with $\rho_{\pi} = 1$. Nevertheless, the nominal interest rate is low for longer due to the slow recovery of inflation.

\paragraph{Optimal policy with a bias in inflation expectations.}
In the main analysis, I have assumed that agents believe that inflation follows a random walk. Under this assumption, inflation expectations and inflation coincide on average. In the following, I relax this assumption and assume that the perceived persistence parameter is less than 1, $\rho_{\pi} < 1$. As discussed earlier, this yields the following inflation-expectations formation
\begin{equation}
\pi^e_{t+1|t} = (1-\rho_{\pi})\bar{\pi} + \rho_{\pi}\pi^e_{t|t-1}+\rho_{\pi}\gamma\left(\pi_t - \pi^e_{t|t-1}\right), \notag
\end{equation}
where $\bar{\pi}$ captures the long-run expectations of the agent. I set $\rho_{\pi} = 0.95$ and compare economies with different $\bar{\pi}$, namely $\bar{\pi}\in \{0\%, 2\%, 4\%\}$ (annualized).

\begin{figure}[ht]
\caption{Mean Bias, Optimal Inflation Target and Welfare}
\centering    
\centering    
\begin{tabular}{cc}
(a) Optimal Inflation Target & (b) Welfare \\
\includegraphics[scale=0.3]{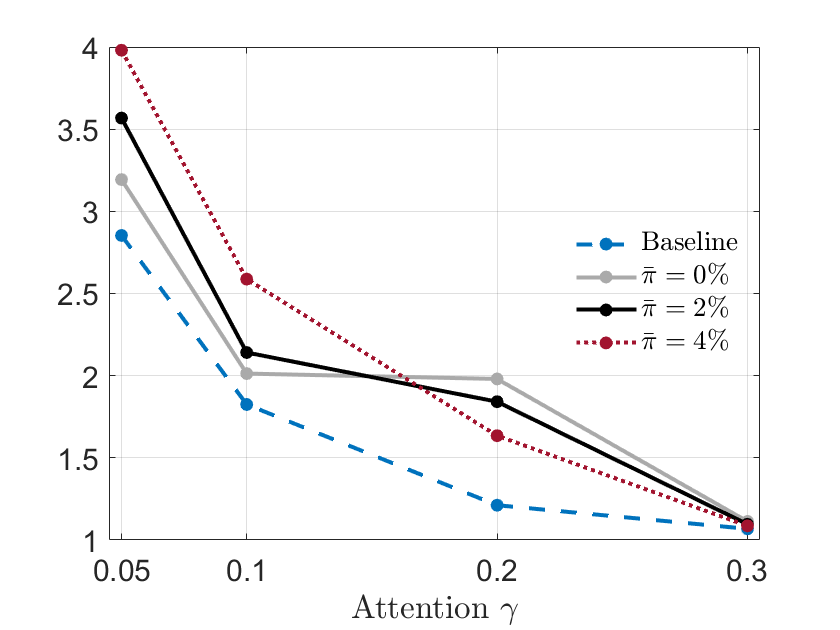}  & \includegraphics[scale=.3]{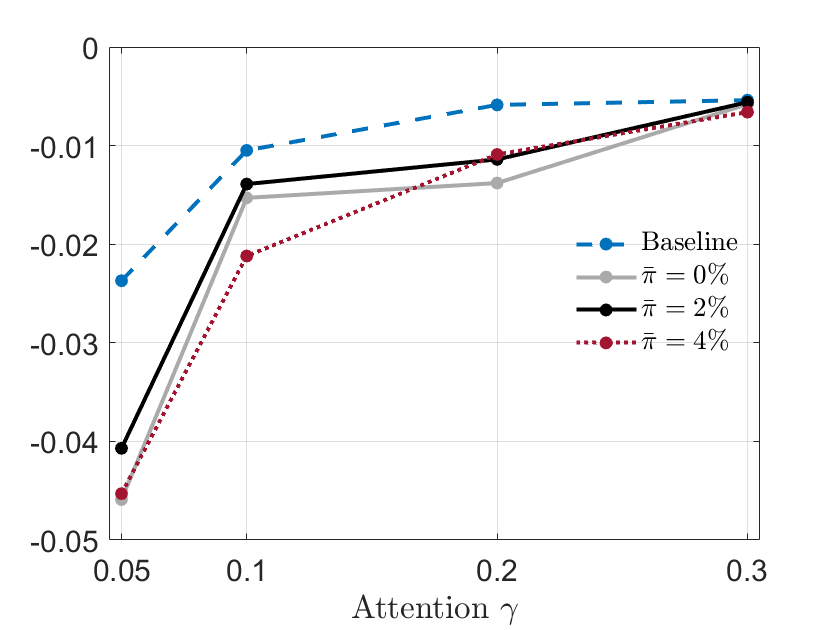}
\end{tabular}
\vspace{0cm}\\ 
 \begin{minipage}{\textwidth}
 \footnotesize{
  Notes: This figure shows the average inflation rate under Ramsey optimal policy (left panel) and welfare (12) (right panel) under Ramsey optimal policy for different levels of attention and different mean beliefs, $\bar{\pi}$. The blue-dashed lines show the results for the case with $\rho_{\pi} = 1$ (which is the baseline case), the gray-dashed-dotted lines show the results for $\rho_{\pi} = 0.95$ and $\bar{\pi} = 0\%$, the black-solid lines for $\rho_{\pi} = 0.95$ and $\bar{\pi} = 2\%$, and the red-dotted lines for $\rho_{\pi} = 0.95$ and $\bar{\pi} = 4\%$.}%
 \end{minipage}
\label{fig:bias}
\end{figure}

Figure \ref{fig:bias} shows the optimal inflation target (left panel) and welfare (12) (right panel) under Ramsey optimal policy for different levels of attention and different mean beliefs, $\bar{\pi}$. The blue-dashed lines show the results for the case with $\rho_{\pi} = 1$ (which is the baseline case discussed above), the gray-dashed-dotted lines show the results for $\rho_{\pi} = 0.95$ and $\bar{\pi} = 0\%$, the black-solid lines for $\rho_{\pi} = 0.95$ and $\bar{\pi} = 2\%$, and the red-dotted lines for $\rho_{\pi} = 0.95$ and $\bar{\pi} = 4\%$.

We see that introducing a mean bias in general leads to an increase in the optimal inflation target and additional welfare losses, independent of $\bar{\pi}$. This mainly comes from the fact that $\rho_{\pi}$ is now below 1, which dampens the degree of updating captured by $\gamma$. Thus, once the economy gets stuck at the ELB and the policymaker tries to decrease real rates by increasing inflation expectations, actual inflation needs to increase more strongly. Therefore, a lower $\rho_{\pi}$ can exacerbate attention traps when they occur. 

Interestingly, the relationship between the optimal target and $\bar{\pi}$ is non-monotonic in the level of attention. While, for example, at $\gamma = 0.2$, the optimal target is highest at $\bar{\pi} = 0\%$, it is highest at $\bar{\pi} = 4\%$ when $\gamma = 0.05$. To understand this, we can write the unconditional average inflation expectations as
\begin{equation}
\pi^e = \frac{(1-\rho_{\pi})\bar{\pi} + \rho_{\pi}\gamma\pi}{1-\rho_{\pi}(1-\gamma)}. \label{avg_pie}\notag
\end{equation}
The following Lemma sheds light on how $\bar{\pi}$ matters for average inflation expectations and how this depends on the level of attention, $\gamma$.
\begin{lemma}\label{lemma_bias}
For the case $\rho_{\pi} = 1$, average inflation expectations move one-for-one with average inflation, independent of $\gamma$:
\begin{equation}
\pi^e = \pi.  \notag
\end{equation}
For the case $0<\rho_{\pi} < 1$, average inflation expectations move less than one-for-one with average inflation
\begin{equation}
0 < \frac{\partial \pi^e}{\partial \pi} = \frac{\rho_{\pi}\gamma}{1-\rho_{\pi}(1-\gamma)} < 1,\notag
\end{equation}
and the strength of this dependency increases with $\gamma$
\begin{equation}
\frac{\partial ^2\pi^e}{\partial\pi\partial\gamma} > 0.\notag
\end{equation}
Average inflation expectations move less than one-for-one with $\bar{\pi}$
\begin{equation}
0 < \frac{\partial \pi^e}{\partial \bar{\pi}} = \frac{(1-\rho_{\pi})}{1-\rho_{\pi}(1-\gamma)} < 1,\notag
\end{equation}
and the strength of this dependency decreases with $\gamma$
\begin{equation}
\frac{\partial ^2\pi^e}{\partial\bar{\pi}\partial\gamma} < 0. \notag
\end{equation}
\end{lemma}
So, as attention falls, there are several opposing forces at work. On the one hand, the effect of $\bar{\pi}$ on average inflation expectations becomes stronger and thus, also exerts more pressure on actual inflation via the Phillips Curve. On the other hand, increasing the inflation target---average inflation---has a smaller effect on average inflation expectations at low levels of attention. Thus, to increase inflation expectations in this case, the inflation target needs to increase more strongly, which is of course costly. Comparing the optimal inflation targets in Figure \ref{fig:bias}, we see that at low levels of attention the first effect dominates. If $\bar{\pi}$ is relatively high, the inflation target is high. 

\clearpage

\subsection{Figures to Section 4.3.5}\label{app:fullatt}
\begin{figure}[h]
\caption{Full Attention, $\gamma = 1$}
\centering    
\centering    
\begin{tabular}{c}
(a) Optimal Inflation Target \\
\includegraphics[scale=0.19]{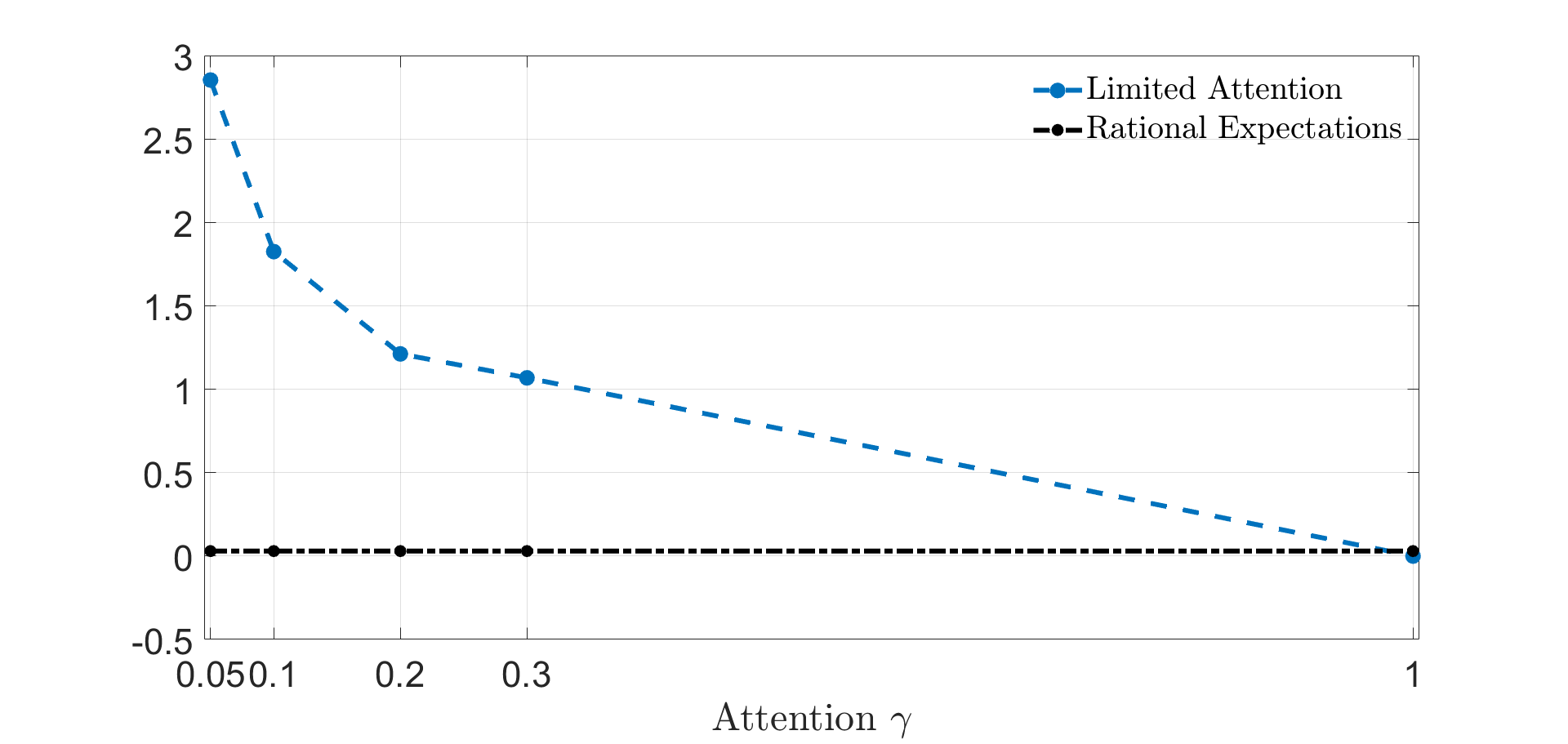}  \\
(b) Inflation Volatility \\
\includegraphics[scale=0.19]{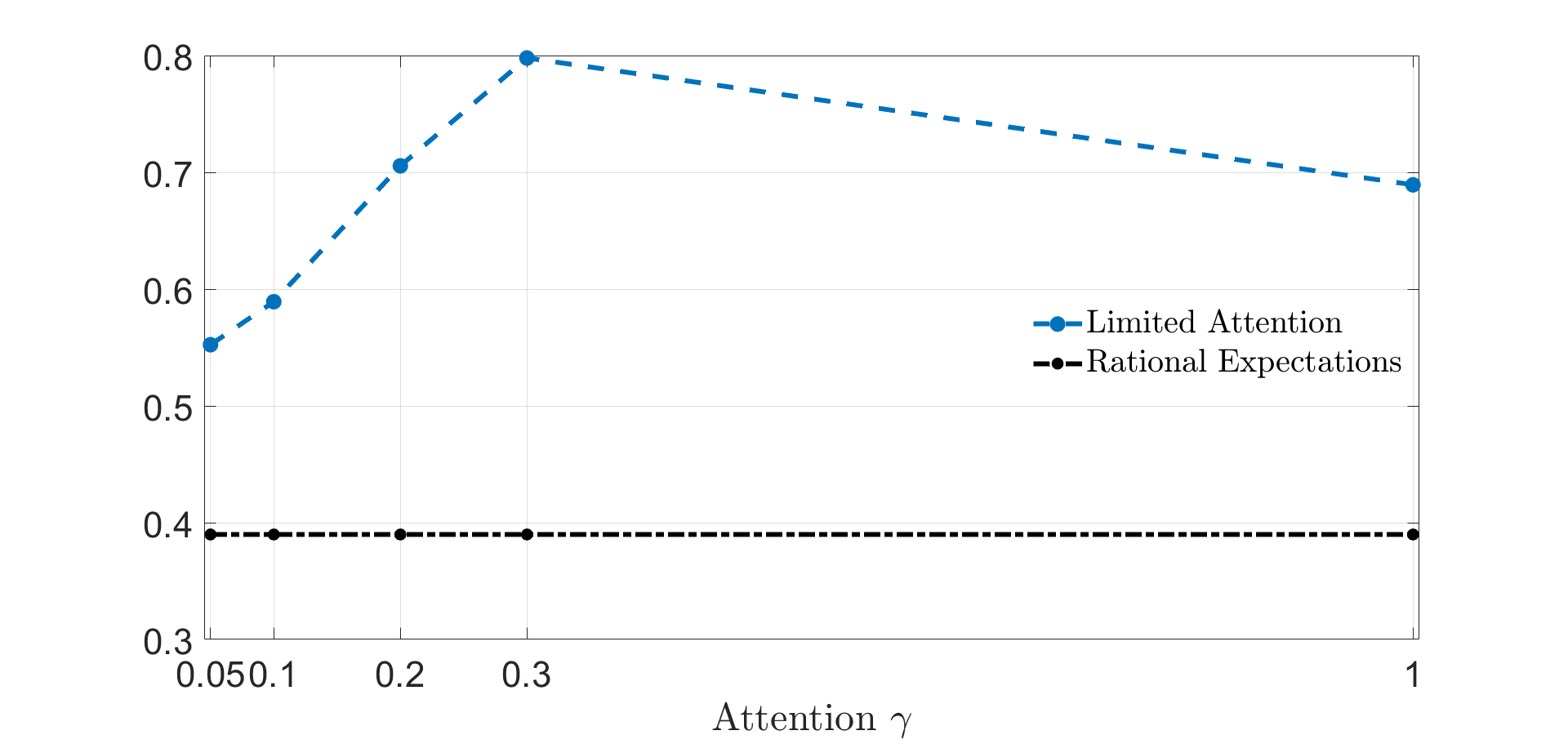}  \\
(c) Welfare \\
\includegraphics[scale=0.19]{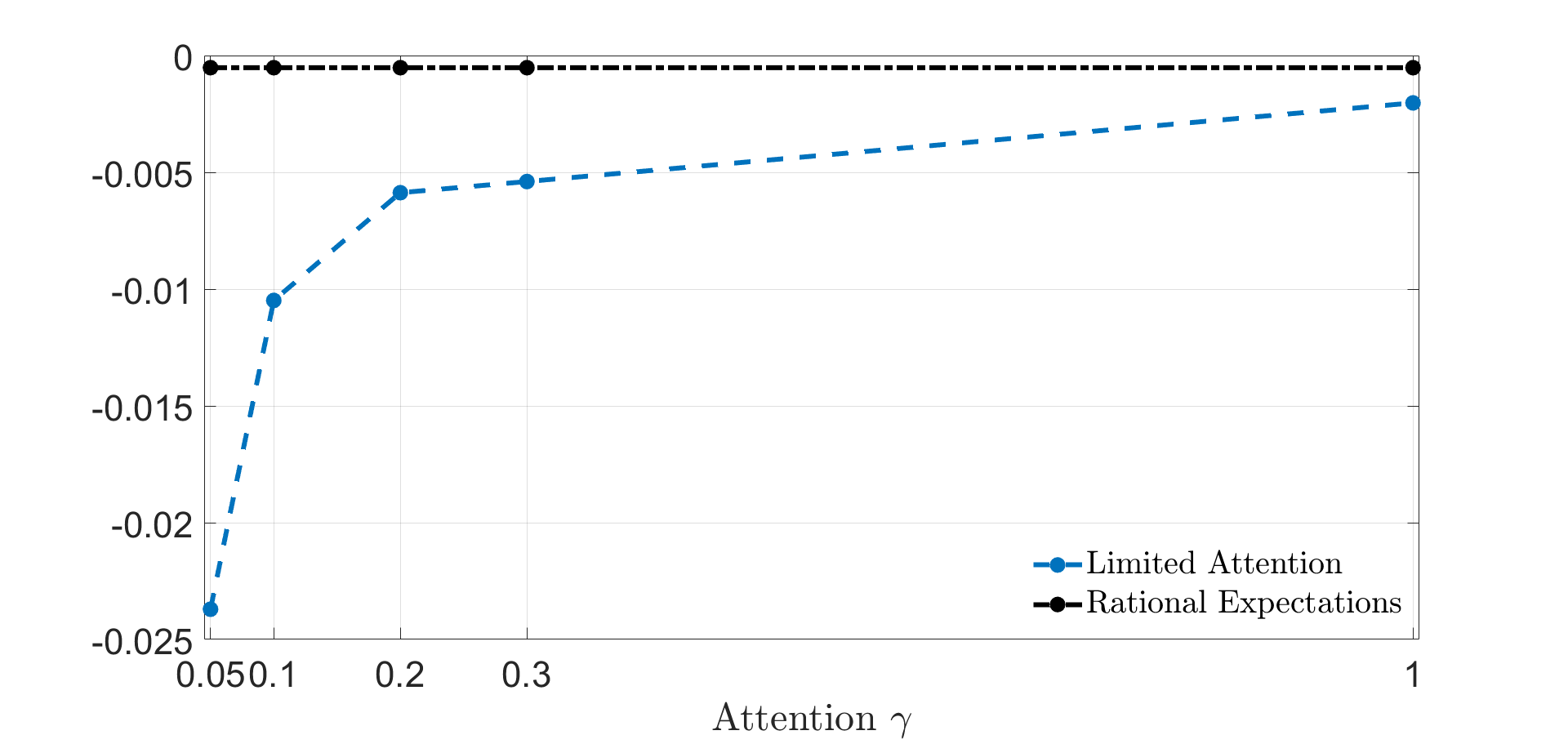}  \\

\end{tabular}
\vspace{0cm}\\ 
 \begin{minipage}{\textwidth}
 \footnotesize{
 Notes: This figure shows the optimal inflation target (panel (a)), inflation volatility (panel (b)) and welfare (panel (c)) under Ramsey optimal policy for different levels of attention, including full attention, i.e., $\gamma = 1$ and compares it to the full-information rational expectations counterparts (black-dashed-dotted lines).}%
 \end{minipage}
\label{fig:fullatt}
\end{figure}

\clearpage
\newpage



\end{document}